\documentclass[aps,amsmath,amssymb,nofootinbib]{revtex4}
\usepackage{graphicx,color}
 \graphicspath{{./}{./figures/}}
\usepackage{dcolumn}
\usepackage{bm}

\definecolor{labelkey}{cmyk}{.4,.2,0,0}

\newcommand{\be}{\begin{equation}}
\newcommand{\ee}{\end{equation}}
\newcommand{\bea}{\begin{eqnarray}}
\newcommand{\eea}{\end{eqnarray}}

\newcommand{\nn}{\nonumber }
\newcommand{\Tr}{{\rm Tr}}

\newcommand{\st}{{\sf t}}
\newcommand{\sP}{{\sf P}}

\begin{document}

\title{Exact solution for a random walk in a time-dependent 1D random environment: \\
the point-to-point Beta polymer}

\author{Thimoth\'ee Thiery and Pierre Le Doussal} \affiliation{CNRS-Laboratoire
de Physique Th{\'e}orique de l'Ecole Normale Sup{\'e}rieure, PSL Research University, 24 rue
Lhomond,75231 Cedex 05, Paris, France}

\begin{abstract}
We consider the Beta polymer, an exactly solvable model of directed polymer on the square lattice,
introduced by Barraquand and Corwin (BC) in \cite{BarraquandCorwinBeta}. We study the statistical properties of its point to point partition sum. The problem is equivalent to a model of a random walk in a time-dependent (and in general biased) 
1D random environment.
In this formulation, we study the sample to sample fluctuations of the transition probability distribution function
(PDF) of the random walk. 
Using the Bethe ansatz we obtain exact formulas for the integer moments, and Fredholm determinant formulas for the Laplace transform of the directed polymer partition sum/random walk transition probability. The asymptotic analysis of these formulas at large time $t$ is performed both (i) in a diffusive vicinity, $x \sim t^{1/2}$, of the optimal direction 
(in space-time) chosen by the random walk, where the fluctuations of the PDF are found to be Gamma distributed; (ii) in the large deviations regime, $x \sim t$, of the random walk, where the fluctuations of the logarithm of the PDF are found to grow with time as $t^{1/3}$ and to be distributed according to the Tracy-Widom GUE distribution. Our exact results complement those of BC
for the cumulative distribution function of the random walk in regime (ii), and in regime (i) they unveil a
novel fluctuation behavior.  We also discuss the crossover regime between (i) and (ii), 
identified as $x \sim t^{3/4}$. Our results are confronted to extensive numerical simulations of the model.
\end{abstract}

\maketitle

\section{Introduction and main results}

\subsection{Overview}

Random walks in random media is a subject of great interest in physics and mathematics. A
lot of works have been devoted to the case of time-independent quenched media
especially in the context of anomalous diffusion (see \cite{BouchaudGeorges1990} for a review, \cite{FisherLeDoussalMonthus2001,solomon1975}). The case of a time-dependent random medium,
with short range correlations both in space and time, has attracted less attention in physics in this context,
mainly since diffusive behavior of the walk at large time is expected in that case. 
However the problem is still non-trivial and can exhibit interesting properties.
For example, the trajectories of a set of identical walkers diffusing independently in the {\it same
realization} of the random environment, exhibit non-trivial space-time correlations, e.g.
typically they tend to stick together. This, and other properties, such as large deviations, have been
studied recently in mathematics \cite{Schertz,RAS,RASY,Deuschel}. On the other hand, there has been much work of the problem of directed polymers (DP), i.e. 
the statistical mechanics of directed paths in a short-range correlated random potential pioneered in \cite{Kardar1987}.
In this framework, recent outstanding progresses have been achieved, notably thanks to
the discovery of exactly solvable models on a square lattice in $D=1+1$ dimension.
This has allowed to put forward a remarkable universality in the DP problem,
connected to the 1D Kardar-Parisi-Zhang (KPZ) universality class \cite{KPZ}
(for review see e.g. \cite{SpohnReview,CorwinReview,HalpinReview}), in particular the emergence 
of the universal Tracy-Widom distributions \cite{TW1994} in the large scale fluctuations of
the DP free energy. These integrable models include last passage percolation with geometric weights \cite{Johansson2000}, 
the so-called log-Gamma polymer \cite{logsep1,usLogGamma,logsep2,logboro}, the Strict-Weak polymer \cite{StrictWeak,StrictWeak2}, the Inverse-Beta polymer \cite{usIBeta}, the Bernoulli-Geometric polymer \cite{ThieryStationnary}
and the Beta polymer \cite{BarraquandCorwinBeta}. Among those the Beta polymer, introduced and first studied by
Barraquand and Corwin (BC)\footnote{Note that random walk in time-dependent Beta distributed random environment already appeared  in \cite{LeJan}. There the authors notably considered the Beta-TDRWRE on the discrete circle $\mathbb{Z}/(N\mathbb{Z})$ and showed the large scale $N \to \infty$ convergence of the process defined by the motion of independent random walkers in the same environment to so-called Brownian sticky flows on the unit circle.}, has the peculiarity that it can 
also be interpreted as a a random walk in a time dependent random environment (TD-RWRE). 
This is due to a very specific choice of local weight which satifies conservation
of probability. The connection between DPs and TD-RWRE was already remarked in \cite{BalazsRassoulAghaSeppalainen2006}, and more generally we note that the interpretation of statistical mechanics models on special varieties in
parameter space in terms of a stochastic process has been previously used in other 
contexts (for review see e.g. \cite{GeorgesLD,3dIsing} and references therein). The Beta polymer provides
a remarkable first example of an exactly solvable TD-RWRE. We note however that
extracting physical properties from the exact solution still represents a technical challenge, 
as is often the case in integrable systems. Thus the existence of an integrable model
of TD-RWRE is far from being the end of the story, and we expect a variety of
interesting result to come in the future from the analysis of the Beta polymer. 

The goal of this paper is indeed to pursue this program by obtaining
exact results on the sample to sample fluctuations of the probability distribution function (PDF) of the random walk
in a given environment, equivalent to
the point to point polymer partition sum. We thus complement the results of
BC in \cite{BarraquandCorwinBeta}, where the statistics of the half-line to point DP partition sum was studied, equivalent to the
cumulative distribution function (CDF) of the random walk (note that it far from
trivial to relate the two observables). More precisely in \cite{BarraquandCorwinBeta} it was notably shown that in the large deviations regime of the RWRE, the fluctuations of the logarithm of the CDF of the random walk scale as $t^{1/3}$ and are distributed according to the Tracy-Widom GUE distribution. Here we will study the fluctuations of the PDF of the random walk (RW) both in the large deviations regime (close in spirit to the results obtained in \cite{BarraquandCorwinBeta} for the CDF fluctuations) {\it and} in the diffusive regime around the optimal direction chosen by the RWRE.

\subsection{Main results and outline of the paper}

We recall in Sec.~\ref{Sec:Model} the definition of the Beta polymer model with parameters $\alpha,\beta>0$ and introduce our notations for the point to point partition sum, $Z_t(x)$. The latter refers to the partition sum for directed polymers of length $t\in \mathbb{N}$, with starting point $x=0$ and endpoint $x\in \mathbb{N}$ (see Fig.\ref{figlattice} for the coordinate system used in this work). Thanks to the interpretation of the Beta polymer as a random walk in a random environment (RWRE), as detailed below, this point to point partition sum can also be interpreted as a transition probability distribution function (PDF) for a directed RWRE where time is reversed and the starting point (resp. endpoint) of the polymer is the endpoint (resp. starting point) of the RW. Our results can thus be interpreted, and are of interest, using both interpretations. In Sec.~\ref{Sec:BA} we use the coordinate Bethe ansatz to obtain an exact formula (\ref{momentFormula01}) for the integer moments of the partition sum:
\bea \label{intro:momentFormula}
&& \!\!\!\!\!\!\!\!\!\!\!\!\!\!\!\!\!\!\!\!\!\!\!\!\!\!\!\!\!\!\!\!\!  \overline{Z_t(x)^n}  = (-1)^n  \frac{\Gamma(\alpha+\beta+n)}{\Gamma(\alpha + \beta  )} 
\prod_{j=1}^{n}  \int_{- \infty}^{+\infty} \frac{dk_j}{2 \pi}
\prod_{1\leq i < j  \leq n} \frac{(k_i-k_j)^2}{(k_i-k_j)^2 + 1} \prod_{j=1}^{n} 
\frac{(i k_j + \frac{\beta - \alpha}{2} )^t}{(i k_j + \frac{\alpha+\beta}{2})^{1 +x} (i k_j - \frac{\alpha+\beta}{2})^{ 1-x + t} }  \ . 
\eea
Where here and throughout the rest of the paper the overline $\overline{()}$ represents the average over the random environment. Using this formula we obtain in Sec.~\ref{Sec:Diff} a Fredholm determinant formula (\ref{kernel3}) for the Laplace transform of the partition sum: 
\bea
&& \overline{e^{- u Z_t(x)}} = \frac{1}{\Gamma(\alpha + \beta) } 
\int_0^{+\infty} dw w^{-1 + \alpha + \beta } e^{-w} g_{t,x}(u w) \quad , \quad 
g_{t,x}(u) = {\rm Det}\left(I+u \hat K_{t,x}(q_1,q_2)\right) 
\eea 
with
\bea \label{intro:kernel3}
\hat K_{t,x}(q_1,q_2) =  - \frac{2}{\pi} \frac{(1+i q_1 (\alpha-\beta))^{t-x}}{(1+i q_1 (\alpha+\beta) )^{1+t-x}}  \frac{(1+i q_2 (\alpha-\beta))^{x}}{(1-i q_2 (\alpha+\beta))^{1+x}} \frac{1}{2 + i (q_1^{-1}-q_2^{-1})}      \   , 
\eea
where $K_{t,x} : L^{2}( \mathbb{R} ) \to L^{2}( \mathbb{R} ) $. We also obtain other, equivalent Fredholm determinant formulas (\ref{kernel1}) and (\ref{kernel2}). The asymptotic behavior of these results at large $t$ in a given direction $x=(1/2 + \varphi)t$ with $-1/2<\varphi<1/2$ is found to drastically depends on the chosen angle $\varphi$ as we now detail.

\begin{figure}
\centerline{\includegraphics[width=8.5cm]{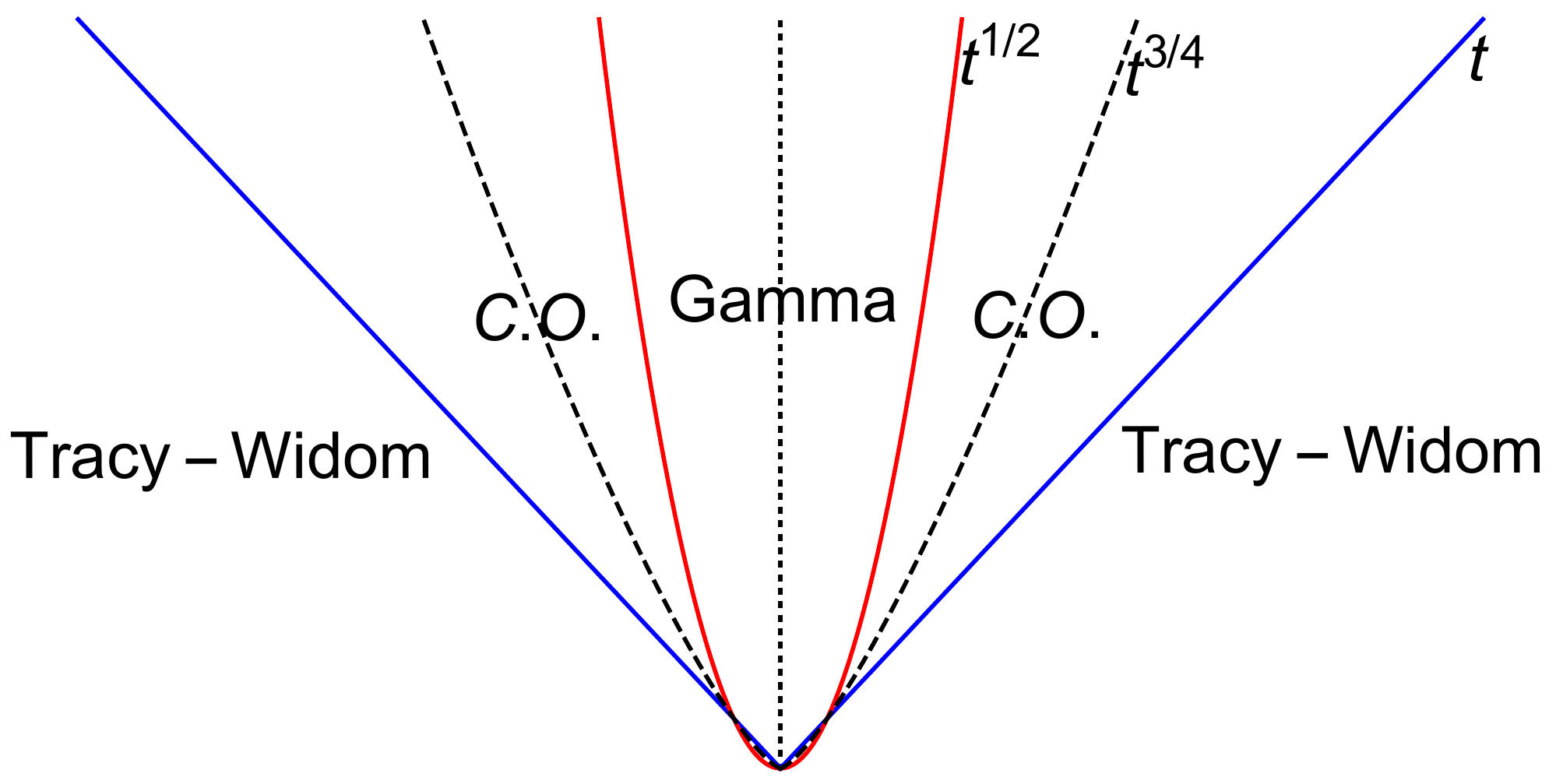}} 
\caption{The different regimes of sample to sample fluctuations of the PDF in the Beta TDRWRE problem around the optimal direction (indicated by a dotted line) for different scaling of the deviation with respect to the optimal direction $\hat x = x- (1/2+\varphi_{opt}) t$. In the diffusive regime $\hat x \sim \sqrt{t}$ the fluctuations of the PDF are Gamma distributed. In the large deviations regime $\hat x \sim t$, fluctuations of the logarithm of the PDF are distributed according to the GUE Tracy-Widom distribution with exponents in agreement with the usual KPZ universality expected in point to point directed polymers problem. These two regimes are connected by a cross-over regime (C.O.) at a scale $\hat x \sim t^{3/4}$.}
\label{fig:cross}
\end{figure}

\medskip

We show that there exists an {\it optimal angle}, $\varphi_{opt} := \frac{\beta -\alpha}{2(\beta + \alpha)}$, defined as the only angle for which $\overline{Z_t(x=(1/2 +\varphi) t)}$ decreases algebraically and not exponentially. It  corresponds to the center of the Gaussian regime in the RWRE interpretation, i.e. the expected direction in space-time chosen by the RW.\\

In a diffusive vicinity around the optimal angle, we show in Sec.~\ref{Sec:Diff} that the fluctuations of an appropriately rescaled partition sum are Gamma distributed. More precisely we show that the rescaled spatial process defined as,
\bea \label{intro:rescaledcalZ}
{\cal Z}_t(\kappa) =  \alpha \sqrt{2 \pi rt} e^{\frac{(r+1)^2}{2r} \kappa^2} Z_t\left(x = (\frac{1}{2} +  \varphi_{opt}(r)) t + \kappa \sqrt{t}  \right)     \   ,
\eea
with $r=\beta/\alpha$, converges at fixed $t$, in the large time limit to a {\it process, constant in $\kappa$, with marginal distribution a Gamma distribution with parameter $\alpha+\beta$}
\bea \label{intro:resrescaledcalZ}
{\cal Z}_{\infty}(\kappa) \sim Gamma(\alpha+ \beta)  \ .
\eea
Let us mention here that we first found this result at the level of one-point observables (i.e. at fixed $\kappa$, ${\cal Z}_{\infty}(\kappa) \sim Gamma(\alpha+ \beta)$) using (\ref{intro:momentFormula}) and (\ref{intro:kernel3}), and were later able to extend it to multi-point correlations in this diffusive regime using results by BC \cite{Barraquand-Corwin-private}. Note that these results would not be expected from the naive application of usual KPZ universality to the point-to-point Beta polymer. This breaking of KPZ universality is due to the RWRE nature of the Beta polymer. Finally, let us also mention that, from the RWRE point of view, although the above results only apply in a diffusive vicinity of a single space-time direction, this spatial region actually contains all the probability in the large time limit.

\medskip

For all other directions $\varphi \neq \varphi_{opt}$ we find in Sec.~\ref{Sec:KPZ} that the fluctuations of an appropriately rescaled free-energy are described by the Tracy-Widom GUE distribution. More precisely we show using a non-rigorous approach that
\begin{equation}\label{intro:asymptoticlim}
\lim_{t \to \infty} Prob\left( \frac{ \log Z_t((1/2+ \varphi) t) + tc_{\varphi}}{\lambda_{\varphi} } <2^{\frac{2}{3}} z \right) = F_2(z) \ ,
\end{equation}
where $F_{2}$ is the cumulative distribution function (CDF) of the GUE Tracy-Widom distribution and the parameters $c_{\varphi}$ and $\lambda_{\varphi} \sim t^{\frac{1}{3}}$ are given by the solutions of a system of transcendental equations in (\ref{SDPeqn}). This regime of fluctuations is the one expected from KPZ universality for point to point directed polymers. In the RWRE interpretation, it corresponds to the fluctuations of the PDF in the large deviations regime. A similar result was proved in \cite{BarraquandCorwinBeta} for the case of the half line to point Beta polymer problem, corresponding to the CDF in the RWRE picture. Quite remarkably we find that (\ref{intro:asymptoticlim}) is formally exactly equivalent to the result of \cite{BarraquandCorwinBeta} if one replaces the half line to point partition sum by the point to point partition sum. In the RWRE picture this shows that the fluctuations of the PDF and of the CDF in the large deviations regime are identical up to order $O(t^{1/3})$ included. 

\medskip
Using the above results, we also briefly discuss in Sec.~\ref{Sec:KPZ} the crossover between the two regimes (\ref{intro:resrescaledcalZ}) and (\ref{intro:asymptoticlim}) and identify the crossover scale as associated with deviations of order $O(t^{3/4})$ around the optimal direction (see Fig.~\ref{fig:cross} for a summary of the different regimes). Finally in Sec.~\ref{Sec:nested}, we compare our results with the approach used by BC \cite{BarraquandCorwinBeta, Barraquand-Corwin-private} and check in Sec.\ref{Sec:Numerics} our results using extensive numerical simulations of the Beta polymer. Two appendix contain technical details.

\section{Model and earlier work} \label{Sec:Model}

\subsection{The Beta polymer}

\begin{figure}
\centerline{\includegraphics[width=12cm]{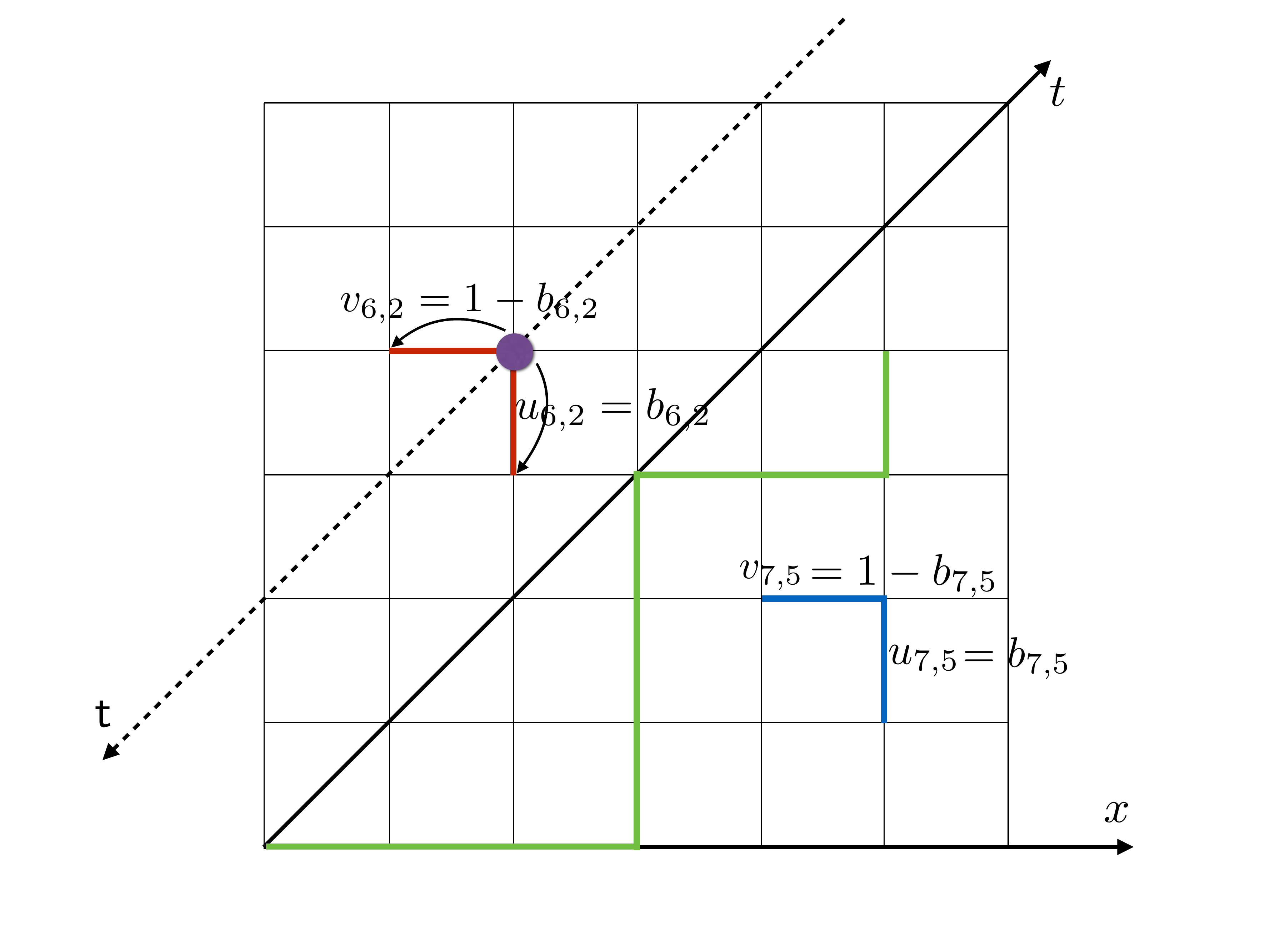}} 
\caption{The Beta polymer and its RWRE interpretation. The random Boltzmann weights live on the edges of the lattice. Blue and Red: two couples of Boltzmann weights. Boltzmann weights leading at the same vertex are correlated: the vertical one $u$ is a Beta RV $b$, whereas the horizontal one $v$ is $1-u$. Green: one admissible directed polymer path from $(t,x)=(0,0)$ to $(t,x) = (9,5)$. In the RWRE interpretation, a particle (purple dot above) performs a directed random walk in time $\st = -t$ and the Boltzmann weights are used as hopping probabilities.}
\label{figlattice}
\end{figure}

Let us now define the point to point Beta polymer problem. We consider the square lattice with coordinates $(t,x)$ as in Fig.~\ref{figlattice}: $x$ is just a regular (euclidean) coordinate along the horizontal axis whereas $t$ goes along the diagonal of the square lattice. On each vertex (i.e. site) of the square lattice lives a random variable (RV) $b_{t,x} \in [0,1]$ that is distributed as a Beta RV with parameters $\alpha, \beta >0$. That is,
\bea \label{distBeta}
b_{t,x} \sim b \sim Beta(\alpha , \beta)  \quad , \quad P^{Beta}_{\alpha,\beta}(b ) db = \frac{\Gamma(\alpha+\beta)}{\Gamma(\alpha) \Gamma(\beta)} b^{-1+\alpha} (1-b)^{-1+\beta} db     \   ,
\eea
where here $P^{Beta}_{\alpha,\beta}$ is the probability distribution function (PDF) of $b$, $\Gamma$ is the Euler's Gamma function and $\sim$ means `distributed as'. We suppose that the different RV on different vertex are uncorrelated. To each vertex $(t,x)$ and RV $b_{t,x}$ we associate two random Boltzmann weights $w_e$ on the vertical and horizontal edges arriving at $t,x$ as 
\bea \label{uv}
&& w_e := u_{t,x} = b_{t,x} \quad \text{if } e= (t-1 , x ) \to (t, x)  \text{ is the vertical edge leading to } (t,x)     \ ,    \nn \\
&& w_e := v_{t,x} = 1-b_{t,x} \quad \text{if } e= (t-1 , x-1 ) \to (t, x)  \text{ is the horizontal edge leading to } (t,x)  \ .
\eea

Hence in the Beta polymer, the Boltzmann weights live on the edges (i.e. bonds) of the square lattice and are correlated only when the edges lead to the same site. Noting generally $u$ (resp. $v$) the Boltzmann weights on vertical (resp. horizontal) edge and $v$, we have $u+v=1$ and $u \sim Beta(\alpha , \beta)$. Given a random environment defined by a drawing of the Boltzmann weights on each edge of the square lattice, the partition sum of the point to point Beta polymer with starting point $(0,0)$ and endpoint $(t,x)$ is defined as
\bea
Z_t(x) = \sum_{\pi : (0,0) \to (t,x)} \prod_{e \in \pi }w_e .
\eea
Where here the sum is over all directed (up/right) paths $\pi$ from $(0,0)$ to $(t,x)$: such a path can only jump to the right following the edge $(t,x) \to (t+1,x+1)$ (in which case the encountered Boltzmann weight is $v_{t+1,x+1}$) or upward following the edge $(t,x) \to (t+1,x)$ (in which case the encountered Boltzmann weight is $u_{t+1,x}$). Equivalently, the partition sum can be defined recursively as for $t \geq 0$,
\bea \label{Zrec}
&& Z_{t+1}(x) = u_{t+1, x} Z_t(x) + v_{t+1, x} Z_t(x-1) \nn \\
&& Z_{t=0}(x) = \delta_{x,0}    .
\eea

\subsection{Relation to a random walk in a random environment} \label{Sec:BetaRWRE}

As already noticed in \cite{BarraquandCorwinBeta}, given a random environment specified by a drawing of the $(u_{tx} , v_{tx} = 1- u_{tx})$, the partition sum of the point to point Beta polymer can also be interpreted as a transition probability for a directed random walk (RW) in the same random environment. We now recall the construction. Let us first introduce a new time coordinate $\st$ as
\bea
\st = -t \ .
\eea
An let us also note
\bea
{\sf p}_{\st ,x} = u_{t,x} \in [0,1] \ .
\eea

A random walk in the Beta distributed random environment is then defined as follows: we note $X_{\st}$ the position of a particle at time $\st$. The particle then performs a RW 
on $\mathbb{Z}$ with the following transition probabilities
\bea
&& X_{\st} \to X_{\st+1}=X_{\st}  \text{ with probability } {\sf p}_{\st ,X_{\st}} =  u_{t= -\st ,X_{\st}}    \   ,   \nn \\
&&  X_{\st} \to X_{\st+1}=X_{\st}-1  \text{ with probability } 1 - {\sf p}_{\st ,X_{\st}} =  v_{t=-\st ,X_{\st}}    \   .   \nn \\
\eea
Hence, the correlations between the Boltzmann weights in the Beta polymer, $u+v=1$, allows to define a RW \footnote{Note that with this choice of coordinates, at each time step $\st \to \st+1$ the particle either stays at the same position (with probability ${\sf p}_{\st ,x}$), or decreases by one unity (with probability $1-{\sf p}_{\st ,x}$). Alternatively one obtains a more symmetric formulation using the coordinate $\tilde{x}= -t +2 x = \st +2x$ and noting $\tilde{{\sf p}}_{\st}(\tilde x) = {\sf p}_{\st ,x = \frac{\tilde x -\st}{2}} $ and $\tilde{X}_{\st} = \st +2 X_t$. The particle then performs a RW on $\mathbb{Z}$ and, at each time step $\st \to \st+1$, the particle either jump by one unity to the right (i.e. $\tilde{X}_{\st+1}=\tilde{X}_{\st}-1$ with probability $\tilde{{\sf p}}_{\st}(\tilde{X}_{\st})$) or to the left (i.e. $\tilde{X}_{\st+1}=\tilde{X}_{\st}+1$ with probability $1-\tilde{{\sf p}}_{\st}(\tilde{X}_{\st})$).} on $\mathbb{Z}$ in a time-dependent random environment defined in terms of the hopping probabilities ${\sf p}_{\st ,x}$ (see Fig.~\ref{figlattice}). In the RWRE language, the partition sum of the Beta polymer is
\bea \label{Eq14}
Z_t(x) = \sP( X_0 = 0 | X_{\st = -t} = x)  \ ,
\eea
the probability, given that a particle starts at position $x$ at time $\st = -t \leq 0$, that it arrives at position $0$ at time $\st = t = 0$. In this language, the recursion equation for the polymer partition sum (\ref{Zrec}) reads:
\bea \label{backwardKo}
&&  \sP(X_0 = 0 | X_{\st -1} = x) = p_{\st-1 x} \sP( X_0 = 0| X_{\st } = x)  
+ (1 - p_{\st-1 x})\sP( X_0 = 0| X_{\st } = x-1)\nn \\
&& \sP( 0,0 | 0, x) = \delta_{x,0}    .
\eea
This equation thus relates the probability for a RW to arrive at the same point starting from different, neighboring points. As such it can be thought of as a Backward equation for the probability $\sP( X_0 = 0 | X_{\st } = x)$.  Note that the starting point of the polymer corresponds in this language to the endpoint of the RW and vice-versa. In the rest of the paper, following this mapping, we will sometimes refer to the random walk in a random environment (RWRE) interpretation of the Beta polymer.

\medskip

{\it Remark:} The RW defined above is a random walk in a one dimensional ($\mathbb{Z}$) time-dependent random environment. Equivalently, it can be thought of as a directed random walk in a two-dimensional ($\mathbb{Z}^2$) quenched static random environment.

\subsection{Relation to the problem and notations of Ref. \cite{BarraquandCorwinBeta}} \label{Sec:LinkWithBC}

The Beta polymer and its RWRE interpretation were introduced in \cite{BarraquandCorwinBeta} where the half-line to point problem was considered. The half-line to point partition sum can be defined recursively as 
\bea \label{ZrecHL}
&& Z^{HL}_{t+1}(x) = u_{t+1 x} Z^{HL}_t(x) + v_{t+1 x} Z^{HL}_t(x-1) \nn \\
&& Z^{HL}_{t=0}(x) = \theta(x)   ,
\eea
where $\theta$ is the Heaviside function ($\theta(0)=1$), and the random environment encoded in the variables $(u_{tx}, v_{tx})$ is the same as before. In the RWRE language,
\bea
Z_t^{HL}(x) = \sP( X_0 \geq 0 | X_{\st = -t} = x)
\eea
is the probability that the particle arrives at a position larger than $0$ at time $\st = 0$, knowing it started from a position $x$ at time $\st  = -t \leq 0$. Hence it is the CDF of the RWRE while the point to point is the PDF, with
respect to the arrival point $X_0$. 

Let us now give here a dictionary between this paper and Ref. \cite{BarraquandCorwinBeta}. In the latter, the half-line to point partition sum was denoted as $\tilde Z (t,n)$. $t$ represents the length of the path and corresponds to our $t$ variable. At each step $n$ can either stay identical or increase by one unit. When $n$ stays constant the polymer encounters a Boltzmann weight distributed as $Beta(\alpha , \beta)$, whereas when $n$ increases the polymer encounters a Boltzmann weight distributed as $1-Beta(\alpha , \beta)$. The parameters $\alpha$ and $\beta$ are also parametrized as $\mu = \alpha$ and $\nu = \alpha+ \beta$. Finally, the initial condition in \cite{BarraquandCorwinBeta} is $\tilde Z(t=0,n) = \theta(x-1) $. From this we conclude that the results obtained in \cite{BarraquandCorwinBeta} for $\tilde Z (t,n)$ identify to results for $Z^{HL}_{t}(x)$ in our notations by either correspondences
\bea
&&( n, \alpha,\beta) \to (x+1 , \alpha , \beta)  \nn \\
&& \text{or, }  \\
&&( n, \alpha,\beta) \to (t-x+1 , \beta , \alpha)  \ . \nn 
\eea
These two choices being related by a reflection along the diagonal of the square lattice.

\section{Bethe Ansatz solution of the Beta polymer} \label{Sec:BA}

In this section we compute the integer moments of the point-to-point Beta polymer using the replica Bethe Ansatz. The following differs from the results of \cite{BarraquandCorwinBeta} by two aspects:

(i) The boundary conditions are different: in \cite{BarraquandCorwinBeta} the chosen boundary conditions were polymers with one end fixed and one end free on a half-line (half-line to point problem, see also Sec.~\ref{Sec:LinkWithBC}).

(ii) In \cite{BarraquandCorwinBeta} the replica Bethe Ansatz solution of the model was made directly on the infinite line, naturally leading to a so-called nested contour integral formula for the moments of the Beta polymer. Here, as in \cite{usIBeta,usLogGamma,we,dotsenko}, we use a different strategy by imposing to the Bethe eigenfunctions artificial boundary conditions on a line of length $L$, and studying the solutions of the associated Bethe equations in the limit $L \to \infty$. In doing so we obtain moments formulas with all integrals on the same contour, and unveil the repulsive nature of the model, a physical aspect that distinguish further this model from the log-Gamma and Inverse-Beta polymer.

\subsection{Bethe ansatz on a line with periodic boundary conditions}

The moments of the Beta polymer random Boltzmann weights are obtained from (\ref{distBeta}) and (\ref{uv}) as
\bea \label{new1}
\overline{u^{n_1} v^{n_2}} =  \frac{(\alpha)_{n_1} (\beta)_{n_2} }{ (\alpha + \beta)_{n_1+n_2} } \ ,
\eea
where here and throughout the rest of the paper the overline $\overline{()}$ denotes the average over disorder. We consider, for $n \geq 1$,
\bea
\psi_t(x_1,\cdots,x_n) := \overline{Z_t(x_1)  \cdots Z_t(x_n)} \ ,
\eea
which is a symmetric function of its arguments. Using the recursion (\ref{Zrec}), one shows \cite{usIBeta} that $\psi_t$ satisfies a linear recursion relation of the form
\bea
\psi_{t+1} (x_1,\cdots,x_n)= \left( T_n \psi_t \right)(x_1,\cdots,x_n) \ , 
\eea
where $T_n$ is a linear operator called the transfer matrix. The precise form of $T_n$ was given in Eq.(13) of \cite{usIBeta} for arbitrary moments $\overline{u^{n_1} v^{n_2}}$. The latter is unimportant for our purpose and we will only use here one of the conclusion of \cite{usIBeta}: the symmetric eigenfunctions of the transfer matrix, that are the symmetric solutions of the spectral problem
\bea
T_n \psi_\mu  = \tilde \Lambda_\mu \psi_\mu \ ,
\eea
 can be obtained using a coordinate Bethe ansatz, with for $x_1 \leq  \dots \leq x_n$, $\psi_{\mu}(x_1, \dots , x_n) = \tilde \psi_{\mu}(x_1, \dots , x_n) := \sum_{\sigma \in S_n} A_{\sigma} \prod_{i=1}^n z_i^{x_{\sigma(i)}}$, 
 and the other sectors obtained using that the function is fully symmetric. 
 The sum is over all permutations of the $n$ variables $z_i$ which parameterize the eigenstates.  
 We note that this fact was also already known from \cite{BarraquandCorwinBeta}. In the notations of \cite{usIBeta}, (see Eq. (18) and (19) there) the $S$-matrix $S(z_i , z_j)$ is given by $S(z_i , z_j) = -\frac{{\sf c}+ {\sf b} z_j+ {\sf a} z_i z_j - z_i}{{\sf c} + {\sf b} z_i+ {\sf a} z_i z_j - z_j}$ with
\bea
{\sf a} = \frac{\overline{u^2} - (\overline{u})^2}{ (\overline{u})(\overline{v})}= \frac{1}{1+\alpha+\beta} \quad , \quad  {\sf b}=\frac{2 \overline{ u v} -(\overline{u})( \overline{v} ) }{(\overline{u} )( \overline{v} )} = \frac{-1+ \alpha+\beta}{1+\alpha+\beta} \quad , \quad {\sf c} =\frac{\overline{v^2} - (\overline{v})^2}{ ( \overline{u} )(\overline{v})}= \frac{1}{1+\alpha+\beta} \ .
\eea
We note that ${\sf a}+{\sf b}+{\sf c}=1$ which is the hallmark 
of a stochastic model with conservation of probability. Here the moment problem
corresponds to a zero-range process as in \cite{povo}. 
We recall that the Bethe ansatz solvability of the two particles problem imposes that the quotient of two amplitudes $A_{\sigma}$ related to each other by the transposition of $i \leftrightarrow j$ is $ \frac{A_{ \dots ji \dots }}{A_{ \dots ij \dots }} = S(z_i,z_j)$. Let us now introduce
\bea 
&& c  = \frac{4}{\alpha+\beta} > 0  \nn \\
&& z_j = e^{ i \lambda_j}   \hspace{0.3 cm},\hspace{0.3 cm} \tilde t_j = - i \cot (\frac{\lambda_j}{2}) =\frac{z_j+1}{z_j-1}  \quad , \quad z_j = -  \frac{1 + \tilde t_j}{1- \tilde t_j} \ .
\eea
Using these notations, the $S$-matrix of the Beta polymer reads
\bea \label{SBeta}
S(z_i , z_j )  = \frac{ 2 \tilde{t}_j - 2 \tilde{t}_i -c}{2 \tilde{t}_j - 2 \tilde{t}_i +c } \ .
\eea 
It has a structure very similar the one of the Inverse-Beta and log-Gamma polymer \cite{usIBeta,usLogGamma} that we recall here for comparison (see Eq.~(45) and (46) in \cite{usIBeta}):
\bea \label{SIBeta}
S^{IB}(z_i , z_j )  = \frac{ 2 t_j - 2 t_i + \bar c}{2 t_j - 2 t_i - \bar c } \quad , \quad t_j  = i \tan(\frac{\lambda_j}{2}) = \frac{z_j-1}{z_j+1} \ . 
\eea
That is, the $S$-matrix of the Beta polymer (\ref{SBeta}) is identical to the $S$-matrix of the Inverse-Beta polymer (\ref{SIBeta}) with the change $\bar c>0 \to - c$ with $c>0$ and $t_j \to \tilde{t}_j$. We will see in the following that this change is responsible for the emergence of a repulsive property of the Bethe ansatz 
for the Beta polymer. Here $c>0$ can be interpreted as a repulsive interaction parameter, while in the Inverse-Beta case $\bar c>0$ was interpreted as an attractive interaction parameter. Apart from this, the similarity between the Bethe ansatz applied to the Inverse-Beta and Beta polymer will help us in finding the Bethe ansatz solution of the Beta polymer using the changes $\bar c \to -c$ and $t_i \to \tilde{t}_i$ (since the $S$-matrix completely controls the form of the Bethe eigenfunction up to a multiplicative constant). In particular, by analogy with the notations used for the Inverse-Beta polymer (see Eq.(47) in \cite{usIBeta}) we will write the Bethe eigenfunctions of the present model as
\begin{equation}\label{brunetansatz2}
\tilde \psi_\mu(x_1,\cdots,x_n) = \sum_{\sigma \in S_n} \tilde A_\sigma \prod_{\alpha=1}^n z_{\sigma(\alpha)}^{x_\alpha}  \hspace{0.3 cm},\hspace{0.3 cm} \tilde A_\sigma = \prod_{1 \leq \alpha < \beta \leq n } (1- \frac{c}{2(\tilde{t}_{\sigma(\alpha)} - \tilde{t}_{\sigma(\beta)})} )  \ .
\end{equation}
Imposing periodic boundary conditions on a line of length $L$, i.e. $\psi_{\mu}(x_1,\cdots , x_j +L ,\cdots,x_n) = \psi(x_1,\cdots,x_n)$ (which is immaterial \cite{usIBeta} in the computation of moments as long as $t <L$) 
leads to Bethe equations of the form (see Eq. (48) of \cite{usIBeta}): 
\begin{equation} \label{BEbeta}
e^{i \lambda_{i}L} = \prod_{1 \leq j \leq n, j \neq i} \frac{2 \tilde t_i- 2 \tilde t_j-c}{2 \tilde t_i-2 \tilde t_j+c} \ .
\end{equation}

\subsection{Resolution of the Bethe equations in the large $L$ limit: repulsion and free particles}

In the large $L$ limit, contrary to the case of the log-Gamma and the Inverse-Beta polymer, the Bethe roots $\lambda_{\alpha}$ of this model are all real and are distributed as for a model of free particles. The particles do not form bound states, also called strings in the Bethe ansatz literature. In this sense the moment problem of the Beta polymer is similar to the repulsive phase of the Lieb-Liniger (LL) model
\footnote{at this stage this is only a formal similarity. In fact we have not
found a continuum limit of the Beta polymer which would identify to the repulsive LL model.}, while the moment problems of the Inverse-Beta polymer (see \cite{usIBeta}), the log-Gamma polymer (see \cite{usLogGamma}) and the continuum directed polymer (see \cite{we}) were similar to the attractive phase. To see this explicitly, let us use a proof by contradiction and consider the possibility of forming a 2-string. The logarithm of the Bethe equation for two particles ($n=2$ of (\ref{BEbeta})) is:
\bea \label{logBE}
\lambda_1 = \frac{2 \pi I_1}{L} - \frac{i}{L} \left( \log(2(\tilde t_1 - \tilde t_2) - c) -\log(2( \tilde t_1 -  \tilde t_2) + c )    \right)   \ ,  \nn \\
\lambda_2 =  \frac{2 \pi I_2}{L} - \frac{i}{L} \left( \log(2(\tilde t_2 - \tilde t_1) - c ) -\log(2(\tilde t_2 - \tilde t_1) +c )    \right)    \ .
\eea
Note that $Im(\lambda_1) + Im(\lambda_2) = 0$ (a property related to the translational invariance of model). Let us consider the possibility of having $Im(\lambda_1) \neq 0$ in the large $L$ limit. Since everything on the right hand side of (\ref{logBE}) is proportional to $1/L$, this means that the $\tilde t_i$ variables must flow exponentially fast to the singularity of the logarithm at $0$. Let us e.g. suppose
\bea \label{logBE2}
2(\tilde t_1 - \tilde t_2) - c = O(e^{-\delta L} )
\eea
with $\delta >0$. Taking the large $L$ limit of (\ref{logBE}), we obtain
\bea
 -Im( \lambda_2) = Im( \lambda_1) \simeq_{L \to \infty}  - \frac{1}{L}( -\delta L) =  + \delta >0  \ ,
\eea
consistency with (\ref{logBE2}) in the $L \to \infty$ limit thus implies
\bea
-2 i (\cot( Re(\lambda_1) + i \delta ) -\cot( Re(\lambda_2) - i \delta ) ) = c >0 .
\eea
However, this last equality cannot be satisfied since $c>0$ and the imaginary part of $cot( x + i y)$ has a sign opposite to the sign of $y$. {\it Hence it is impossible to form a bound state of two particles and $c$ can be interpreted as a repulsive interaction parameter.} Generalizing this phenomenon to arbitrary $n$, the large-$L$ limit of the Bethe equations is particularly simple, namely
\bea
\lambda_i = \frac{2 \pi I_{i}}{L} + O(\frac{1}{L^2}) \ .
\eea
That is, to first order in $1/L$ the particles behave as free particles as already announced. In the following we will parametrize the particle quasi-momentas by the $\tilde t_i$ variables. These are pure imaginary numbers that we write (to maintain the analogy with \cite{usIBeta}, see Eq.(50) there):
\bea
\tilde t_{\alpha} = i \frac{k_\alpha}{2}  \quad , \quad k_\alpha \in \mathbb{R}  \ .
\eea

\medskip

{\it Interpretation of the repulsive nature of the model:}

One interpretation of the repulsive nature of the model can be traced back to the relation
\bea \label{conservationProba}
u + v = 1
\eea
that holds in this model for Boltzmann weight for edges arriving on the same vertex. In this model, if it is favorable for the polymer to travel through the edge carrying the disorder $u$, then it means that $u$ is large. Hence $v$ must be small and it is not favorable for the polymer to travel through the edge carrying $v$. Consider now
two replicas (second moment problem). Clearly the transition $(x-1,x) \to (x,x)$ for these two replica
is not favorable. This can be interpreted as a nearest neighbor repulsion for these two
replicas (i.e particles). Of course the transition $(x-1, x-1) \to (x,x)$ remains favorable 
(on site attraction). The balance between the two processes however seems to favor the
repulsive nature of the model. In the Inverse-Beta (resp. Log-Gamma) polymer, (\ref{conservationProba}) is replaced by $v-u = 1$ (resp. $v=u$) and the model is attractive.

Another way to see this is through the fact that (\ref{conservationProba}) precisely permits to interpret the Beta polymer as a RWRE (in this language (\ref{conservationProba}) is the conservation of probability on each vertex). In this interpretation time is reversed (see Sec.~\ref{Sec:BetaRWRE}) and if two particles stay together at time $\st$ in the random environment, then they are more likely to stay on the same site at time $\st+1$ and the transition $(x, x) \to (x,x-1)$ is not favored. The attraction between particles in the RWRE language becomes a repulsion in the polymer language when time is reversed.

\subsection{Bethe ansatz toolbox}

{\it Scalar product and norm in the large $L$ limit}

Using the already discussed analogy between the Bethe ansatz for the Inverse-Beta and Beta polymer, we easily conclude from Eq.(49) of \cite{usIBeta} that the eigenfunctions of the Beta polymer are orthogonal with respect to the following weighted scalar product:
\bea \label{wps2}
 &&\langle \phi , \psi \rangle = \sum_{ (x_1 , \cdots , x_n ) \in  \{0,\cdots, L-1\}^n } \frac{1}{\prod_{x} \tilde{h}_{ \sum_{\alpha=1}^n \delta_{x,x_\alpha}}} \phi^*(x_1,\cdots,x_n)  \psi(x_1,\cdots,x_n)    \ , \nn \\
 && \tilde h_n= \prod_{ k = 0} ^{ n-1} \frac{4}{ 4+ k c } = (\alpha+\beta)^n \frac{\Gamma(\alpha+\beta)}{\Gamma(\alpha+ \beta + n)} \ .
\eea
We also obtain the formula for the norm of a general eigenstate of $n$ particles in the large $L$ limit (see Eq.(52) of \cite{usIBeta}, here adapted for the only relevant case here, that is for the case without string states)
\bea \label{norme}
||\mu||^2 = \langle \tilde \psi_\mu, \tilde \psi_\mu \rangle = n! L^{n}   \prod_{1\leq i < j  \leq n} \frac{ (k_i-k_j)^2 + c^2 }{(k_i-k_j)^2}  + O(L^{n-1}) \ .
\eea 

{\it Quantization in the large $L$ limit}

The model being repulsive, the sum over eigenstates is computed using the free-particle quantization (similar to Eq.(51) of \cite{usIBeta}):
\bea
\sum_{\lambda_{\alpha} } = \frac{L}{2 \pi } \int_{ - \pi }^{\pi} d \lambda_{\alpha} =  \frac{L}{2 \pi } \int_{ - \infty }^{\infty} \frac{4 d k_{\alpha} }{4 + k_{\alpha}^2} \ .
\eea 

{\it Energy-momentum}

We will also need the eigenvalue associated with the unit translation-operator on the lattice:
\bea
\prod_{\alpha = 1}^n z_\alpha =   \prod_{\alpha = 1}^n  \frac{2 + i k_{\alpha} }{  -2 + i k_{\alpha}  } \ ,
\eea
as well as the eigenvalue associated with the translation in time
$\tilde \Lambda_{\mu} = \prod_{i=1}^n (\bar u + \bar v z_i^{-1})$ which gives 
\bea
\tilde \Lambda_{\mu} = \prod_{\alpha =1}^{n}  ( \frac{\alpha}{\alpha+\beta} +  \frac{\beta}{\alpha + \beta} (-1) \frac{1- \tilde t_{\alpha}}{1+\tilde t_{\alpha}}) = \prod_{\alpha =1}^{n} \left(  \frac{2(\alpha - \beta) + i(\alpha + \beta) k_{\alpha}}{2(\alpha + \beta) + i (\alpha +  \beta) k_{\alpha} }   \right)  \ .
\eea

\subsection{A large contour-type moment formula}

We have now all the ingredients to compute the integer moments of the partition sum. The initial condition is:
\bea
Z_{t=0}(x) = \delta_{x,0}  \Longrightarrow \psi_{t=0}(x_1,\cdots,x_n) = \prod_{i=1}^n \delta_{x_i, 0} \ .
\eea
We use the scalar product (\ref{wps2}) to decompose it onto the Bethe eigenstates:
\begin{equation}
\psi_{t}(x_1 , \cdots , x_n) = \sum_{ \mu} \frac{ n!  }{  \tilde h_n  || \psi_\mu || ^2}  (\tilde \Lambda_{\mu})^t \psi_\mu (x_1 , \cdots , x_n) \ .
\end{equation}
In particular,
\bea
\overline{Z_t(x)^n} = \sum_{ \mu} \frac{ ( n!)^2 }{\tilde h_n || \psi_\mu || ^2}  (\tilde \Lambda_{\mu})^t  \left( \prod_{\alpha=1}^n z_{\alpha} \right)^x
 \ .
\eea
Replacing in this expression each terms by its value in the large $L$ limit, one obtains:
\bea
&& \!\!\!\!\!\!\!\!\!\!\!\!\!\!\!\!\!\!\!\!\!\!\!\!\!\!\!\!\!\!\!\!\! \overline{Z_t(x)^n} = \frac{L^n}{(2 \pi)^n} \frac{1}{n!} \prod_{i=1}^n  \int_{-\infty}^{+\infty} \frac{4 d k_i }{4 + k_i^2}  \frac{ (n!)^2}{ (\alpha + \beta)^n } \frac{\Gamma(\alpha+\beta+n)}{\Gamma(\alpha+ \beta )}  \frac{1}{n! L^n} \prod_{1\leq i < j  \leq n} \frac{(k_i-k_j)^2 }{(k_i-k_j)^2 + c^2} \nn \\
&&
\prod_{j=1}^n \left(  \frac{2(\alpha - \beta) + i(\alpha + \beta) k_{j}}{2(\alpha + \beta) + i (\alpha +  \beta) k_{j} }   \ \right)^{t}  \left(  \frac{2 + i k_{j} }{  -2 + i k_{j}  }  \right)^x \ . 
\eea
Rescaling $k \to - \frac{4}{\alpha + \beta } k = -c k $ and rearranging, we obtain
\bea \label{momentFormula01}
&& \!\!\!\!\!\!\!\!\!\!\!\!\!\!\!\!\!\!\!\!\!\!\!\!\!\!\!\!\!\!\!\!\!  \overline{Z_t(x)^n}  = (-1)^n  \frac{\Gamma(\alpha+\beta+n)}{\Gamma(\alpha + \beta  )} 
\prod_{j=1}^{n}  \int_{- \infty}^{+\infty} \frac{dk_j}{2 \pi}
\prod_{1\leq i < j  \leq n} \frac{(k_i-k_j)^2}{(k_i-k_j)^2 + 1} \prod_{j=1}^{n} 
\frac{(i k_j + \frac{\beta - \alpha}{2} )^t}{(i k_j + \frac{\alpha+\beta}{2})^{1 +x} (i k_j - \frac{\alpha+\beta}{2})^{ 1-x + t} }  \ . 
\eea 

Which is our main result for the positive integer moments of the point-to-point partition sum of the Beta polymer. Let us now make a remark and introduce a more general formula.

\medskip

{\it Remark and multi-points moment formula} \label{subsec:comments}

Let us first note that the formula (\ref{momentFormula01}) has a different structure compared to other formulas obtained using the replica Bethe ansatz on other exactly solvable models of DP (by this we mean using the same replica Bethe ansatz as the one used in this paper, i.e. using periodic boundary conditions and solving the Bethe equations in the large $L$ limit as e.g. in \cite{usIBeta,usLogGamma,we,dotsenko}). Indeed, in other cases, one obtains a formula which contains a discrete summation over strings configurations, corresponding to the sum over all eigenstates of an attractive quantum problem (see e.g. Eq.(60) of \cite{usIBeta}) and that is expressed as the sum of different $n_s$-dimensional integrals with $1 \leq n_s \leq n$.  Here the repulsive nature of the model leads to a simpler formula since the $n^{th}$ moment is expressed as a {\it single} $n$-dimensional integral. We note however that such type of moments formulas already appeared in the literature. In a more general context (see e.g. in the context of Macdonald processes \cite{BorodinMacdo}), it is usual to encounter moments formulas of one of the following three types:
\begin{enumerate}
\item{The first type consists in expressing the $n^{th}$ moments of an observable of a stochastic process as one $n$-dimensional nested-contours integral with $n$ contours chosen to avoid some poles of the integrand and arranged in a so-called nested fashion (see Fig.~\ref{fig:nestedcontours} at the end of the paper for an example). In the following we will refer to this type of formula as the nested-contours type.}
\item{In the second type of formula all the contours of the nested-contours integral formula are deformed onto the largest one. If there are no poles encountered along this deformation, one then obtains a formula with $n$ integrals on the same contour as in (\ref{momentFormula01}). In the following we will refer to this type of formula as the large-contours type }
\item{In the third type of formula all the contours of the nested contours integral formula are deformed onto the smallest one. In doing this operation, one generally encounters multiple poles of the integrand and one has to keep track of the resulting residues. This process of ``book-keeping'' residues then lead to formula for the $n^{th}$ moment as a sum of $n_s$-dimensional integrals with $1\leq n_s \leq n$ integration variables on one contour, corresponding to the summation over strings configurations evoked above. In the following we will refer to this type of formula as the small-contours or string type.} 
\end{enumerate}
The specificity of the replica Bethe ansatz applied to the Beta polymer case is thus the fact that we directly obtained a large-contours formula for the moments (see (\ref{momentFormula01})), rather than a string type formula as could have been naively expected. For the Beta-polymer half-line to point problem considered in \cite{BarraquandCorwinBeta}, Barraquand and Corwin obtained directly a nested contour integral formula for the moments of the partition sum. They also adapted their approach to study the moments of the point-to-point problem and obtained a nested-contours integral formula for the moments $\overline{Z_t(x)^n}$ which was recently brought to our attention by them \cite{Barraquand-Corwin-private}. We show in Sec.~\ref{Sec:nested} how their work compares to ours. In particular we establish the equivalence between our formula (\ref{momentFormula01}) and a nested-contours type formula. Before going further, let us also mention here that with their approach it is also possible to derive rigorously another formula of type $2$ valid for the {\it multi-points moments} of the partition sum that we now display: if $0 \leq x_1 \leq \cdots \leq x_n $, then,
\bea \label{momentFormula07}
&& \!\!\!\!\!\!\!\!\!\!\!\!\!\!\!\!\!\!\!\!\!\!  \overline{Z_t(x_1) \cdots Z_t(x_n)}  =  (-1)^n \frac{\Gamma(\alpha+\beta+n)}{\Gamma(\alpha + \beta  )} 
\prod_{j=1}^{n}  \int_{\mathbb{R}} \frac{dk_j}{2 \pi  }
\prod_{1\leq i < j  \leq n} \frac{k_i-k_j}{k_i-k_j+i } \prod_{j=1}^{n} 
\frac{(i k_j + \frac{\beta - \alpha}{2} )^t}{(i k_j + \frac{\alpha+\beta}{2})^{1 +x_j} (i k_j - \frac{\alpha+\beta}{2})^{ 1-x_j + t} }  \ .
\eea
This formula will be used below to extract interesting information about the correlations of
the fluctuations in the diffusive regime. 

\section{Cauchy-type Fredholm determinant formula for the Laplace transform and asymptotic analysis in the optimal direction of the RWRE} \label{Sec:Diff}

\subsection{ The issue of the first site} \label{subsecFirstWeight}

In the following we will consider the sequence, with $n \in \mathbb{N}$ 
\bea \label{firstSite1}
Z_n = \frac{ \Gamma(\alpha + \beta  )}{\Gamma(\alpha+\beta+n)}   \overline{ Z_t(x)^n}     \  ,
\eea
and the associated generating function
\bea  \label{defg}
g_{t,x}( u ) = \sum_{n=0}^{\infty} \frac{(-u)^n}{n!} Z_n    \  ,
\eea
the reason being that only the latter can be simply expressed as a Fredholm determinant. Note that the sum in (\ref{defg}) converges since $0<Z_n < \overline{ Z_t(x)^n}$ and $0<Z_t(x) <1$ (since $Z_t(x)$ can be interpreted as a probability, see Sec.~\ref{Sec:BetaRWRE}). The situation here is also quite different compared to other exactly solvable models of DP: in the continuum case the growth of the moment is too fast to obtain a convergent generating function, while in the log-Gamma and Inverse-Beta case only a finite number of moments exist. Here all the moments exist and do determine the PDF of $Z_t(x)$. Note that it is a priori not clear whether $Z_n$ are the moments of a (positive or not) random variable $\tilde Z_t(x)$ (one can e.g. check that this is not the case at $t=0$). If it is the case, then $Z_t(x)$ is given in law by the product $w_{00} \tilde Z_t(x)$ where $w_{00}$ is a random variable (independent of $\tilde Z_t(x)$) distributed as

\bea \label{distGamma}
 w_{00} \sim Gamma(\alpha + \beta) \quad , \quad P_{w_{00}} (w) = \frac{1}{\Gamma(\alpha + \beta) } w^{-1 + \alpha + \beta } e^{-w} \theta(w) \quad, \quad \int dw w^n P_{w_{00}}(w) 
 = \frac{\Gamma(\alpha+\beta+n)}{\Gamma(\alpha+\beta)}  \ ,
 \eea
 where $P_{w_{00}}(w)$ is the PDF of $w_{00}$. Note that the situation is here also different compared to the Inverse-Beta case. In the latter, one needs to add a Boltzmann weight on the first site to obtain a partition sum whose Laplace transform can be expressed as a Fredholm determinant \cite{usIBeta}. Here we are formally removing a Boltzmann weight on the first site. In any case the Laplace transform of the original partition sum can be obtained from $g_{t,x}( u )$ using
\bea \label{FirstSiteLT}
\overline{e^{-u Z_t(x) }} = \langle g_{t,x}( u w_{00}) \rangle_{w_{00}} = \int_0^{+\infty} dw g_{t,x}( u w) P_{w_{00}}(w)  \ .
\eea 

Note that alternatively, using the Hypergeometric function defined $\forall u \in \mathbb{C}$ by $_0 F_1(\alpha+\beta ; u) = \sum_{n=0}^{\infty}  \frac{\Gamma(\alpha + \beta)}{\Gamma(\alpha + \beta+n)}\frac{(-u)^n}{n!}$, the generating function $g_{t,x}(u)$ can be rewritten as
\bea
g_{t,x}(u) = \overline{_0 F_1(\alpha+\beta ; - u Z_t(x))}     \ .
\eea

\subsection{Cauchy type Fredholm determinant formulas} \label{subsecFredholm}

Starting from (\ref{momentFormula01}), it can be shown (see Appendix \ref{app:Fredholm}) that the generating function $g_{t,x}(u)$ can be written as a Fredholm determinant:
\begin{equation}\label{kernel1bis} 
 g_{t,x}(u) = {\rm Det} \left( I  +  u K_{t,x} \right)  ,
\end{equation}
with the kernel:
\begin{eqnarray} \label{kernel1} 
&&  K_{t,x}(v_1,v_2) =   \int_{- \infty}^{+\infty}   \frac{dk}{ \pi}  e^{ -  2 i k(v_1-v_2) -  (v_1+v_2) }
\frac{(i k + \frac{\beta - \alpha}{2} )^t}{(i k + \frac{\alpha+\beta}{2})^{1 +x} (i k - \frac{\alpha+\beta}{2})^{ 1-x + t} }   , 
 \end{eqnarray}
and $ K_{t,x} : L^2 ( \mathbb{R}_+) \to L^2 ( \mathbb{R}_+) $. Note that the integral on $k$ defining (\ref{kernel1}) converges $\forall (t,x) \in \mathbb{N}^2$ (at large $k$ the integrand decays as $1/k^2$). We can also write a simpler expression for the Kernel, writing (\ref{kernel1}) as a product of operators
\bea
&& K_{I,J}(v_1,v_2) = \int_{p} A(v_1 , p) B(p ,v_2) \nn \\
&&  A(v_1 , p) = -  \frac{2}{\pi} \theta(v_1) e^{-v_1 (1 + i p)}  \frac{(p+i(\alpha-\beta))^{I-1}}{(p-i (\alpha+\beta))^I}  \nn \\
&& B(p , v_2) = \theta(v_2) e^{-v_2 (1 - i p)} \frac{(p+i(\alpha-\beta))^{J-1}}{(p+i (\alpha+\beta) )^J} 
\eea
where we have reintroduced the euclidean coordinate of the square lattice $I = 1+x$ and $J = 1+t-x$ (in this coordinate system the starting point of the polymer is $(I,J)=(1,1)$) and performed the change of variables $  k = p/2$. We then use ${\rm Det} \left( I  +  A B \right)  = {\rm Det} \left( I  +  B A \right)$, leading to 
\begin{equation} \label{kernel2bis}
 g_{I,J}(u) = {\rm Det} \left( I  +  u \tilde{K}_{I,J} \right)
\end{equation}
with the kernel: 
\bea \label{kernel2}
\tilde K_{I,J}(p_1,p_2) = &&  \int_{v } B(p_1 , v) A(v ,p_2)  \nn \\
= && \int_{v >0 } e^{-v (1 - i p_1)} \frac{(p_1+i(\alpha-\beta))^{J-1}}{(p_1+i (\alpha+\beta) )^J}  \times (-  \frac{2}{\pi} )e^{-v (1 + i p_2)}  \frac{(p_2+i(\alpha-\beta))^{I-1}}{(p_2-i (\alpha+\beta))^I}   \nn \\
= &&  -\frac{2}{\pi} \frac{(p_1+i(\alpha-\beta))^{J-1}}{(p_1+i (\alpha+\beta) )^J}  \frac{(p_2+i(\alpha-\beta))^{I-1}}{(p_2-i (\alpha+\beta))^I} \frac{1}{2 + i (p_2-p_1)} 
\eea
and $ \tilde K_{I,J} : L^2 ( \mathbb{R}) \to L^2 ( \mathbb{R}) $. Performing the change of variables $p \to 1/q$ this is also equivalent to
\begin{equation} \label{kernel3bis}
 g_{I,J}(u) = {\rm Det} \left( I  +  u \hat{K}_{I,J} \right)
\end{equation}
with the kernel $ \hat K_{I,J} : L^2 ( \mathbb{R}) \to L^2 ( \mathbb{R}) $:
\bea \label{kernel3}
\hat K_{I,J}(q_1,q_2) =  - \frac{2}{\pi} \frac{(1+i q_1 (\alpha-\beta))^{J-1}}{(1+i q_1 (\alpha+\beta) )^J}  \frac{(1+i q_2 (\alpha-\beta))^{I-1}}{(1-i q_2 (\alpha+\beta))^I} \frac{1}{2 + i (q_2^{-1}-q_1^{-1})} 
\eea
A final expression which will be preferred in the following. We will also equivalently use the coordinate system $(t,x)$ and the notation $\hat K_{t,x} = \hat K_{I= 1+x , J=1+t-x}$. In Appendix~\ref{app:Proba} we use the above Fredholm determinant formulas to obtain a formula for the PDF of $Z_t(x)$.

\medskip

{\it Remark:} \label{subsec:comments2}

Note that the formulas (\ref{kernel1bis}), (\ref{kernel2bis}) and (\ref{kernel3bis}) have the distinctive feature that the Laplace transform variable $u$ simply multiplies the kernel inside the Fredholm determinant. This should be contrasted with 
formulas obtained using similar replica Bethe ansatz calculations for other exactly solvable directed polymers models: in these other cases the Laplace transform variable $u$ appears inside the kernel in 
a non-trivial manner. 
In a more general context, (see e.g. \cite{BorodinMacdo}) this 
appears as a simple consequence of the fact that the moments formula we obtained is a so-called large-contours type formula (see the discussion in \ref{subsec:comments}). Indeed, it is known in the literature that such formulas lead after summation to Fredholm determinant formulas with the variable $u$ appearing in front of the kernel (usually referred to as Cauchy-type formulas). In this context, the other usually encountered Fredholm determinant formulas are known as Mellin-Barnes type formulas and are obtained after summation of small contours moments formulas (see \ref{subsec:comments}).

It is usually assumed that performing the asymptotic analysis on Cauchy-type formulas is extremely difficult (see in particular the work of Tracy and Widom on the ASEP \cite{TracyWidom2008,TracyWidom2009} on which we will comment more later).  We will however see in the following that our Cauchy-type formula (\ref{kernel3}) is well suited to perform the asymptotic analysis in a specific spatial direction (which will actually turns out to be the optimal direction chosen by the RWRE). In the other directions however, this will not be the case and we will first obtain an alternative Fredholm determinant formula for $g_{t,x}(u)$ of the Mellin-Barnes type in
order to carry out the asymptotics.

\subsection{Asymptotic analysis of the first-moment: definition of the optimal direction and of the asymptotic regimes} \label{subsec:trace}

Let us now perform an asymptotic analysis in the large time limit with
\bea \label{ballscaling}
t \gg 1 \quad , \quad x=(1/2+ \varphi) t \quad , 
\eea
i.e.  $I = 1+x = 1 + (1/2+ \varphi) t$ and $J = 1+t-x= 1 + (1/2-\varphi) t$. We first consider the trace of $\hat{K}_{t,x}$, or equivalently $Z_1$:

\bea \label{sdp1}
-Z_1 = Tr(\hat K_{t,x}) = \frac{-1}{\pi} \int dq  \frac{(1+i q (\alpha-\beta))^{t}}{(1-i q (\alpha+\beta))^{1+x} (1+i q (\alpha+\beta) )^{1+t-x}} \ . 
\eea
Note that this integral can be performed exactly, the result being, as expected, $Z_1= \frac{1}{\alpha + \beta } \overline{Z_t(x) } =\frac{\alpha^{t-x} \beta^{x}}{(\alpha+\beta)^{t+1}} C^x_t$ where $C^x_t$ is the binomial coefficient. Note that in terms of RWRE, the mean value of the partition sum is the mean value of the PDF transition probability. The latter can also be interpreted as the PDF of a RW in an averaged environment: the {\it annealed} PDF, defined as
\bea \label{Pann}
\sP_{{\rm ann}}(X_0 = 0 | X_{-t} = x) := \overline{ \sP(X_0 = 0 | X_{-t} = x)} = \overline{Z_t(x) }
\eea
is the transition PDF for a RW defined as in (\ref{backwardKo}) with $p_{\st ,x}$ replaced by its average: $p_{\st,x} \to \overline{p_{\st,x}} = \overline{u} = \alpha/(\alpha+\beta)$. Note also that by translational invariance of the averaged environment we have $\sP_{{\rm ann}}(X_0 = 0 | X_{-t} = x) = \sP_{{\rm ann}}(X_t = -x | X_{0} = 0) = \overline{Z_t(x) } $. The asymptotic analysis could easily be performed on this exact formula but the goal here is to understand how the properties of the asymptotic regime emerge from the integral formula (\ref{sdp1}). A simple calculation shows that the integral on $q$ in  (\ref{sdp1}) is dominated by a saddle-point at
\bea \label{qsp}
q_{sp} = -\frac{i (r (2 \varphi -1)+2 \varphi +1)}{\alpha  (r+1) (r (2 \varphi -1)-2 \varphi -1)}
\eea
where we introduced the assymmetry ratio
\bea
r = \beta/\alpha \in \mathbb{R}_+       \  .
\eea
We obtain
\bea \label{sdpZ1}
Z_1  && = \frac{1}{\pi} \left(\frac{2 r \left(\frac{r-2 r \varphi }{2 \varphi +1}\right)^{\varphi }}{(r+1) \sqrt{r-4 r \varphi ^2}}\right)^t \int dq \frac{1}{1+ \alpha^2(1+r)^2 q_{sp}^2} e^{t (q -q_{sp})^2 \frac{\alpha ^2 (r+1)^2 (-2 r \varphi +r+2 \varphi +1)^4}{32 r^2 \left(4 \varphi ^2-1\right)} + tO((q-q_{sp})^3)}  \nn \\
&& = \frac{1}{\alpha(1+r)} \sqrt{\frac{2}{\pi t (1 -4 \varphi^2)}} \left( \frac{2}{\sqrt{1- 4 \varphi^2}}  \left(\frac{1-2 \varphi }{2 \varphi +1}\right)^{ \varphi }  \frac{ r^{(1/2 +\varphi)}}{1+r} \right)^t  \left( 1 + O(1/\sqrt{t}) \right) \ .
\eea
where from the first to the second line we rescaled $q-q_{sp} \to (q-q_{sp})/\sqrt{t}$. Note that $|q_{sp}| <1/(\alpha +\beta) = 1/(\alpha(1+r))$ and the implicit deformation of contours in (\ref{sdpZ1}) is legitimate. We thus obtain that the first moment $ \overline{Z_t(x = (1/2 + \varphi)t )}  = (\alpha + \beta) Z_1$  decays exponentially with time at a rate $\log(\psi_r(\varphi))$ where $\psi_r(\phi) =  \frac{2}{\sqrt{1- 4 \varphi^2}}  \left(\frac{1-2 \varphi }{2 \varphi +1}\right)^{ \varphi }  \frac{ r^{(1/2 +\varphi)}}{1+r} $. Note that $\psi_r(\phi)$ is always smaller than $1$ except at its maximum, the optimal angle, defined by
\bea \label{OptimalAngle}
\varphi_{opt}(r) = \frac{r-1}{2 (r+1)} \in ]-1/2 , 1/2 [ \ ,
\eea
for which $\psi_r(\varphi_{opt})=1$ ans $q_{sp}=0$. In this specific direction $\overline{Z_t(x)}$ actually decreases only algebraically as
\bea
Z_1 && = \frac{1}{\alpha+\beta} \overline{Z_t(x)} \simeq_{\varphi = \varphi_{opt}} \frac{1}{ \alpha \sqrt{2 \pi rt } } \left( 1 + O(1/\sqrt{t}) \right) \ .
\eea 
 In terms of RWRE, we thus have that the annealed PDF (\ref{Pann}) decreases exponentially in all directions, except in the direction $\varphi_{opt}$ where $\sP_{{\rm ann}}(X_t = -x | X_{0} = 0)  \simeq \frac{(1+r)}{\sqrt{2 \pi r t}} \left( 1 + O(1/\sqrt{t}) \right)  $ . The optimal angle thus appears as the most probable space-time direction taken by a RW in an averaged environment. Let us now show that it is at the center of a region where $\sP_{{\rm ann}}(X_t = -x | X_{0} = 0)$ is a Gaussian distribution. To see this explicitly, let us now consider a diffusive perturbation around the optimal direction as
\bea \label{diffscaling}
x = (\frac{1}{2} + \varphi_{opt}(r) )t + \kappa \sqrt{t} \ .
\eea
Inserting this scaling in (\ref{sdp1}), it is easily seen that the large time behavior of $Z_1=-Tr(\hat K_{t,x})$ is still controlled by the same-saddle point around $q_{sp} =0$. We now obtain, rescaling again $q \to q/\sqrt{t}$,
\bea \label{sdpdiff}
Z_1 && = -Tr(\hat K_{t,x})  =  \int_{\mathbb{R}} \frac{dq}{\pi \sqrt{t}  } \frac{1}{1+\alpha^2(1+r)^2 \frac{q^2}{t}} e^{- 2 r \alpha^2  q^2  + 2 i ((1+r) \alpha \kappa q  +   O(\frac{1}{\sqrt{t}}))}  \nn \\ 
&&  =  \frac{1}{\alpha \sqrt{2 \pi rt }}  e^{- \frac{(r+1)^2}{2r}  \kappa^2} \left( 1 + O(\frac{1}{\sqrt{t}}) \right)  \ .
\eea
Or equivalently, in terms of the annealed PDF,
\bea \label{sdpdiff2}
x = (\frac{1}{2} + \varphi_{opt}(r) )t + \kappa \sqrt{t} \quad \Longrightarrow \quad \sP_{{\rm ann}}(X_t = -x | X_{0} = 0) = \frac{(1+r)}{\alpha \sqrt{2 \pi rt }}  e^{- \frac{(r+1)^2}{2r}  \kappa^2}  \left( 1 + O(\frac{1}{\sqrt{t}})  \right) \ .
\eea
And the annealed PDF, that is the transition PDF of the RW in an averaged environment, is in the large time limit in the diffusive scaling (\ref{diffscaling}) Gaussian distributed with a diffusion coefficient $D_{{\rm ann}} = \frac{r}{2 (r+1)^2}$. Note that this spatial region actually contains all the probability in the large time limit. Note also that (\ref{sdpdiff2}) can be seen as a consequence of the central limit theorem in an averaged environment. Finally, note that the saddle-point position $q_{sp}$ (see (\ref{qsp})) is $0$ only in the optimal direction. Following the change of variables from the quasi-momentas $\lambda$ of the particles in the Bethe wavefunction (\ref{brunetansatz2}) to the variable $q$ in the kernel (\ref{kernel3}), we note that $q \sim 0$ corresponds to $\lambda \sim 0$.

\medskip
In the following we will refer to the regime described by the scaling (\ref{diffscaling}) as the Gaussian regime, or, to avoid confusion with other sources of randomness, as the diffusive vicinity of the optimal direction. Indeed, although the scaling (\ref{diffscaling}) corresponds to the Gaussian regime for the RW in an averaged environment (see (\ref{sdpdiff2})), in the following we will be interested in the sample to sample fluctuations of the RWRE PDF. As we will show these fluctuations will turn out to be Gamma (and not Gaussian) distributed. The other regime described by the scaling (\ref{ballscaling}) with $\varphi \neq \varphi_{opt}$ will be referred to, using the RWRE language, as the large deviations regime. Let us now investigate the consequences of the existence of the saddle-point (\ref{qsp}) beyond the first moment, i.e. on the full Fredholm determinant.

\subsection{Asymptotic analysis in the diffusive vicinity of the optimal direction on the Cauchy-type Fredholm determinant} \label{saddlepointCauchy}

The saddle-point performed on the trace of the kernel (\ref{kernel3}) is under control for arbitrary $\varphi$. A natural question is now to understand whether one can use the same saddle-point on the full Fredholm determinant (57) and (\ref{kernel3}). In most cases (more precisely that is whenever $q_{sp} \neq 0$) the answer is negative. Indeed, if $q_{sp} \neq 0$, $Im(q_{sp}) \neq 0$ and the evaluation of e.g. the $Tr(\hat K \circ \hat K)$ term (or equivalently $Z_2$) using the saddle point involves a shift of the integrals on $q$  to ensure that the contour of integration on $q$, initially equal to $\mathbb{R}$, passes upon the saddle-point at $q_{sp}$. However the presence of the term $ \frac{1}{2 + i (q_2^{-1}-q_1^{-1})}$ present in (\ref{kernel3}) forbids that shift 
and we cannot use the saddle-point (i.e. it is impossible to deform the contours of integrations without crossing poles). If one tries to perform the saddle-point analysis on the form of the kernel (\ref{kernel2}) the difficulty is different: the trace is still dominated by the saddle point at $p_{sp} = \frac{1}{q_{sp}}$ and it seems a priori easy to shift all contours of integrations to avoid crossing the poles of the term $ \frac{1}{2 + i (p_2-p_1)}$ (using a simple translation of all contours). However in this case $|Im(p_{sp})| >(\alpha + \beta)$ so that the difficulty is now to avoid crossing the pole at $p = \pm i (\alpha + \beta)$. One way to do so is to try to close the contours of integration on the upper half plane, but in this case one inevitably crosses the poles of the term $ \frac{1}{2 + i (p_2-p_1)}$. This difficulty thus appears as a true property of the asymptotic analysis of the kernels (\ref{kernel1}), (\ref{kernel2}), (\ref{kernel3}): the saddle-point suggested by the study of the trace cannot be used for higher order terms. We will come back to this point in Section \ref{Sec:KPZ}.

\medskip

In the optimal direction however (and in its diffusive vicinity (\ref{diffscaling})), we noticed that $q_{sp} =0$. In this case there is no shift to perform and {\it all the terms (corresponding to all the moments) in the series expansion of the Fredholm determinant can be evaluated using the saddle-point}, as we will see below. We thus now write
\bea \label{diffscaling2}
x = (\frac{1}{2} + \varphi_{opt}(r) )t + \kappa \sqrt{t} \ ,
\eea
and consider the saddle point at $q_{sp}=0$ for the evaluation of the Fredholm determinant, using the expansion
\bea
&& {\rm Det} ( I  + u \hat K) = e^{  \Tr \ln(I + u \hat K) } = 1  + u \Tr \hat K + \frac{u^2}{2} ( (\Tr \hat K)^2 - \Tr \hat K \circ \hat K)  \\
&&  + \frac{u^3}{3!} ((\Tr \hat K )^3 - 3 \Tr \hat K \Tr \hat K \circ \hat K + 2 \Tr \hat K \circ \hat K \circ \hat K) + \dots \nonumber 
\eea
In this expansion, all the terms that involve powers of the Kernel, such as $\Tr  \hat K \circ \hat K $ and $\Tr  \hat K \circ \hat K \circ \hat K$ contain terms such as $\frac{q_1 q_2}{q_1 q_2 + (q_2 - q_1)}$. Using the saddle-point and rescaling $q_i \to q_i / \sqrt{t}$, these terms bring out additional factors of $1/\sqrt{t}$ and are subdominant. Hence we have
\bea
{\rm Det} ( I  + u \hat K) && = 1  + u \Tr \hat K + \frac{u^2}{2} (\Tr \hat K)^2 + \frac{u^3}{3!} (\Tr \hat K )^3 + \dots + O(1/\sqrt{t})  \nonumber  \\
&& = e^{ u \Tr \hat K}+ O(1/\sqrt{t})  
\eea

If one rescales $u$ as $u = \alpha \sqrt{2 \pi rt } e^{ \frac{(r+1)^2}{2r}  \kappa^2}  \tilde u$ such that $u Z_1= - u \Tr \hat K \simeq_{t \to \infty} \tilde u = O(1)$, then we obtain the following convergence (that holds at each order in the expansion in $\tilde{u}$):
\bea
g_{t,x}(\alpha \sqrt{2 \pi rt }  e^{ \frac{(r+1)^2}{2r}  \kappa^2}\tilde u)={\rm Det} ( I  + \alpha \sqrt{2 \pi rt } e^{ \frac{(r+1)^2}{2r}  \kappa^2}  \tilde u \hat K)  \simeq_{t \gg 1} e^{-\tilde{u}} + O(1/\sqrt{t}) \ .
\eea 
Note here that the relatively slow growth of moments shows that the convergence in moments imply the convergence of Laplace transform, and hence in distribution. Let us now define for convenience a rescaled partition sum as
\bea \label{rescaledcalZ}
{\cal Z}_t(\kappa) =  \alpha \sqrt{2 \pi rt} e^{\frac{(r+1)^2}{2r} \kappa^2} Z_t\left((\frac{1}{2} +  \varphi_{opt}(r)) t + \kappa \sqrt{t}  \right)  \ .
\eea
We thus find that, using (\ref{FirstSiteLT}),
\bea
\overline{e^{ - \tilde{u} {\cal Z}_t(\kappa) }} =  \langle g_{t,x}( \alpha \sqrt{2 \pi rt } e^{ \frac{(r+1)^2}{2r}  \kappa^2} \tilde u w_{00}) \rangle_{w_{00}} = \langle e^{-   \tilde u w_{00} } \rangle_{w_{00}} + O(1/\sqrt{t})
\eea
i.e.
\bea \label{ResultOptimalDirection}
{\cal Z}_{t= \infty} (\kappa)  \sim Gamma(\alpha+ \beta) \ ,
\eea
and corrections are of order $O(1/\sqrt{t})$. The form of the Gamma distribution was recalled in (\ref{distGamma}). The rescaled partition sum (\ref{rescaledcalZ}) is thus distributed as a Gamma random variable in the large $t$ limit at fixed $\kappa$. Using the convergence in law (\ref{ResultOptimalDirection}), we obtain that the positive integer moments of the partition sum are
\bea
x = (\frac{1}{2} + \varphi_{opt}(r) )t + \kappa \sqrt{t} \quad \Longrightarrow \quad \overline{Z_t(x)^n} = \frac{\Gamma(\alpha + \beta + n)}{\Gamma(\alpha + \beta )}  \frac{e^{-n \frac{(r+1)^2}{2r}  \kappa^2}}{(\alpha \sqrt{2 \pi rt } )^n}  +  O(1/(\sqrt{t})^{n+1})     \     .
\eea
Using again (\ref{ResultOptimalDirection}), we obtain the first two moments of the directed polymer free-energy at large $t$ in the diffusive scaling (\ref{diffscaling2}):
\bea \label{resdiffusivescaling}
&& \overline{\log\left( Z_t\left((\frac{1}{2} +  \varphi_{opt}(r)) t + \kappa \sqrt{t}\right) \right)}  = -\frac{1}{2} \log (2 \pi  r t)  -\log (\alpha )+\psi (r \alpha +\alpha )  - \frac{(r+1)^2}{2r} \kappa^2+  O(1/\sqrt{t}) \nn \\
&& \overline{\log\left( Z_t\left((\frac{1}{2} +  \varphi_{opt}(r)) t + \kappa \sqrt{t}\right) \right)^2}^c = \psi'(\alpha +\alpha  r)  +  O(1/\sqrt{t}) \ ,
\eea
where $\psi= \Gamma'/\Gamma$ is the diGamma function and $\overline{()}^c$ denotes the connected average over disorder. Those results are quite different from what could naively be expected from the usual KPZ universality. Notably we find that here the free-energy is not extensive and its fluctuations are of order $1$ and not $t^{1/3}$. These unusual results (in the context of directed polymers) are linked to the fact that the Beta polymer is also a RWRE. We note that all the fluctuations of the free-energy (which are small) are entirely due to the presence of the fictitious Boltzmann weight $w_{00}$ whose distribution is subtly encoded in the algebraic content of the model (that is it comes out of the structure of the Bethe ansatz). The question of the universality of this behavior for TD-RWRE in other type of random environments deserves further investigations in the future.

\subsection{Multi-point correlations in a diffusive vicinity of the optimal direction} \label{subsec:Multi}

We now go further and study the asymptotic limit for multi-point correlations in a diffusive vicinity of the optimal direction. We consider the formula for the multi-points moments (\ref{momentFormula07}) with, for $i=1 , \cdots , n$
\bea
x_i =(\frac{1}{2} +  \varphi_{opt}(r)) t + \kappa_i \sqrt{t} \quad , \quad \kappa \in \mathbb{R} \quad , \quad t \to \infty \ ,
\eea
and here $\kappa_1 \leq \cdots \leq \kappa_n$ for formula  (\ref{momentFormula07}) to apply. As before, let us first perform the change of variables $k_j \to 1/2q_j$. We obtain a formula equivalent to (\ref{momentFormula07}) as 
\bea \label{multipointasympt1}
&& \!\!\!\!\!\!\!\!\!\!\!\!\!\!\!\!\!\!\!\!\!\!  \overline{Z_t(x_1) \cdots Z_t(x_n)}  =  \frac{\Gamma(\alpha+\beta+n)}{\Gamma(\alpha + \beta  )} 
\prod_{j=1}^{n}  \int_{\mathbb{R}} \frac{dq_j}{\pi  }
\prod_{1\leq i < j  \leq n} \frac{q_j-q_i}{q_j-q_i + 2 i q_i q_j } \prod_{j=1}^{n} 
\frac{(q_j  - i(\beta - \alpha) )^t}{(q_j - i  (\alpha+\beta))^{1 +x_j} (q_j + i (\alpha+\beta))^{ 1-x_j + t} }  \ .
\eea
Using the same saddle-point calculation around $q \simeq 0$ we now obtain, changing $q_j \to q_j/ \sqrt{t}$ in (\ref{multipointasympt1}),
\bea \label{multipointasympt2}
\!\!\!\!\!\!\!\!\!\!\!\!\!\!\!\!\!\!\!\!\!\!  \overline{Z_t(x_1) \cdots Z_t(x_n)}  && =  \frac{\Gamma(\alpha+\beta+n)}{\Gamma(\alpha + \beta  )} 
\prod_{j=1}^{n}  \int_{\mathbb{R}} \frac{dq_j}{\pi \sqrt{t}  }
\prod_{1\leq i < j  \leq n} \frac{q_j-q_i}{q_j-q_i + \frac{2 i}{\sqrt{t}} q_i q_j } \nn \\
&& \times  \prod_{j=1}^{n} \frac{1}{1+\alpha^2(1+r)^2 \frac{q_j^2}{t}} e^{- 2 r \alpha^2  q_j^2  + 2 i ((1+r) \alpha \kappa_j q    + O(\frac{1}{\sqrt{t}})}  \nn \\
\overline{Z_t(x_1) \cdots Z_t(x_n)} &&  = \frac{\Gamma(\alpha+\beta+n)}{\Gamma(\alpha + \beta  )}  \left(  \frac{1}{\alpha \sqrt{2 \pi rt }} \right)^n e^{- \frac{(r+1)^2}{2r} \sum_{j=1}^n \kappa_j^2} + O(\frac{1}{(\sqrt{t})^{n+1}})
\eea
As before, taking into account the interactions between particles (encoded in the $\prod_{1\leq i < j  \leq n} \frac{q_j-q_i}{q_j-q_i + 2 i q_i q_j }$ term) just leads to $O(1/\sqrt{t})$ corrections to this leading behavior. Hence we now obtain that in the diffusive vicinity of the optimal direction, the rescaled spatial process defined in (\ref{rescaledcalZ}) converges in the large time limit to a {\it constant process with marginal distribution a Gamma distribution}
\bea \label{resrescaledcalZ}
{\cal Z}_{\infty}(\kappa) \sim Gamma(\alpha+ \beta)  \ ,
\eea
where here the equality now holds in the sense of the full spatial process, extending the one-point result (\ref{ResultOptimalDirection}). The fact that all different rescaled partition sum in (\ref{rescaledcalZ}) share the same fluctuations suggests that these fluctuations are not influenced by the last edges visited by the typical polymer path. Correspondly, in the RWRE language, the fluctuations of the probability to arrive at the site $(0,0)$ starting from infinity in a diffusive vicinity of the optimal direction are Gamma distributed
and not sensitive to the first edges visited by the RWRE.

\medskip

On the other hand, for a fixed starting point, the probability for the RW to arrive at different end points in the diffusive regime are all distributed as Gamma random variables from (\ref{ResultOptimalDirection}), but these Gamma random variables are a priori different, and we now show that they must be. Indeed, applying a general theorem from \cite{RAS} to the Beta TD-RWRE, we know that for a given environment, with probability $1$ (here probability refers to the disorder distribution), the random walk rescaled diffusively converges to a Gaussian distributed RV with a quenched diffusion coefficient equal to the annealed diffusion coefficient $D_{quenched} = D_{ann} = \frac{r}{2(r+1)^2}$. That is, with probability $1$, we have
\bea \label{Eq.83}
\sP \left(\frac{X_{\st} + (1/2 +\varphi_{opt}(r)) \st}{\sqrt{\st}} \in [ \kappa, \kappa+ d \kappa ]  | X_0 =0 \right) \sim_{\st \to \infty , d\kappa \ll 1 } = \frac{(\alpha+\beta)}{\alpha \sqrt{2 \pi r  }}  e^{- \frac{(r+1)^2}{2r} \kappa^2}  d \kappa .
\eea
(Here the apparent change of sign comes from (\ref{Eq14})). In (\ref{Eq.83}) the disordered environment seems erased and no traces of Gamma fluctuations are found in the diffusive region. This result however only concerns the probability for a RW to arrive in a vicinity of order $\sqrt{\st}$ of a given point, while our result (\ref{resrescaledcalZ}) really gives the probability to arrive at one point. This means that the Gamma RVs at each arrival points must be different in some way so that, when summing their contribution in a with of order $\sqrt{\st}$ , they are effectively averaged out to lead to (\ref{Eq.83}) (here the $\alpha + \beta$ in the numerator is now interpreted as $\overline{\cal Z}_{\infty}(\kappa)$). Our result is thus not inconsistent with the general result of \cite{RAS} if this averaging can take place \cite{BCprivate2}. Note finally that the Gamma variables on different end points can still be correlated, but only on a width of order $t^{\delta}$ with $\delta < 1/2$.

\medskip

We have thus now obtained a rather complete understanding of the fluctuations of the partition sum of the Beta polymer / RWRE transition probability in a diffusive vicinity of the optimal direction. Let us now investigate the large deviations regimes.

\section{Asymptotic analysis in the large deviations regime: KPZ universality}\label{Sec:KPZ}

In this section we show how the usual KPZ universality is hidden in the Beta polymer in the large time limit in all directions $\varphi \neq \varphi_{opt}$. In most of this section we use heuristic arguments that will be supported in the next section (see Sec.~\ref{Sec:nested}) by using results of BC.

\subsection{Recall of the results of Barraquand-Corwin \cite{BarraquandCorwinBeta}}

Let us first recall the results of \cite{BarraquandCorwinBeta} that will be of interest in the following. In the half-line to point problem, one shows that, in any direction $\varphi < \varphi_{opt}(r)$,
\bea
\lim_{t\to \infty} Proba\left( \frac{\log Z_t^{HL}((1/2 + \varphi) t ) + I(\varphi) t}{t^\frac{1}{3} \sigma(\varphi)} \leq y \right) = F_{2}(y)
\eea
where $F_{2}$ is the cumulative distribution function of the Tracy-Widom GUE distribution. This was rigorously proven for the case of $\alpha=\beta=1$ (a technical argument) and presented in \cite{BarraquandCorwinBeta} as Theorem 1.15. The constants $I(\varphi)$ and $\sigma(\varphi)$ are solution of a system of transcendental equations which reads (we now introduce the notations of \cite{BarraquandCorwinBeta}, parametrizing $\varphi = -\frac{x(\theta)}{2}$, and $\theta$ implicitly given by the first equation below)
\bea \label{BCresult}
&& x(\theta) = \frac{ \psi'(\theta + \alpha + \beta) + \psi'(\theta) - 2 \psi'(\theta+ \alpha)}{\psi'(\theta) - \psi'(\theta + \alpha + \beta)} \nn \\
&& I(\theta) = \frac{\psi'(\theta + \alpha + \beta) -\psi'(\theta + \alpha )}{\psi'(\theta )-\psi'(\theta + \alpha + \beta)} \left( \psi(\theta+ \alpha + \beta) - \psi(\theta) \right) + \psi(\theta + \alpha + \beta)-\psi(\theta + \alpha) \nn \\
&& 2 \sigma(\theta)^3 = \psi''(\theta+ \alpha) - \psi''( \alpha + \beta+ \theta) + \frac{\psi'(\alpha + \theta) -\psi'(\alpha + \beta + \theta) }{ \psi'(\theta) - \psi'(\alpha + \beta+ \theta)} \left( \psi''(\alpha + \beta+ \theta) - \psi''(\theta) \right) \ .
\eea
Where $\psi = \Gamma'/\Gamma$ is the diGamma function. Here the assumption $\varphi < \varphi_{opt}(r)$ ensures here that one looks at a direction in the large deviations regime of the cumulative distribution of the RW. With the notations of Sec.~\ref{Sec:BetaRWRE}: $Z_t^{HL}((1/2 + \varphi) t ) = \sP(X_0 \geq 0 | X_{\st = -t} = (1/2 + \varphi) t  ) $ and if $\varphi \geq  \varphi_{opt}(r)$ the probability that the Random walk arrives on the half-line $x \geq 0$ remains finite as $t \to \infty$. If $\varphi < \varphi_{opt}(r)$ it decreases exponentially as a function of $t$ with a rate function given by $I(\theta)$, corresponding to the extensive part of the free-energy in the polymer language. For our purpose we note, as already emphasized in \cite{BarraquandCorwinBeta}, that this result also implies that for $\varphi < \varphi_{opt}(r)$ (using that under mild assumptions satisfied here the large deviation rate function of the CDF in a RWRE problem is the same as the one of the PDF, see \cite{RASY})
\bea \label{PointToPointFE}
\lim_{t \to \infty} \frac{\log Z_t((1/2 + \varphi) t )}{t} =_{a.s.} - I(x(\theta))  \ , 
\eea
where as in (\ref{BCresult}) $x(\theta) =- 2 \varphi$. That is the point to point free energy of the Beta polymer is the same as the half line to point free energy in the large deviations regime. Note that the point-to-point free-energy in the region $ \varphi > \varphi_{opt}(r)$ can be obtained by using the symmetry $(x,\alpha,\beta) \to (t-x , \beta , \alpha)$, which amounts at using the result (\ref{BCresult}) for  $ \varphi > \varphi_{opt}(r)$ and $(\alpha,\beta) \to (\beta , \alpha)$.

\medskip
A first challenge in the following will thus be to retrieve using our formulas the result (\ref{PointToPointFE}), and to extend it to obtain a description of the fluctuations of the $\log Z_t(x)$ in the large deviations regime as well.

\subsection{An inherent difficulty and a puzzle}

In this section we put forward an issue that one encounters if one tries to perform the asymptotic analysis $x= (1/2 +\varphi)t$ and $t\gg 1$ of $g_{t,x}(u)$ in a direction $\varphi \neq \varphi_{opt}(r)$ using one of the Fredholm determinant expression derived in Sec.~\ref{subsecFredholm}, e.g. using the kernel (\ref{kernel3}). The problem can be formulated as follows: one one hand we have
\bea \label{problem1}
g_{t,x}(u) ={\rm Det} \left( I + u \hat K_{t,x} \right) \ .
\eea
Starting from this expression, the most natural idea is to perform an asymptotic analysis by rescaling  $u$ in some way while keeping the Fredholm determinant structure in (\ref{problem1}) intact. Following the computation of Sec.\ref{subsec:trace}, we know that imposing the rescaling on $u$ to be such that the trace of the kernel converges implies
\bea \label{rescaling1}
u \sim \sqrt{t} \left( \frac{2}{\sqrt{1- 4 \varphi^2}}  \left(\frac{1-2 \varphi }{2 \varphi +1}\right)^{ \varphi }  \frac{ r^{(1/2 +\varphi)}}{1+r} \right)^{-t}  \ .
\eea
On the other hand however, we know that we want to obtain (\ref{PointToPointFE}) in the large time limit. This result suggests that the leading part of the proper rescaling of $u$ should be
\bea \label{rescaling2}
 u  \sim e^{I(\theta) t + o(t)} \ ,
\eea
in order to properly take into account the non-zero free-energy of the DP. The two scalings (\ref{rescaling1}) and (\ref{rescaling2}) are not mutually consistent. 
The rescaling (\ref{rescaling1}) ensures the convergence of the trace of the kernel, as a result of the saddle-point that controls the trace (see Sec.~\ref{subsec:trace}), but cannot be used for higher order terms in the Fredholm determinant (see Sec.~\ref{saddlepointCauchy}) and is in apparent contradiction with the free-energy (\ref{PointToPointFE}). The rescaling (\ref{rescaling2}) on the other hand does not ensure the convergence of the trace. The series expansion in $u$ defined by the Fredholm determinant expression (\ref{problem1}) thus does not appear well suited to perform the asymptotic analysis. As we will se in the following, the way out of this dilemma will be to recast this series expansion as, schematically
\bea \label{problemsolution}
&& {\rm Det} ( I  + u \hat K_{t,x}) = {\rm Det} ( I  +  \check K_{t,x}(u)) \ ,
\eea
i.e. finding a new Fredholm determinant expression for $g_{t,x}(u)$ where $u$ appears non-trivially in the expression of the kernel $\check K_{t,x}(u)$. It is interesting to note that the first Cauchy-Type Fredholm determinant formula that appeared in the literature around KPZ was in the work of Tracy and Widom on the ASEP \cite{TracyWidom2008}. There it was also emphasized that the asymptotic analysis of this type of formula was extremely difficult, and the solution was later found in \cite{TracyWidom2009} and involved a transformation of Fredholm determinants similar as (\ref{problemsolution}). On the other hand in our case it is interesting to note that the Cauchy-type Fredholm determinant formula did appeared very well suited to perform the asymptotic analysis in the diffusive regime of the TD-RWRE.

\subsection{A formal formula for the moments of the Beta polymer in terms of strings}

From other studies on other exactly solvable models of directed polymer and related models (see also the discussion in Sec.~\ref{subsec:comments2}), we know that alternative Fredholm determinant formulas can be obtained by starting from strings-type/small-contours moments formulas for $\overline{Z_t(x)^n}$. In Appendix \ref{app:puzzle} we explain how one can use the known relations between different exactly solvable models of directed polymers on the square lattice to arrive at the following conjecture
\bea \label{BetaWithStrings}
&& \overline{(Z_t(x))^n}  \hspace{0.2cm} ``=" \hspace{0.2cm}  \frac{ \Gamma(\alpha+\beta+n)}{ \Gamma(\alpha+\beta)} n! \sum_{n_s=1}^n  \frac{1}{n_s!} \sum_{(m_1,..m_{n_s})_n} 
\prod_{j=1}^{n_s}  \int_{L^n} \frac{dk_j}{2 \pi}
\prod_{1\leq i < j  \leq n_s} \frac{4(k_i-k_j)^2 + (m_i - m_j)^2}{4(k_i-k_j)^2 + (m_i + m_j)^2} \nn \\
&&
\prod_{j=1}^{n_s} \frac{1}{m_j}  \left( \frac{  \Gamma(-\frac{m_j}{2} + \alpha+\beta+i k_j ) }{  \Gamma(\frac{m_j}{2} +  \alpha+\beta + i k_j )  } \right)^{1 +x}  \left( \frac{   \Gamma(-\frac{m_j}{2}  +i k_j )}{ \Gamma(\frac{m_j}{2} + i k_j ) } \right)^{ 1-x + t} \left( \frac{ \Gamma ( \beta +i k_j+\frac{m_j}{2})}{\Gamma( \beta +i k_j-\frac{m_j}{2}   )}  \right)^t \ .
\eea
where here $\sum_{(m_1,\cdots,m_{n_s})_n} $ means summing over all $n_s$-uplets $(m_1 , \cdots, m_{n_s})$ such that $ \sum_{i=1}^{n_s} m_i  = n$. In this formula the integration $\int_{L} \frac{dk_j}{2 \pi}$ is actually not a real integration (hence the presence of quotes around the equality sign) as we now detail. This formula is only valid in the sense of a specific residue expansion of the integrand. It consists in recursively taking,  always with a plus sign, starting with the integration over $k_{n_s}$ and then iterating up to $k_1$, the residues of the integrand coming from the $   \Gamma(-\frac{m_j}{2}  +i k_j )/ \Gamma(\frac{m_j}{2} + i k_j ) $ term as well as the residue at $k_j = k_l + \frac{i}{2} (m_l + m_j)$ and iterating. Note that when one takes the residue at $k_j = k_l + \frac{i}{2} (m_l + m_j)$, it creates new residues for the ``integration'' on $k_l$ in the term $ \Gamma(-\frac{m_j}{2}  +i k_j )/ \Gamma(\frac{m_j}{2} + i k_j ) $. Note also that the position of these residues do not depend on $\alpha$ and $\beta$, permitting an easy implementation using e.g. Mathematica.

\medskip

This formula is strictly speaking a conjecture, and only arises from an educated guess (see Appendix \ref{app:puzzle}). Its validity was tested against direct checks for small values of $x,t$ and $n$. Unfortunately, we could not find a contour of integration $L$ which makes this formula correct as a true contour integral formula. From the discussion in Sec.~\ref{subsec:comments} we know that the correct way to find such a formula is first to use a nested contour integral representation of $\overline{(Z_t(x))^n}$. This is done in Sec.~\ref{Sec:nested} where we obtain a formula similar to this one.

\subsection{A formal Fredholm determinant and KPZ universality}

The formal formula (\ref{BetaWithStrings}) is very close to the formula (60) obtained in \cite{usIBeta} for the Inverse-Beta polymer. (See Appendix \ref{app:puzzle} for the comparison). As such, we can follow the same steps as done in \cite{usIBeta} and obtain a new Fredholm determinant formula $g_{tx}(u) =  {\rm Det} \left( I + \check K_{t,x} \right)$ with
\begin{eqnarray}\label{Fredholmdet2}
 \check K_{t,x}(v_1,v_2) = && \int_{L}   \frac{dk}{ \pi}  \frac{-1}{2i} \int_C \frac{ds}{ \sin( \pi s ) }   u^s  e^{ -  2 i k(v_1-v_2) -  s (v_1+v_2) } \\
 &&  \left( \frac{  \Gamma(-\frac{s}{2} + \alpha+\beta + i k ) }{  \Gamma(\frac{s}{2}  + \alpha+ \beta +i k )  } \right)^{1 +x} \left( \frac{   \Gamma(-\frac{s}{2} +i k )}{ \Gamma(\frac{s}{2}  + i k ) } \right)^{ 1-x + t} \left( \frac{ \Gamma ( \beta +i k+\frac{s}{2})}{\Gamma( \beta +i k-\frac{s}{2})} \right)^t \nonumber  \ .
\end{eqnarray}
As before, this formula is formal since the integration on $L$ is actually not an integration. A similar (but not formal) formula can be derived using an approach based on nested contour integrals (see Sec.~\ref{Sec:nested}). Let us however ignore this for now and try to perform a saddle-point analysis on (\ref{Fredholmdet2}) by sending $t\to \infty$ with $x = (1/2 + \varphi)t$ and proceed as if (\ref{Fredholmdet2}) was well defined using a true integration. The analysis is then strictly similar to the one made in \cite{usIBeta} from which we borrow the notations and to which we refer the reader for more details. The ``integration'' in (\ref{Fredholmdet2}) is dominated in the large $t$ limit by the factor
\bea
\exp \left \{ t \left( G_\varphi (i k +s/2) -   G_\varphi (i k - s/2)  \right) \right\} \ ,
\eea
where
\bea
G_{\varphi}(x) = \log \Gamma(\beta+x ) - (1/2- \varphi) \log \Gamma(x) - (1/2+\varphi) \log \Gamma(\alpha+\beta+x) \ .
\eea 
We look for a cubic saddle-point at $(s,k) =(0 , - i k_{\varphi})$. Its position is implicitly defined by
\bea
\psi'(\beta + k_{\varphi}) - (\frac{1}{2} - \varphi ) \psi'(k_{\varphi})  - (\frac{1}{2} + \varphi ) \psi'(\alpha+ \beta+ k_{\varphi})
\eea
With the same notations as in \cite{usIBeta} we define a rescaled free energy $ f_{t}(\varphi)$ as 
\bea \label{RescalingLarget}
&& F_{t}(\varphi) = - \log Z_t(x =  (1/2+\varphi) t )  = c_\varphi t + \lambda_{\varphi} f_{t}(\varphi) \nn \\
&& c_\varphi = - G_{\varphi}'( k_\varphi ) \quad , \quad  \lambda_\varphi = \left( \frac{ t G_{\varphi}'''( k_\varphi ) }{8} \right)^{\frac{1}{3}}  \ .
\eea
Following the exact same steps as in \cite{usIBeta} (see also Appendix C there for a discussion that can be adapted to our setting to consider the effect of the additional weight $w_{00}$ introduced in Sec.~\ref{subsecFirstWeight}) we obtain
\begin{equation}\label{asymptoticlim}
\lim_{t \to \infty} Prob\left( \frac{ \log Z_t((1/2+ \varphi) t) + tc_{\varphi}}{\lambda_{\varphi} } <2^{\frac{2}{3}} z \right) = F_2(z) \ ,
\end{equation}
where $F_{2}$ is the cumulative distribution function of the Tracy-Widom GUE distribution and the parameters $c_{\varphi}$ and $\lambda_{\varphi} \sim t^{\frac{1}{3}}$ are given by (\ref{RescalingLarget}) with
\bea \label{SDPeqn}
&& \varphi = \frac{\psi'(\beta + k_{\varphi})- \frac{1}{2} \left(\psi'(k_{\varphi}) + \psi'(\alpha+ \beta+ k_{\varphi}) \right)}{\psi'(\alpha+ \beta+ k_{\varphi}) -\psi'(k_{\varphi}) } \nn \\
&&  c_{\varphi} = - G_{\varphi}'( k_\varphi ) = \left(\varphi +\frac{1}{2}\right) \psi(k_{\varphi}+\alpha +\beta )-\psi(k_{\varphi}+\beta )+\left(\frac{1}{2}-\varphi \right) \psi(k_{\varphi})  \nn \\
 &&  \frac{8 \lambda_{\varphi}^3}{t}=  G_{\varphi}'''( k_\varphi ) =-\left(\varphi +\frac{1}{2}\right) \psi''(k_{\varphi}+\alpha +\beta )+\psi''(k_{\varphi}+\beta )-\left(\frac{1}{2}-\varphi \right) \psi''(k_{\varphi}) \ .
\eea

The above system of equations is expected to be valid for $\varphi_{opt}(r) = \frac{\beta- \alpha}{2 (\alpha + \beta)} < \varphi < 1/2$, a limitation which is not visible from our formal derivation but that we now explain. A first hint is to consider the limit $\varphi = \pm 1/2$. In the limit $\varphi \to -\frac{1}{2}$, $Z_t = \prod_{i=1}^t u_i$, the product of $t$ independent $u$ variables. In the limit $\varphi \to -\frac{1}{2}$, $Z_t = \prod_{i=1}^t v_i$. This implies $-\overline{\log Z_t(0)}/t = \psi(\alpha+\beta) - \psi(\alpha)$ and $-\overline{\log Z_t(t)}/t = \psi(\alpha+\beta) - \psi(\beta)$. The second limit is correctly reproduced by (\ref{SDPeqn}) using $k_{\varphi} \to_{\varphi \to 1/2^-} 0^+$, whereas the first one is not. Furthermore, the two first equations in (\ref{SDPeqn}) (i.e. those that determines the extensive part of the free-energy) are also equivalent to the two first equations in (\ref{BCresult}) using the symmetry $(\varphi, \alpha , \beta ) \to (-\varphi , \beta , \alpha) $ on (\ref{BCresult}) with the identification $\theta = k_{\varphi}$, $I(\theta) = - G_{\varphi}'( k_\varphi )$. To see this more explicitly rewrite $c_{\varphi}$ in (\ref{SDPeqn}) as
\bea
c_{\varphi} = - G_{\varphi}'( k_\varphi ) && = \left(\varphi +\frac{1}{2}\right) \psi(k_{\varphi}+\alpha +\beta )-\psi(k_{\varphi}+\beta )+\left(\frac{1}{2}-\varphi \right) \psi(k_{\varphi}) \nn \\
&&=  (\varphi-1/2) \left( \psi(k_{\varphi}+\alpha +\beta ) -  \psi(k_{\varphi}) \right) + \psi(k_{\varphi}+\alpha +\beta ) - \psi(k_{\varphi}+\beta ) \nn \\
c_{\varphi} = - G_{\varphi}'( k_\varphi )  && = \frac{\psi'(\alpha+ \beta+ k_{\varphi}) - \psi'(\beta + k_{\varphi})}{ \psi'(k_{\varphi})- \psi'(\alpha+ \beta+ k_{\varphi})} \left( \psi(k_{\varphi}+\alpha +\beta ) -  \psi(k_{\varphi}) \right) + \psi(k_{\varphi}+\alpha +\beta ) - \psi(k_{\varphi}+\beta ) 
\eea
where from the first to the second line we have only rearranged the terms, and from the second to the last we have inserted $\varphi$ as given by (\ref{SDPeqn}). Applying the identification mentioned above shows the identity between the two first equations in (\ref{BCresult}) and those in (\ref{SDPeqn}).

This fact was expected since the extensive part of the free-energy should be the same in the half-line (in the large deviation regime) and in the point-to-point Beta polymer problem. The condition of validity of (\ref{BCresult}), $\varphi < \varphi_{opt}(r)$, then becomes our condition of validity after applying the symmetry.

\medskip

The new, non-trivial predictions from our formal computations are that
\begin{itemize}
\item{The fluctuations of the free-energy in the point-to-point Beta polymer problem are also of the GUE Tracy-Widom type with the characteristic $t^{1/3}$ scaling.}
\item{The non-universal constant in front of these fluctuations are exactly the same in the point-to-point and in the half-line to point problem. To see this, notice that the third line of (\ref{SDPeqn}) is also equivalent to the third line of (\ref{BCresult}) using the identification $2 \sigma(\theta)^3=G'''(\varphi)$.}
\end{itemize}

The first prediction could have been expected from the KPZ universality of directed polymer problem, a paradigm which however appears dangerous to apply in the Beta polymer as we saw from the study of the fluctuations in the optimal direction. The second is interesting from the RWRE point of view. Indeed, although it is known on general grounds (see \cite{RASY}) that the large deviations rate functions of the PDF and CDF of a TD-RWRE are identical, a similar identity for the {\it fluctuations} of the logarithm of the PDF and CDF is to our knowledge not known. Here, in the particular example of the Beta TD-RWRE, we have showed that these fluctuations are identical up to order $t^{1/3}$ included. It would be interesting to understand if this holds more generally for other TD-RWRE.

\subsection{Crossover between Gamma and Tracy-Widom fluctuations}  \label{Sec:Cross2}

Let us now use our results (\ref{resdiffusivescaling}) and (\ref{SDPeqn}) to try to gain some information on the crossover between the diffusive regime and the large deviations regime of the RW. Let us first see how both regimes are connected and study the behavior of (\ref{SDPeqn}) around the optimal angle: $\varphi =  \varphi_{opt}(r) + \delta \varphi$. Expanding in $\delta \varphi>0$, the solution of (\ref{SDPeqn}) reads
\bea
&& k_{\varphi} \simeq \frac{\alpha  r}{ (r+1) \delta \varphi  }  +\left(\frac{1}{2}-\alpha  r\right) + O(\delta \varphi) \nn \\
&&  - G_{\varphi}'(k_{\varphi}) \simeq \frac{(1+r)^2}{2 r} \delta \varphi^2 + \frac{(r-1)(r+1)^3}{6 r^2}\delta \varphi ^3 + O(\delta \varphi ^4) \nn \\
&& G_{\varphi}'''(k_{\varphi}) = \frac{ (r+1)^4}{\alpha ^2 r^3}  \left( \delta \varphi^4 + \frac{r^2-1}{r} \delta \varphi^5 + O(\delta \varphi^6) \right)  .
\eea
Hence, at fixed, small angle around the optimal position, $\varphi = \varphi_{opt}(r) + \delta \varphi$, we conclude from (\ref{RescalingLarget}) that
\bea \label{kpzTodiff}
&& \log Z_t(x) \simeq -\frac{(1+r)^2}{2 r} t \delta \varphi^2 + \left( \frac{(r+1)^4}{2 \alpha^2 r^3} \right)^{\frac{1}{3}}  t^{\frac{1}{3}} \delta \varphi^{\frac{4}{3}} \chi_{GUE}  \ ,\nn \\
&& x  =( \frac{1}{2} + \varphi_{opt}(r) + \delta \varphi)t \quad, \quad t \gg 1 \ .
\eea
Where $\chi_{GUE}$ is a RV which is distributed with the GUE Tracy-Widom distribution. On the other hand, we know from Sec.~\ref{subsec:Multi} that, on a diffusive scale around the optimal direction,
\bea \label{diffTokpz}
&& \log Z_t(x) \simeq - \log(\alpha) 
- \frac{1}{2} \log(2 \pi r t) - \frac{(1+r)^2}{2 r} \kappa^2+ \log \chi_{\alpha+\beta} , \nn \\
&& x = ( \frac{1}{2} + \varphi_{opt}(r) )t + \kappa \sqrt{t} \quad, \quad t \gg 1 \ .
\eea
Where $\chi_{\alpha+\beta}$ is a RV distributed with a Gamma distribution of parameter $\alpha+\beta$. Introducing the coordinate $\hat x = x- (\frac{1}{2} + \varphi_{opt}(r))t$ and changing $\delta \varphi \to \hat x /t$ in (\ref{kpzTodiff}) and $\kappa \to \hat x/\sqrt{t}$ in (\ref{diffTokpz}), one sees that both regimes are connected by the angle-dependent term in the extensive part of the free-energy $-\frac{(1+r)^2}{2 r} t \delta \varphi^2 =  - \frac{(1+r)^2}{2 r} \kappa^2 = -\frac{(1+r)^2}{2 r} \frac{\hat x^2}{t}$. Though the following is non-rigorous, it thus appears reasonable to schematically give a more complete picture of fluctuations at large $t$ around the central region as
\bea
&& \log Z_t(x) \simeq  - \frac{1}{2} \log( 2 \pi \alpha \beta t)- \frac{(\alpha + \beta)^2}{2 \alpha \beta} \frac{\hat x^2}{t} + \log \chi_{\alpha+ \beta} + \left( \frac{(\alpha+\beta)^4}{2 \alpha^3 \beta^3} \right)^{\frac{1}{3}} \frac{\hat x^{\frac{4}{3}}}{t} \chi_{GUE}  \nn \\
&&  x = (\frac{1}{2} + \varphi_{opt}(r) )t + \hat x \quad , \quad t \gg 1 \quad , \quad \hat x = o(t) \ .
\eea
The latter being exact at large time in the central region $\hat x = O(1)$ and including the diffusive regime $\hat x \sim \sqrt{t}$, as well as in the beginning of the large deviations regime $\hat x = \delta \varphi t$ with $\delta \varphi \ll 1$. In between, in the crossover region $\sqrt{t} \ll \hat x  \ll t$ the fluctuations should be an interpolation between Gamma and Tracy-Widom fluctuations and the above picture is too simple. Equating the amplitude of these two sources of fluctuations, it predicts the existence of a cross-over scale where the competition between Tracy-Widom and Gamma type fluctuations is maximal as
\bea \label{co} 
\hat x_{c.o.} \sim t^{\frac{3}{4}} \ .
\eea
As summarized in Fig.~\ref{fig:cross}. A way to formalize the identification of this crossover scale
is to introduce an (a priori unknown) scaling function involving e.g. the second cumulant of the
fluctuations, as
\bea
\left( \overline{ (\log Z_t(x))^2 }^c \right)^{1/2} = t^\gamma f(\frac{\hat x}{t^\delta})  
\eea 
for $\hat x= o(t)$, where $\frac{1}{2} < \delta < 1$. Imposing the matching conditions on the two above mentioned
regimes we obtain
\bea
&& t^\gamma f(y=\frac{\hat x}{t^\delta})  \simeq_{y \to 0} \psi'(\alpha + \beta) \\
&& t^\gamma f(y=\frac{\hat x}{t^\delta})  \simeq_{y \to +\infty} 
\left( \frac{(\alpha+\beta)^4}{2 \alpha^3 \beta^3} \right)^{\frac{1}{3}} \frac{\hat x^{\frac{4}{3}}}{t} 
\sqrt{ Var(\chi_{GUE})}
\eea
This implies $\gamma=0$ and $\delta=3/4$ recovering (\ref{co}). A more precise characterization of the fluctuations in this regime and of the scaling function remains to be obtained. Note that since
the formula $\overline{ \log Z_t(x)} - \overline{ \log Z_t(0) }  \simeq - \frac{(\alpha + \beta)^2}{2 \alpha \beta} \frac{\hat x^2}{t}$
was found to hold in both regimes, it is also expected to hold in the crossover regime.
Hence in this regime, writing $\hat x = \omega t^{\frac{3}{4}}$, with a fixed $\omega$, we expect that 
$\overline{ \log Z_t(x)}  \simeq -\frac{\sqrt{t} (r+1)^2 \omega ^2}{2 r} $
at large $t$.

\section{Nested-Contour integral formulas for the point-to-point problem} \label{Sec:nested}

During the late stages of redaction of this work, we were informed \cite{Barraquand-Corwin-private} by Guillaume Barraquand and Ivan Corwin of the existence of a nested contour integral formula for the multi-points moments of the Beta polymer, from which (\ref{momentFormula07}) was derived. This approach also allows to obtain other moments formulas and Fredholm determinant formulas which partially justify the heuristic approach used in the last section. The goal of this section is to make the link between our formulas and theirs.

\subsection{Alternative moments formulas}

{\it A nested contour integral formula:}

Let us start from the moments formula (\ref{momentFormula01}) that we now recall for readability of the reasoning:

\bea \label{momentFormula02}
&& \!\!\!\!\!\!\!\!\!\!\!\!\!\!\!\!\!\!\!\!\!\!\!\!\!\!  \overline{Z_t(x)^n}  = (-1)^n  \frac{\Gamma(\alpha+\beta+n)}{\Gamma(\alpha + \beta  )} 
\prod_{j=1}^{n}  \int_{- \infty}^{+\infty} \frac{dk_j}{2 \pi}
\prod_{1\leq i < j  \leq n} \frac{(k_i-k_j)^2}{(k_i-k_j)^2 + 1} \prod_{j=1}^{n} 
\frac{(i k_j + \frac{\beta - \alpha}{2} )^t}{(i k_j + \frac{\alpha+\beta}{2})^{1 +x} (i k_j - \frac{\alpha+\beta}{2})^{ 1-x + t} }  \ . 
\eea

\begin{figure}
\centerline{\includegraphics[width=8.5cm]{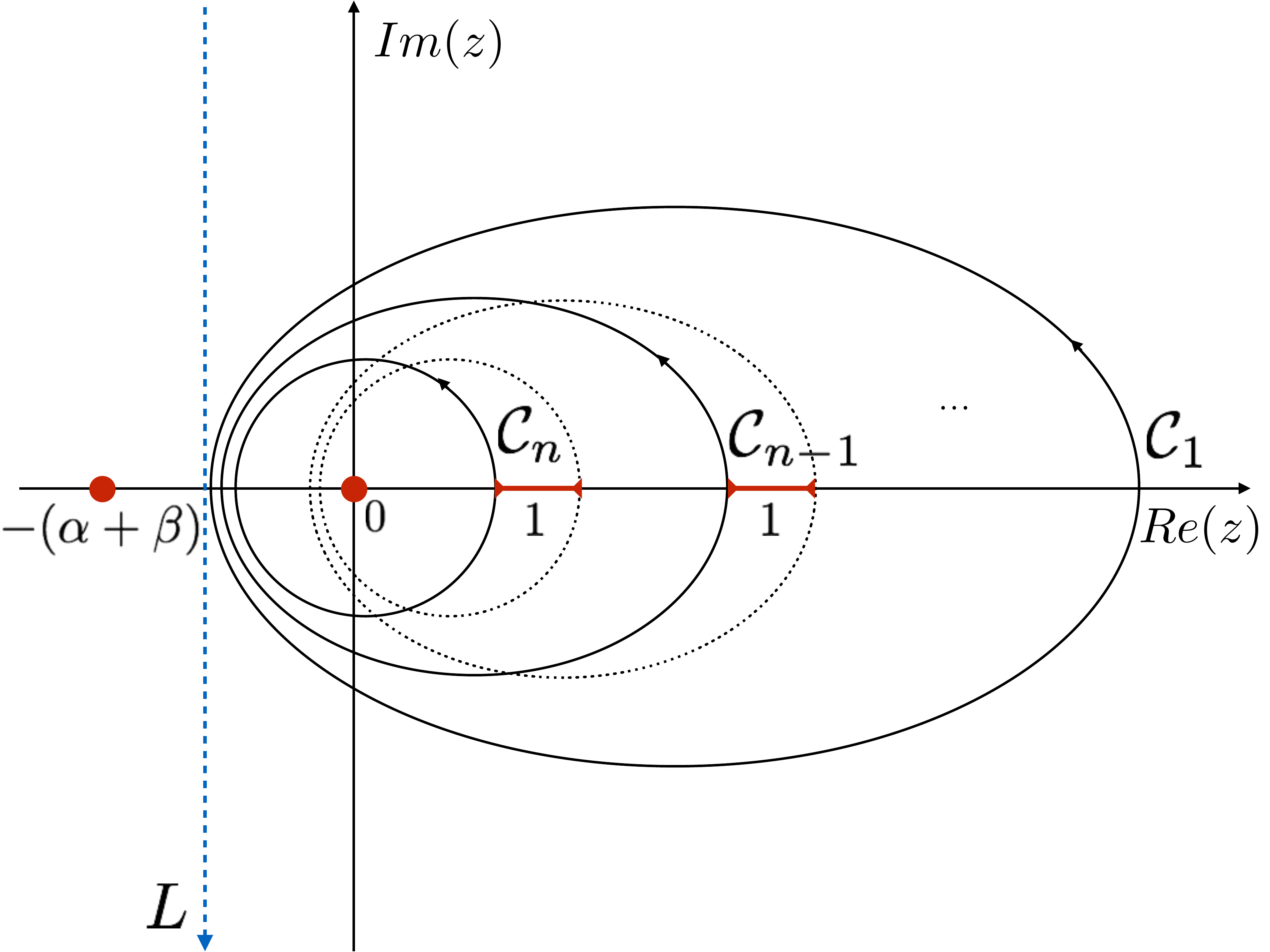}} 
\caption{The different contours involved in (\ref{momentFormula05}) and (\ref{momentFormula06}). In (\ref{momentFormula05}) the $n$ variables $z_i$ are integrated on the vertical line $L$ which is oriented from top to down (dotted blue above). In (\ref{momentFormula06}) the $z_i$ variables are integrated on different contours ${\cal C}_i$ organized in a nested fashion as explained in the text. In the above picture the red colour is used to emphasize the important properties of the nested contours: they enclose $0$ but not $-(\alpha +\beta)$, and are such that the contour ${\cal C}_{j+1} +1$ (dotted contours above) is inside 
${\cal C}_{j}$ $\forall j =2 , \cdots , n$.}
\label{fig:nestedcontours}
\end{figure}

We first note that the first part of the integrand in (\ref{momentFormula02}), i.e. the interaction term between particles, can be rewritten as:
\bea \label{symmetrization}
    \prod_{1\leq i < j  \leq n} \frac{(k_i-k_j)^2}{(k_i-k_j)^2 + 1} =\frac{1}{n!} \sum_{\sigma \in S_n}  \prod_{1\leq i < j  \leq n} \frac{k_{\sigma(i)}-k_{\sigma(j)}}{k_{\sigma(i)}-k_{\sigma(j)}+ i } \ .
\eea
where $S_n$ is the group of permutation of $\{1, \cdots, n\}$. This identity is shown in App.~\ref{app:ProofSymmmetrization}. Inserting (\ref{symmetrization}) in (\ref{momentFormula02}), using that the second part of the integrand 
\bea
 \prod_{j=1}^{n} 
\frac{(i k_j + \frac{\beta - \alpha}{2} )^t}{(i k_j + \frac{\alpha+\beta}{2})^{1 +x} (i k_j - \frac{\alpha+\beta}{2})^{ 1-x + t} }  
\eea
is symmetric by exchange $ k_i \leftrightarrow k_j$, relabelling $k_{\sigma(i)} \to k_i$, and using that the number of elements of $S_n$ is $n!$ we obtain:

\bea \label{momentFormula03}
&& \!\!\!\!\!\!\!\!\!\!\!\!\!\!\!\!\!\!\!\!\!\!\!\!\!\!  \overline{Z_t(x)^n}  = (-1)^n  \frac{\Gamma(\alpha+\beta+n)}{\Gamma(\alpha + \beta  )} 
\prod_{j=1}^{n}  \int_{- \infty}^{+\infty} \frac{dk_j}{2 \pi}
\prod_{1\leq i < j  \leq n} \frac{k_i-k_j}{k_i-k_j+ i }  \prod_{j=1}^{n} 
\frac{(i k_j + \frac{\beta - \alpha}{2} )^t}{(i k_j + \frac{\alpha+\beta}{2})^{1 +x} (i k_j - \frac{\alpha+\beta}{2})^{ 1-x + t} }  \ . 
\eea
Let us now make the change of variables $k_j = -i ( z_j + \frac{\alpha+\beta}{2})$. We obtain
\bea \label{momentFormula04}
&& \!\!\!\!\!\!\!\!\!\!\!\!\!\!\!\!\!\!\!\!\!\!\!\!\!\!  \overline{Z_t(x)^n}  =  \frac{\Gamma(\alpha+\beta+n)}{\Gamma(\alpha + \beta  )} 
\prod_{j=1}^{n}  \int_{L} \frac{dz_j}{2 \pi i }
\prod_{1\leq i < j  \leq n} \frac{z_i-z_j}{z_i-z_j-1 }  \prod_{j=1}^{n} 
\frac{(z_j + \beta )^t}{(z_j + \alpha+\beta)^{1 +x} (z_j )^{ 1-x + t} }  \ ,
\eea
where here the contour $L = - \frac{\alpha+\beta}{2} + i \mathbb{R}$ is oriented from top to down. Note that the poles of the integrand are now located at $z_j = 0$, $z_j = - (\alpha + \beta)< 0$ and $z_j = z_i -1$ if $i<j$. Apart from these, the integrand being analytic in $z_n$, and decaying as $1/z_n^2$ at infinity, the contour of integration on $z_n$ can be transformed into ${\cal C}_n$, a small, positively oriented, circle around $0$ and that excludes $- (\alpha + \beta)$. Note that this transformation of contour is only possible because we have eliminated the poles at $z_j = z_i+1$ for $j<i$ that would have arose if one had performed the change of variable $k_j \to z_j$ directly on (\ref{momentFormula02}) and not on (\ref{momentFormula03}). At this point, one can now recursively close all contours in a so-called nested manner so that, $\forall i < j $, ${\cal C}_i$ is positively oriented, contains ${\cal C}_{j}+1$ (to avoid the pole at $z_j = z_i -1$ for $i<j$) and excludes $-(\alpha + \beta)$ and $0$ (see Fig.~\ref{fig:nestedcontours}). We thus obtain
\bea \label{momentFormula05}
&& \!\!\!\!\!\!\!\!\!\!\!\!\!\!\!\!\!\!\!\!\!\!\!\!\!\!  \overline{Z_t(x)^n}  =  \frac{\Gamma(\alpha+\beta+n)}{\Gamma(\alpha + \beta  )} 
\prod_{j=1}^{n}  \int_{{\cal C}_j} \frac{dz_j}{2 \pi i }
\prod_{1\leq i < j  \leq n} \frac{z_i-z_j}{z_i-z_j-1 }  \prod_{j=1}^{n} 
\frac{(z_j + \beta )^t}{(z_j + \alpha+\beta)^{1 +x} (z_j )^{ 1-x + t} }  \ .
\eea

This formula is almost identical to the formula obtained in Proposition 3.4 of \cite{BarraquandCorwinBeta} for the moments of the partition sum in the half-line to point Beta polymer problem, using $n_j = t-x + 1$ and $\beta \leftrightarrow \alpha$. Actually, suppressing the first factor of Gamma function in (\ref{momentFormula05}) and changing 
\bea
 \prod_{j=1}^{n} 
\frac{(z_j + \beta )^t}{(z_j + \alpha+\beta)^{1 +x} (z_j )^{ 1-x + t} }  \to  \prod_{j=1}^{n} 
\frac{(z_j + \beta )^t}{(z_j + \alpha+\beta)^{x } (z_j )^{ 1-x + t} }  \ ,
\eea
one obtains exactly the same formula. Hence, from a computational point of view, the  half-line to point problem and the point to point problem appears extremely similar for the Beta polymer. This interesting fact and the formula (\ref{momentFormula05}) was first pointed out to us by BC \cite{Barraquand-Corwin-private}. In the end it appears that it is this similarity between the half-line to point and the point to point problem that is responsible for the fact that both problems have exactly the same fluctuations in the large deviations regime (see the discussion below (\ref{SDPeqn})).

\medskip

{\it Multi-points formulas:}

Interestingly, using the techniques used in \cite{BarraquandCorwinBeta} it is also possible to obtain a formula for the multi-point moments of the partition sum of the point-to-point Beta polymer as
\bea \label{momentFormula06}
&& \!\!\!\!\!\!\!\!\!\!\!\!\!\!\!\!\!\!\!\!\!\!\!\!\!\!  \overline{Z_t(x_1) \cdots Z_t(x_n)}  =  \frac{\Gamma(\alpha+\beta+n)}{\Gamma(\alpha + \beta  )} 
\prod_{j=1}^{n}  \int_{{\cal C}_j} \frac{dz_j}{2 \pi i }
\prod_{1\leq i < j  \leq n} \frac{z_i-z_j}{z_i-z_j-1 }  \prod_{j=1}^{n} 
\frac{(z_j + \beta )^t}{(z_j + \alpha+\beta)^{1 +x_j} (z_j )^{ 1-x_j + t} }  \ ,
\eea
where the contours are the same nested contours as those used in (\ref{momentFormula05}) and here $0 \leq x_1 \leq x_2 \leq ... \leq x_n $. The existence of this formula was pointed out to us by BC \cite{Barraquand-Corwin-private}. From this formula, successively un-nesting the contours from the ${\cal C}_i$ to the $L$ in an opposite manner as what was just done (see Fig.~\ref{fig:nestedcontours}), and performing the change of variables $z_j = i k_j - \frac{\alpha+\beta}{2}$, we obtain the formula (\ref{momentFormula07}) that was extensively discussed above. 

\medskip
{\it Moment formula for the Beta polymer in terms of strings}

Finally, it is also possible to obtain a formula for the moments of the point-to-point Beta polymer problem using a decomposition into strings. Proving this formula involves successively shrinking the contours ${\cal C}_i$ in (\ref{momentFormula05}) on ${\cal C}_n$ and keeping track of all the residues encountered in the calculation. This procedure is actually quite tedious but the steps being exactly similar as those performed in \cite{BarraquandCorwinBeta} we can easily adapt them to obtain \footnote{following the above remarks, (\ref{momentformula08}) is an adaptation of Proposition 3.6. of \cite{BarraquandCorwinBeta} with an incorrect minus sign in the determinant there, which has been corrected here}

\bea \label{momentformula08}
&& \!\!\!\!\!\!\!\!\!\!\!\!\!\!\!\!\!\!\!\!\!\!\!\!\!\!\!\!\!\!\!\!\! \overline{Z_t(x)^n} = \frac{\Gamma(\alpha + \beta+ n)}{\Gamma(\alpha + \beta)} n! \sum_{n_s=1}^n  \frac{1}{n_s!} \sum_{(m_1,..m_{n_s})_n} 
\prod_{j=1}^{n_s}  \int_{{\cal C}^n} \frac{dz_j}{2 i \pi} {\rm det}\left( \frac{1}{z_i +m_i - z_j}   \right)_{n_s \times n_s } \nn \\
&&
\prod_{j=1}^{n_s}  
 \left( \frac{  \Gamma(z_j +\alpha + \beta  ) }{  \Gamma(z_j +\alpha + \beta + m_j)  } \right)^{1 +x} \left( \frac{  \Gamma( z_j) }{  \Gamma(z_j + m_j)  } \right)^{1 -x +t }  \left( \frac{ \Gamma ( z_j + \beta+ m_j)}{ \Gamma (  z_j + \beta)}  \right)^t \ ,
\eea
where ${\cal C} $ is a small circle around $0$ of radius $r < 1/2$, excluding $-(\alpha + \beta)$ and as in (\ref{BetaWithStrings}) $\sum_{(m_1,\cdots,m_{n_s})_n} $  means summing over all $n_s$-uplets $(m_1 , \cdots, m_{n_s})$ such that $ \sum_{i=1}^{n_s} m_i  = n$. We note that performing in this formula the shifts $z_j =  i k_j - \frac{\alpha+\beta}{2} - m_j/2$, we obtain, using
\bea
{\rm det} \left[ \frac{1 }{i(k_i -k_j) + (m_i +m_j)/2}  \right]_{n_s \times n_s } = \prod_{i=1}^{n_s} \frac{1}{m_i} \prod_{1 \leq i < j \leq n_s} \frac{4(k_i-k_j)^2 + (m_i - m_j)^2}{4(k_i-k_j)^2 + (m_i +m_j)^2}
\eea

exactly the  integrand inside formula (\ref{BetaWithStrings}), with the difference that here all contours of integrations are well specified and are {\it different} for each $k_i$. It does not seem possible to deform all these contours of integrations onto a single contour $L$ (one would encounters many poles in doing so) as suggested in the formal formula (\ref{BetaWithStrings}) and this explain why we could only find a formal formula. The formula (\ref{momentformula08}) thus appears as the correct interpretation of (\ref{BetaWithStrings}))
as a contour integral.

\subsection{Mellin-Barnes type Fredholm determinant}

Finally, following the same steps as in \cite{BarraquandCorwinBeta} for the half-line to point problem, we obtain from (\ref{momentformula08}) another Fredholm determinant formula $g_{t,x}(u) = {\rm Det}\left(I + K^{MB}_{t,x} \right)$ defined in (\ref{defg}) with the kernel \footnote{similarly (\ref{KernelSmallContour}) is an adaptation of Theorem 1.12. of \cite{BarraquandCorwinBeta} with the minus sign missprint there signaled above corrected here as well}
\bea \label{KernelSmallContour}
&& K^{MB}_{t,x}(v,v') = \frac{1}{(2 \pi i)^2} \int_{1/2 - i \infty }^{1/2 + i \infty} \frac{\pi ds}{\sin(\pi s)} (u)^s  \left( \frac{  \Gamma(v +\alpha + \beta  ) }{  \Gamma(v+\alpha + \beta + s)  } \right)^{1 +x} \left( \frac{  \Gamma( v) }{  \Gamma(v+s)  } \right)^{1 -x +t }  \left( \frac{ \Gamma ( v+ \beta+ s)}{ \Gamma (  v+ \beta)}  \right)^t  \frac{1}{v'-v-s} \ , \nonumber \\
&& 
\eea
where $(v,v') \in {\cal C}_0^2$, with ${\cal C}_0$ a small circle around $0$ of radius $r<1/4$ excluding $-(\alpha + \beta)$ and $-1$. This formula is valid for $u \in \mathbb{C} \backslash \mathbb{R}_-$. Repeating the saddle-point analysis on this kernel would lead to the same result (\ref{SDPeqn}) as the one found heuristically earlier.

\section{Numerical results} \label{Sec:Numerics}

In this section we verify numerically some of the results obtained in the paper. All the results presented in the following are based on numerical simulations of the Beta polymer with parameters $\alpha = \beta =1$. Using a transfer matrix type algorithm, we compute numerically the partition sum for $5 \times 10^5$ different random environments. For each random environment, we store the value of the partition sum for different polymers length $t$ from $t = 90$ to $t=2048$ with a power-law type binning as $t=t_i = \lfloor 128 \sqrt{2}^{i/10} \rfloor$ with $i=1 , \cdots , 10$, and for different positions. The studied positions are chosen as $x= \lfloor t/2 + \hat x\rfloor $ with either $\hat x = \kappa_i t^{\frac{3}{4}}$ with $\kappa_i = 3 i/20$ and $i = 0, \cdots , 20$ (to study the diffusive regime around the optimal direction $\varphi_{opt}(1) = 0$) or $\hat x = \varphi_i t$ with $\varphi_i = i/40$ with $i=0 , \cdots ,20$ (to study the large deviations regime of the TD-RWRE).

\subsection{In the diffusive regime.}

{\it One-point}

Let us first focus on one-point statistics of the rescaled spatial process at finite time ${\cal Z}_{t}(\kappa)$ (see (\ref{rescaledcalZ})). Using the simulations described previously, we obtain numerical approximations of its PDF for different times $t_i$ and diffusion parameters $\kappa_j$, and compare it to the infinite time prediction (\ref{resrescaledcalZ}) which we recall here (for the choice $\alpha=\beta=1$): ${\cal Z}_{\infty}(\kappa)$ is distributed as ${\cal Z} \sim Gamma(2)$. The PDF of ${\cal Z}$ is thus (see (\ref{distGamma}) for the PDF of a Gamma RV with parameter $\alpha+\beta$)
\bea \label{PGamma2}
P_{Gamma(2)}({\cal Z} ) = {\cal Z}  e^{-{\cal Z} } \ .
\eea

\begin{figure}
\centerline{\includegraphics[width=6.5cm]{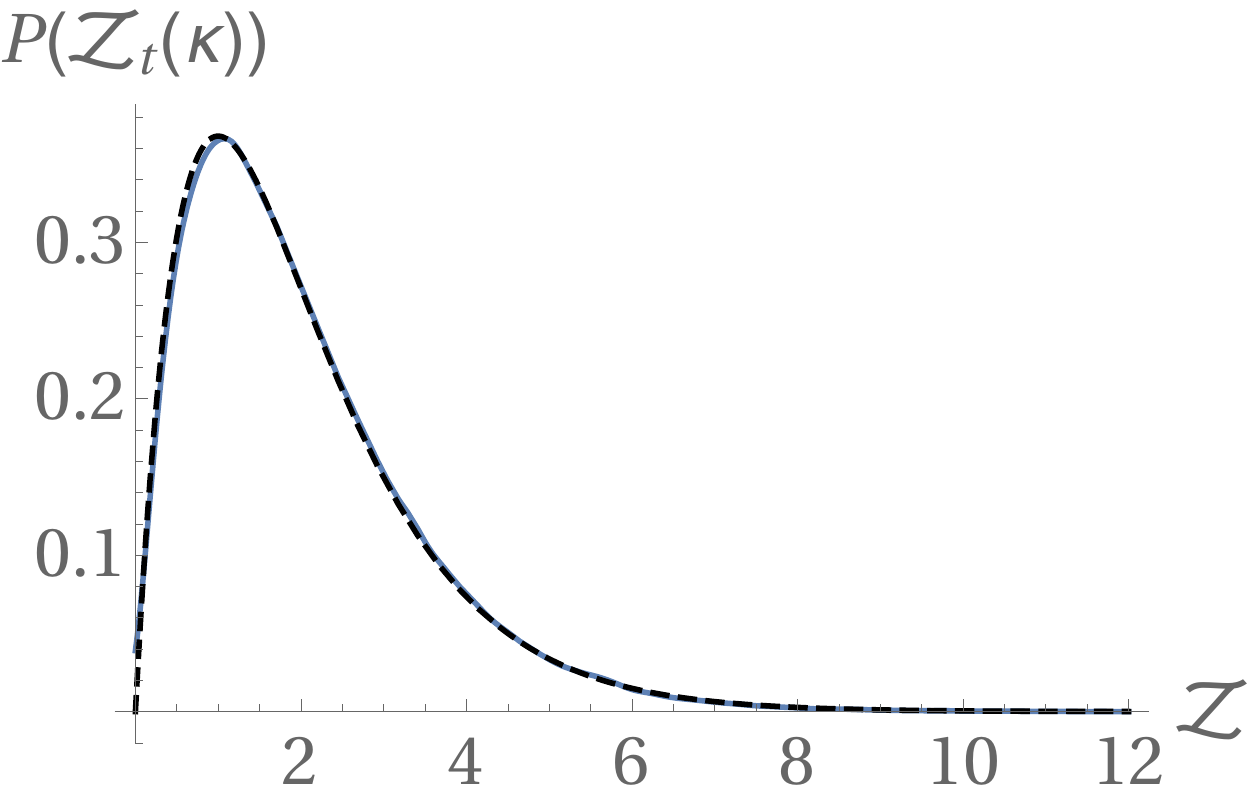} \includegraphics[width=6.5cm]{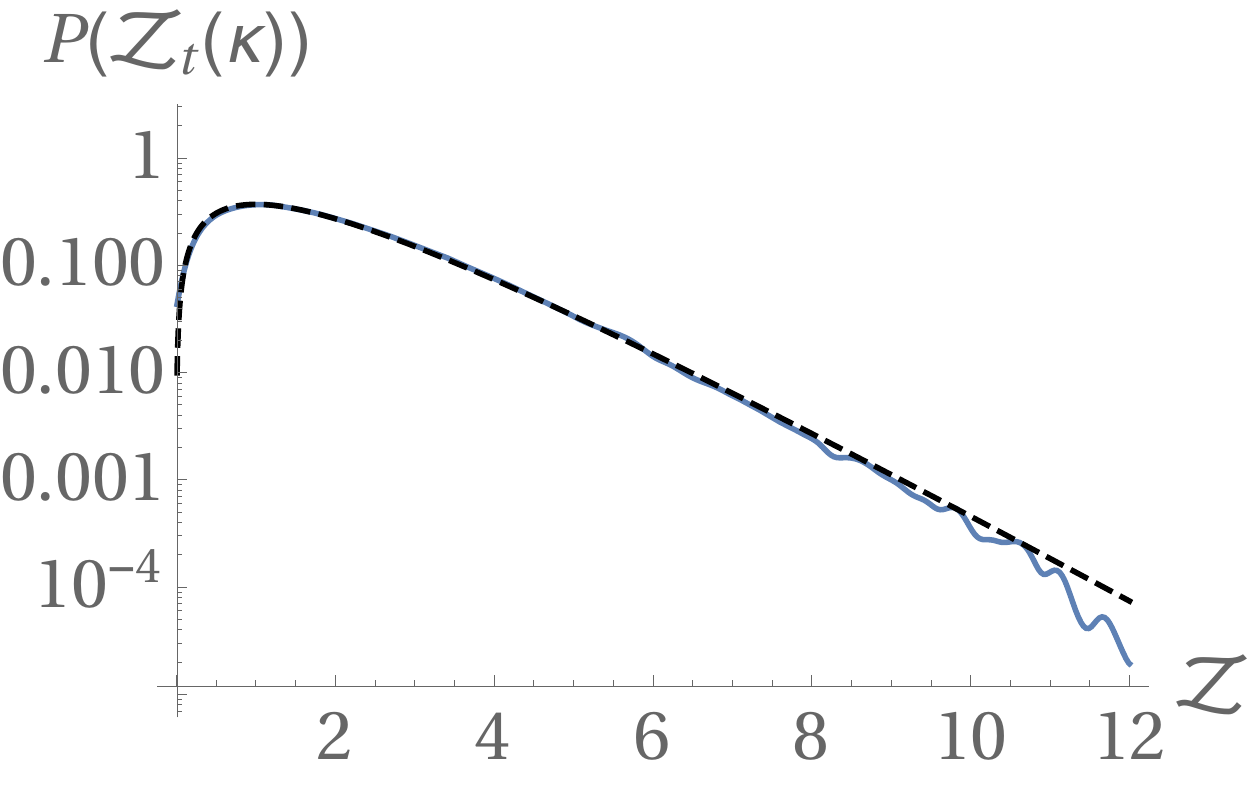}} 
\caption{Left: Blue line: Empirical PDF of ${\cal Z}_{t = 2048}(0)$ obtained from the numerical simulations. Black-dashed line: PDF of a $Gamma(2)$ distributed RV (\ref{PGamma2}). Right: Same figure in a logarithmic scale. There are no fitting parameters.}
\label{fig:diff1}
\end{figure}

\begin{figure}
\centerline{\includegraphics[width=6cm]{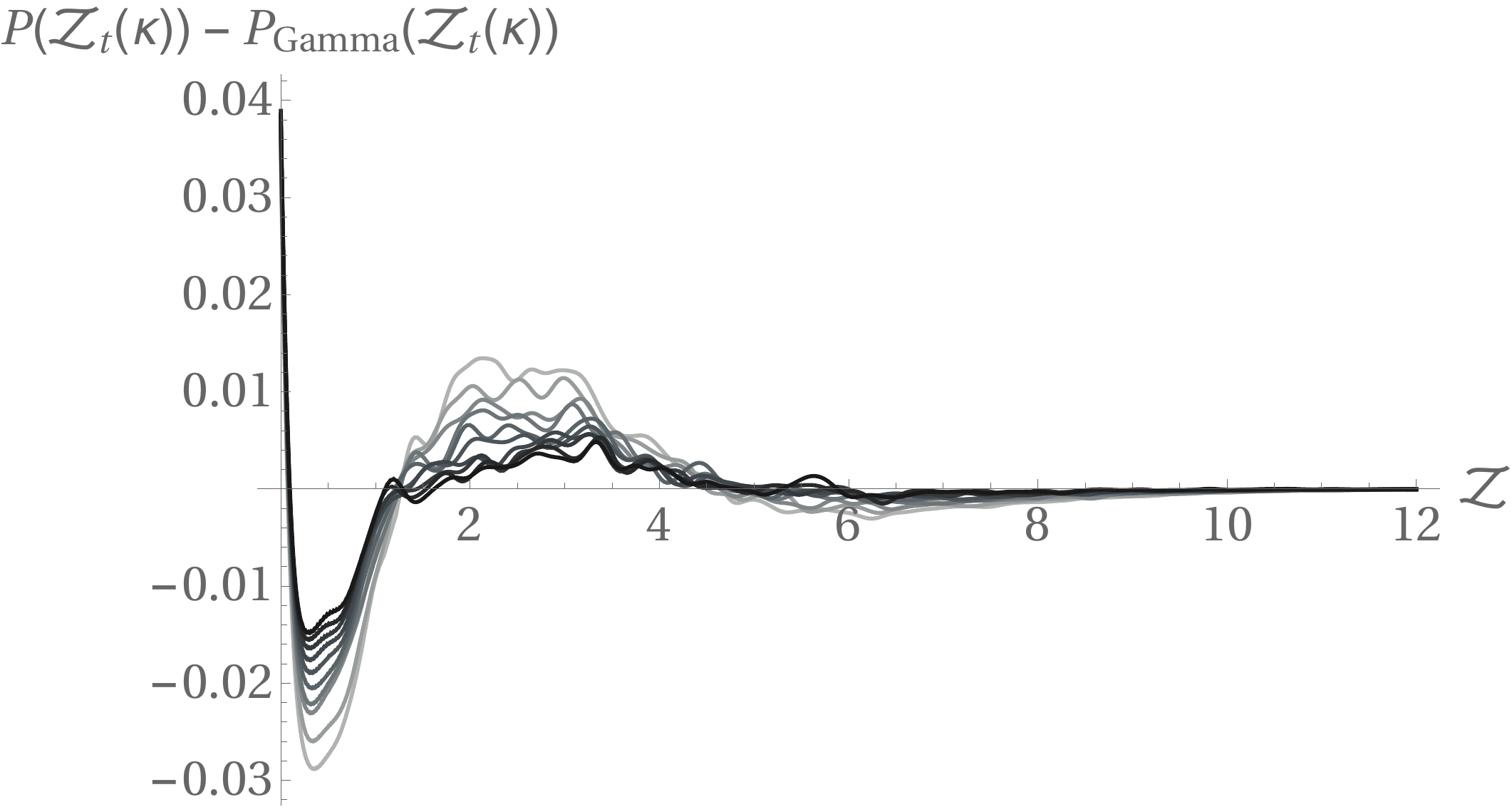} \includegraphics[width=6cm]{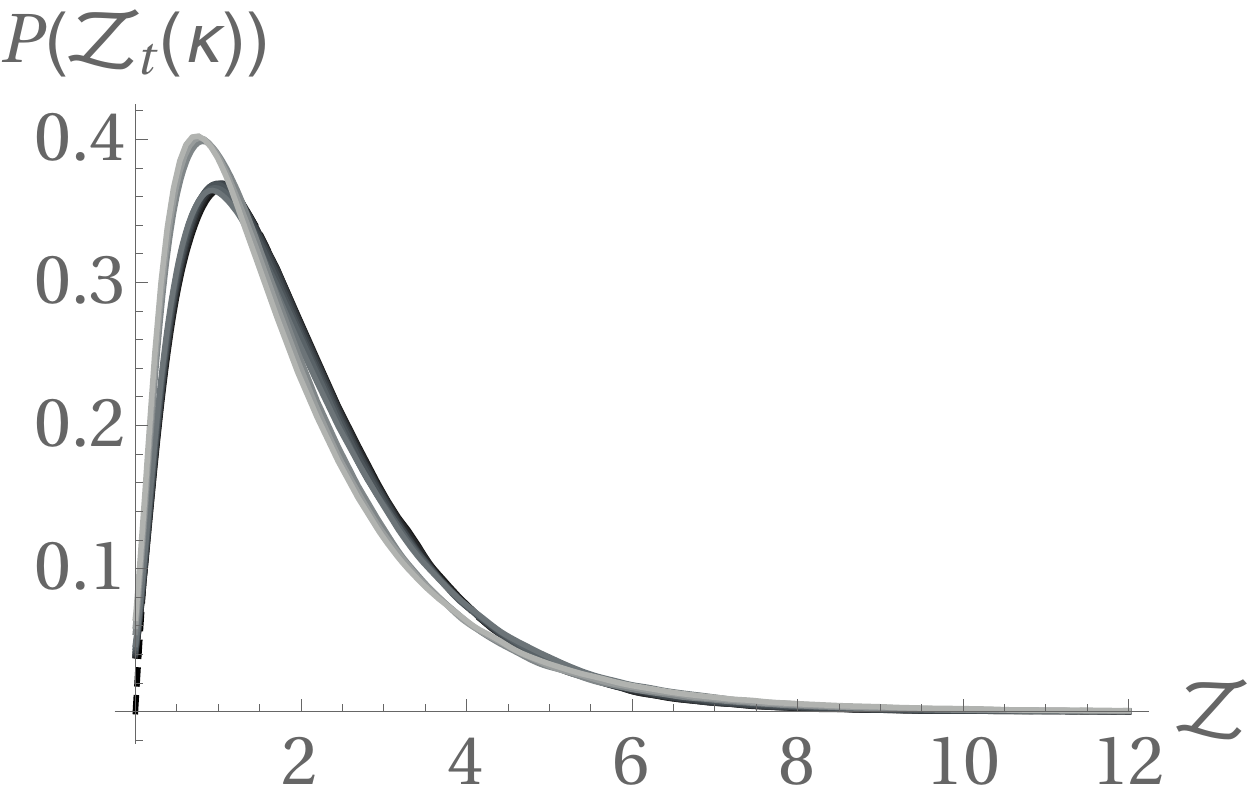} \includegraphics[width=6cm]{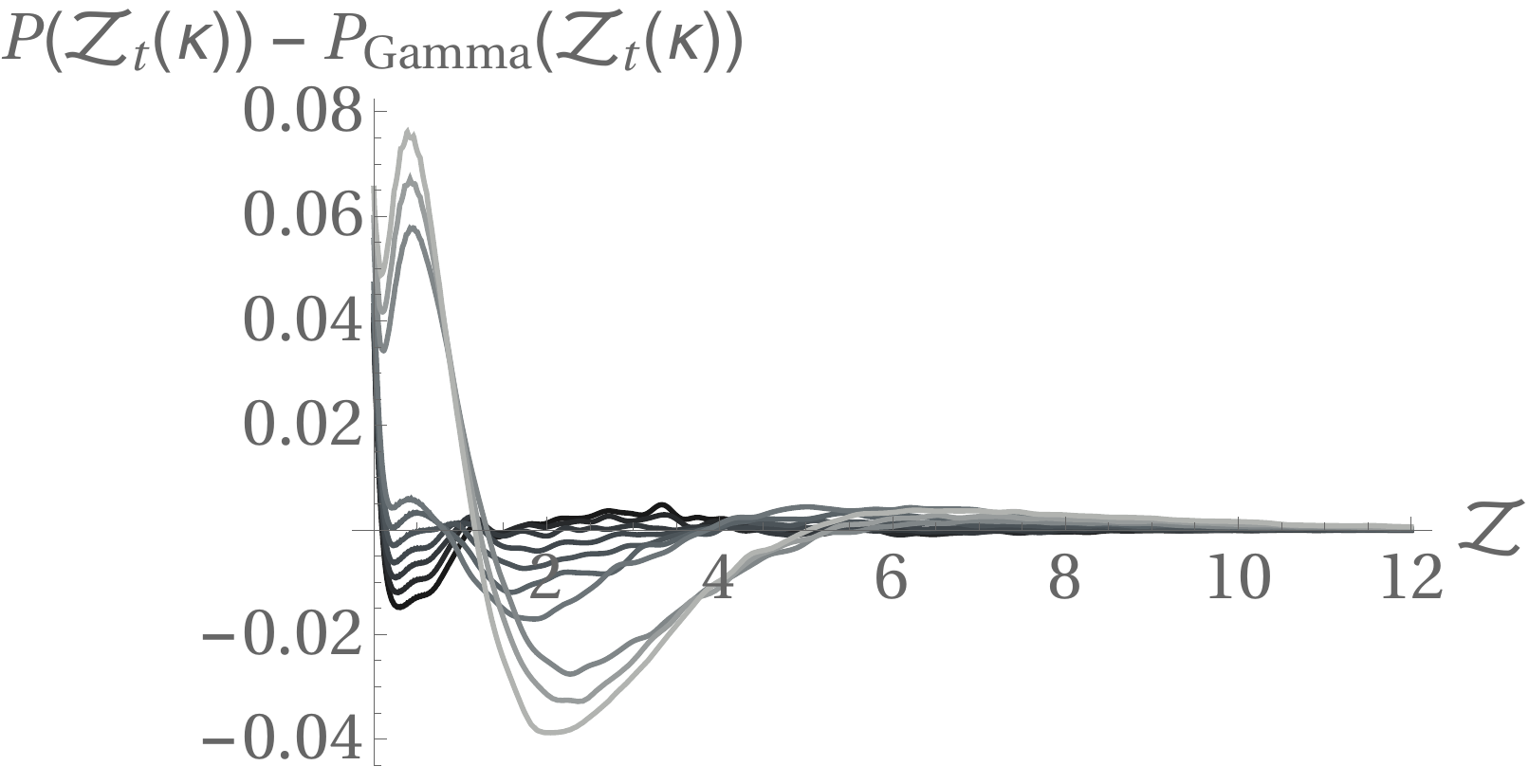}} 
\caption{Left: Difference between the empirical PDF of ${\cal Z}_{t_i}(0)$ and the PDF of a $Gamma(2)$ distributed RV (\ref{PGamma2}) for $i=1,\cdots,10$ (from light gray to black). Middle: Empirical PDF of ${\cal Z}_{t=2048}(\kappa_i)$ for $i=0, \cdots , 9$ (from black to light gray) together with the PDF of a $Gamma(2)$ distributed RV (\ref{PGamma2}) (black-dashed line). Right: Difference between the empirical PDF of ${\cal Z}_{t=2048}(\kappa_i)$ and the PDF of a $Gamma(2)$ distributed RV (\ref{PGamma2}) for $i=0, \cdots , 9$ (from black to light gray). There are no fitting parameters.}
\label{fig:diff2}
\end{figure}

In Fig.~\ref{fig:diff1} we compare the numerically obtained PDF for $t=2048$ and $\kappa = 0$ (i.e. exactly in the central region) in log and linear scale. The agreement is excellent. In Fig.~\ref{fig:diff2} we show how this result vary as a function of the diffusion constant $\kappa$ and the time $t$. As a function of the diffusivity constant $\kappa$. Important deviations from the asymptotic behavior start to appear only for $\kappa \geq \kappa_7 \simeq 0.95 $. To compare this value, note that the mean value of $Z_t( \kappa \sqrt{t})$ is theoretically predicted to converge to a gaussian form with
\bea
\overline{ Z_t( \kappa \sqrt{t}) }  \simeq  \frac{1}{\sigma \sqrt{ 2 \pi t}} e^{ - \frac{\kappa^2}{2 \sigma^2} } \quad , \quad  \sigma =1/2 \ .
\eea
Hence $\kappa_7 / \sigma  \simeq 1.9$ and in terms of probability in the RWRE picture, more than $94\%$ of the accessible positions of the particles appear very well-described by our asymptotic result for $t=2048$.

\medskip

{\it Multi-points}

\begin{figure}
\centerline{\includegraphics[width=6cm]{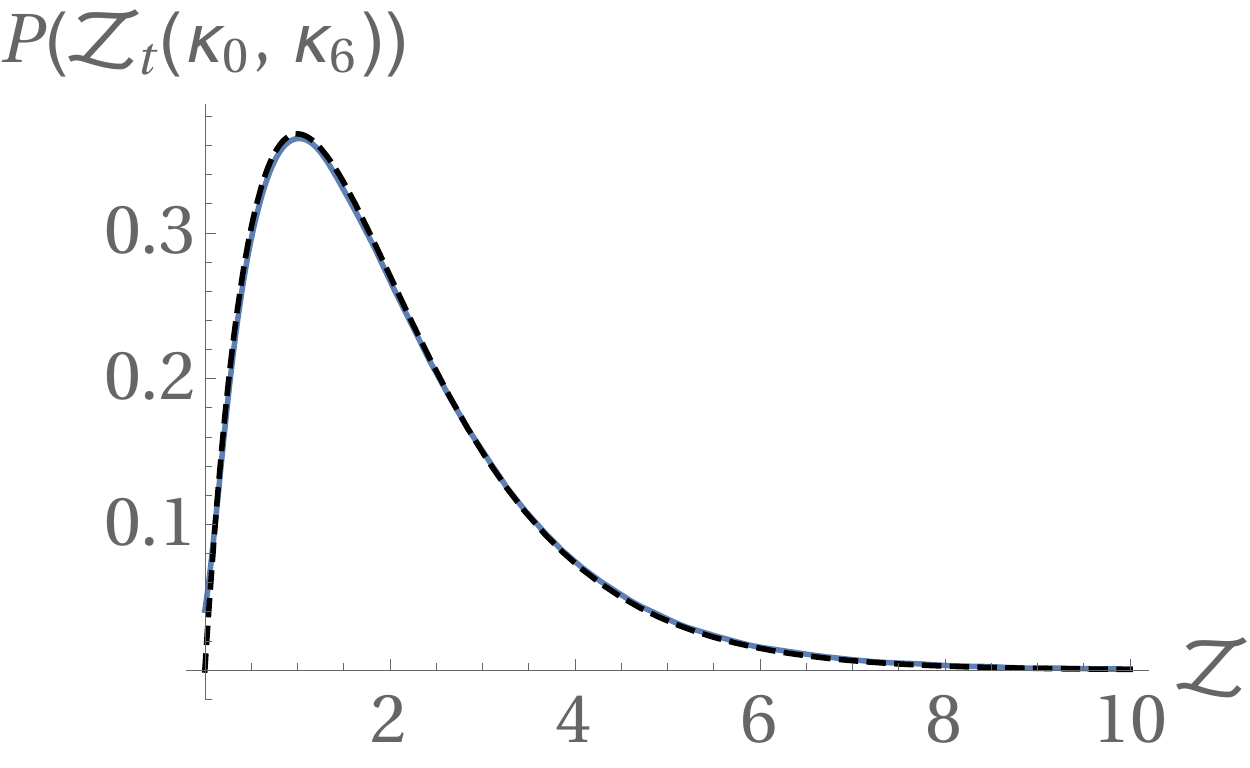} \includegraphics[width=6cm]{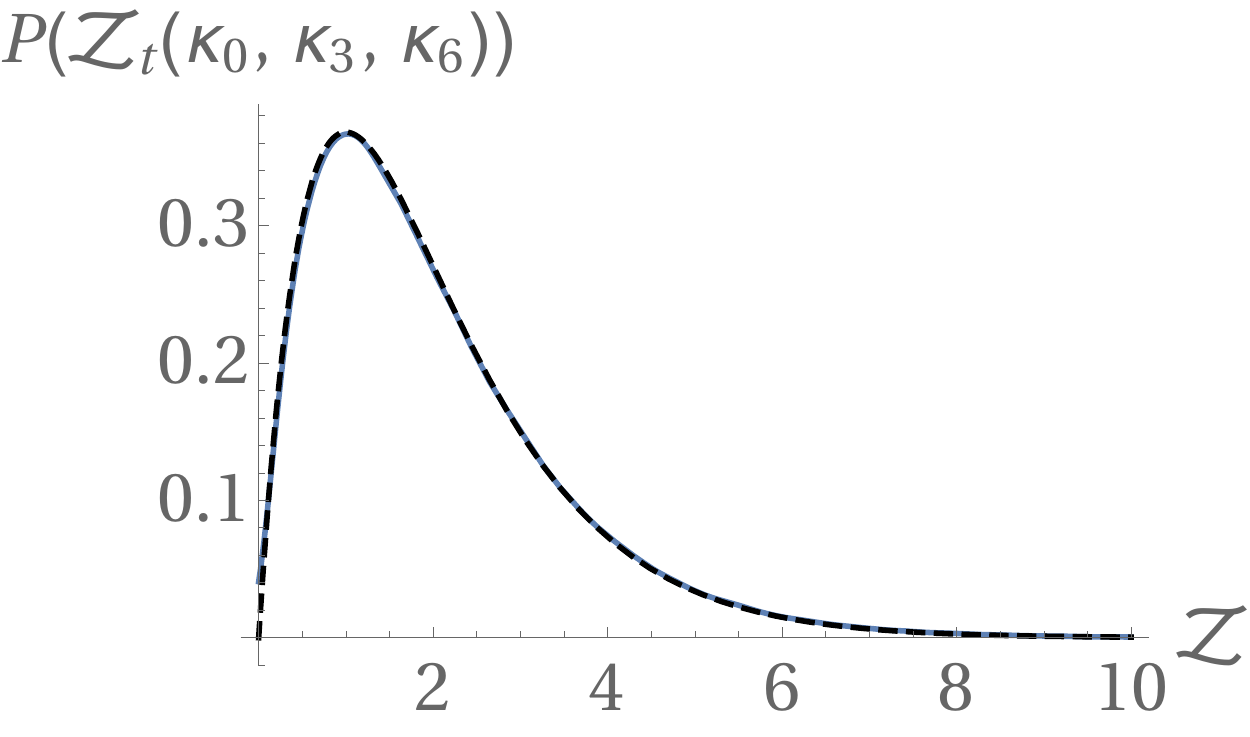} \includegraphics[width=6cm]{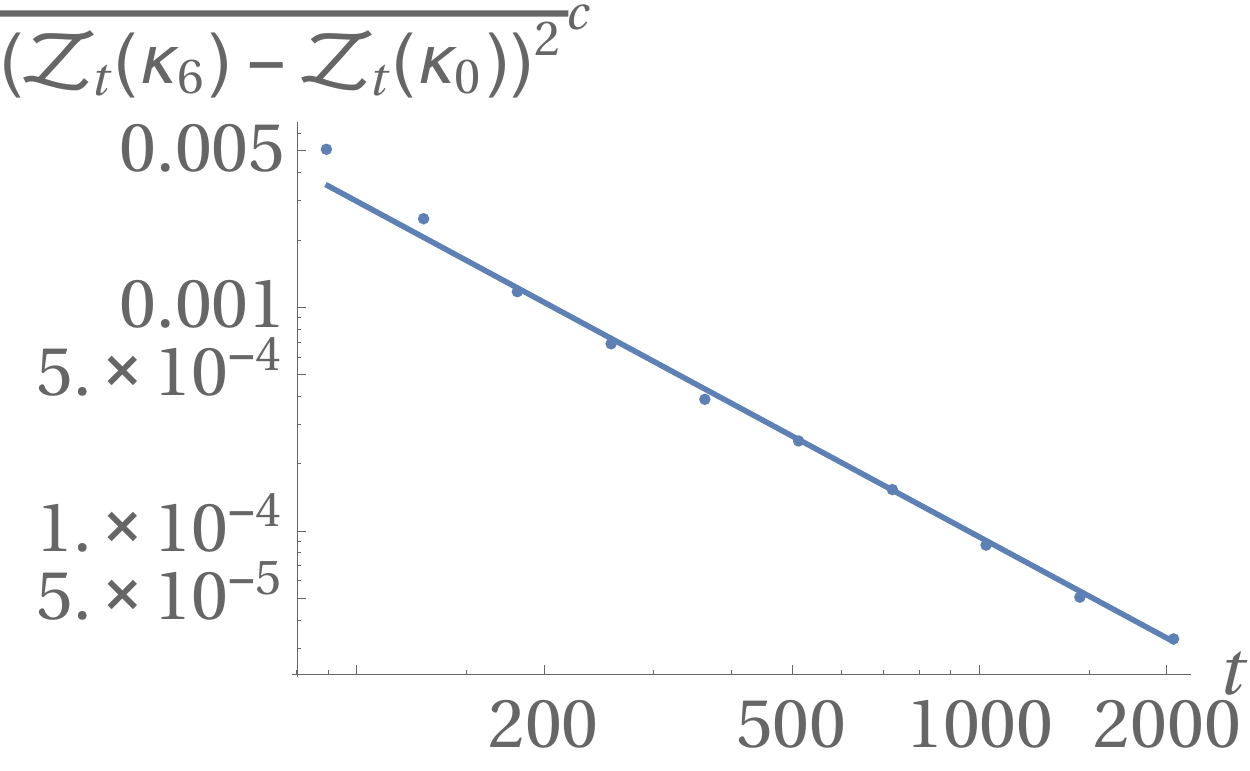}} 
\caption{Left (resp. Middle): Blue line: Empirical PDF of ${\cal Z}_{t=2048}(\kappa_0,\kappa_6) $ (resp. ${\cal Z}_{t=2048}(\kappa_0,\kappa_3,\kappa_6) $). Black-dashed line: PDF of a $Gamma(2)$ distributed RV (\ref{PGamma2}) (here there are no fitting parameters.). Right: Blue dots: Empirical variance $\overline{ \left( {\cal Z}_t(\kappa_6)-{\cal Z}_t(\kappa_0) \right)^2 }^c$ obtained from numerical simulations in a log-log scale. The blue line corresponds to a power-law decay as $1/t^{3/2}$.}
\label{fig:diff3}
\end{figure}

Let us now verify some of the predictions of our results for the multi-points correlations in the diffusive regime. The prediction (\ref{resrescaledcalZ}) is that, up to $O(1/\sqrt{t})$ deviations, ${\cal Z}_t(\kappa)$ converges to a constant process with marginal distribution ${\cal Z}_{\infty} \sim Gamma(2)$. We show here the verification of two consequences of this result. The first one is that, at fixed $\kappa^{(1)} , \dots , \kappa^{(n)}$ for $n$ arbitrary, the RV ${\cal Z}_t(\kappa^{(1)} , \cdots, \kappa^{(n)}) = \left( {\cal Z}_t(\kappa^{(1)}) \cdots {\cal Z}_t(\kappa^{(n)}) \right)^{1/n}$ converges to a $Gamma(2)$ distributed RV. In Fig.~\ref{fig:diff3} we compare this prediction with the numerically obtained PDF of ${\cal Z}_t(\kappa^{(1)} , \cdots, \kappa^{(n)})$ for $n=2$ with $\kappa^{(1)}=\kappa_0$ and $\kappa^{(2)}=\kappa_6$ and $n=3$ with $\kappa^{(1)}=\kappa_0$, $\kappa^{(2)}=\kappa_3$ and $\kappa^{(3)}=\kappa_6$ and $t=2048$ and obtain an excellent agreement. Another implication of this result is that the variance $\overline{ \left( {\cal Z}_t(\kappa^{(1)})-{\cal Z}_t(\kappa^{(2)}) \right)^2 }^c$ must, for arbitrary $\kappa^{(1)}$ and $\kappa^{(2)}$, decay to $0$ at large $t$ faster than $1/\sqrt{t}$ (since corrections to (\ref{resrescaledcalZ}) are $O(1/\sqrt{t})$). We show in Fig.~\ref{fig:diff3} that this is the case for $\kappa^{(1)}=\kappa_0$ and $\kappa^{(2)}=\kappa_7$ and actually measure a faster decay as $1/t^{3/2}$.

\subsection{In the large deviations regime.}

\begin{figure}
\centerline{\includegraphics[width=6cm]{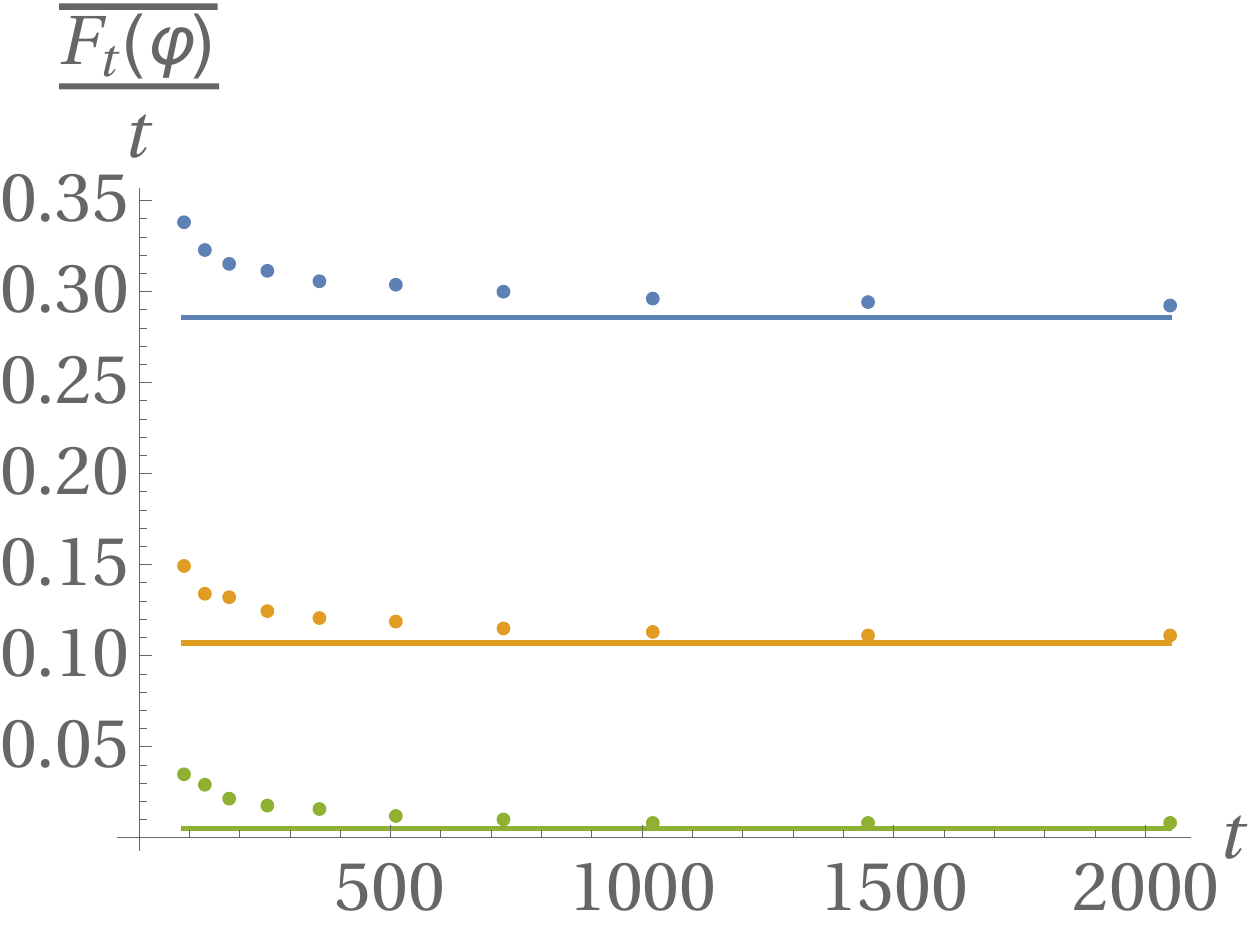} \includegraphics[width=6cm]{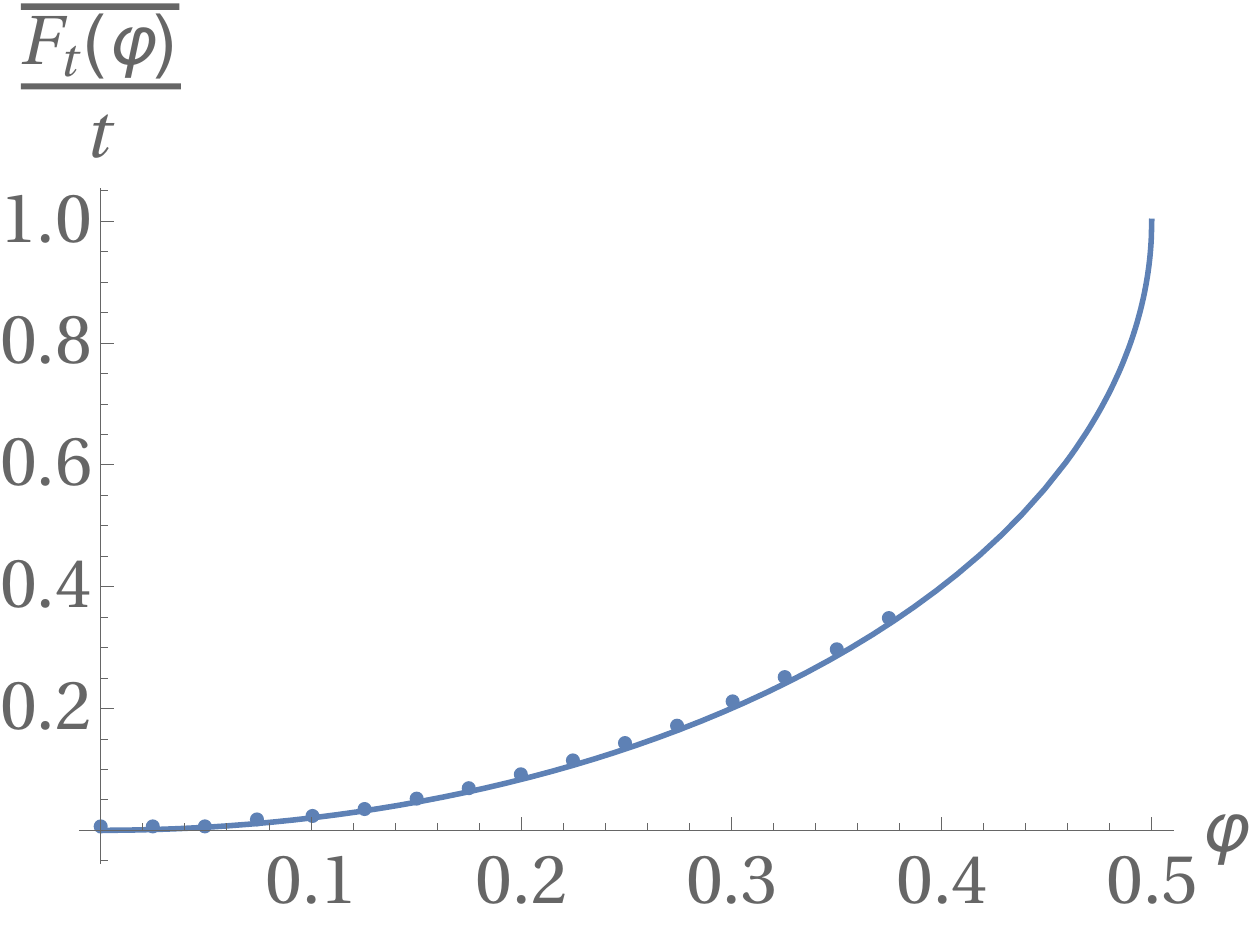}} 
\caption{Left: Dots: Empirical mean value $\overline{F_t(\varphi_i t)}/t$ as a function of $t$ for $i=2,9,14$ (green, orange and blue). Lines: asymptotic prediction (\ref{asymptNum}). Right: Dots: Empirical mean value $\overline{F_{t=2048}(\varphi t)}/t$ as a function of $\varphi$. Line: asymptotic prediction (\ref{asymptNum}). The numerical values for $\varphi \geq 0.4$ are not obtained due to numerical errors caused by the difficulty of dealing both with 'large' ($\sim 1/\sqrt{t}$) values of the partition sum at small $\varphi$ and exponentially small ($ \sim e^{- c_{\varphi} t}$) values of the partition sum at large $\varphi$ that are simply set to $0$ by the algorithm. There are no fitting parameters..}
\label{fig:ball1}
\end{figure}

\begin{figure}
\centerline{\includegraphics[width=6cm]{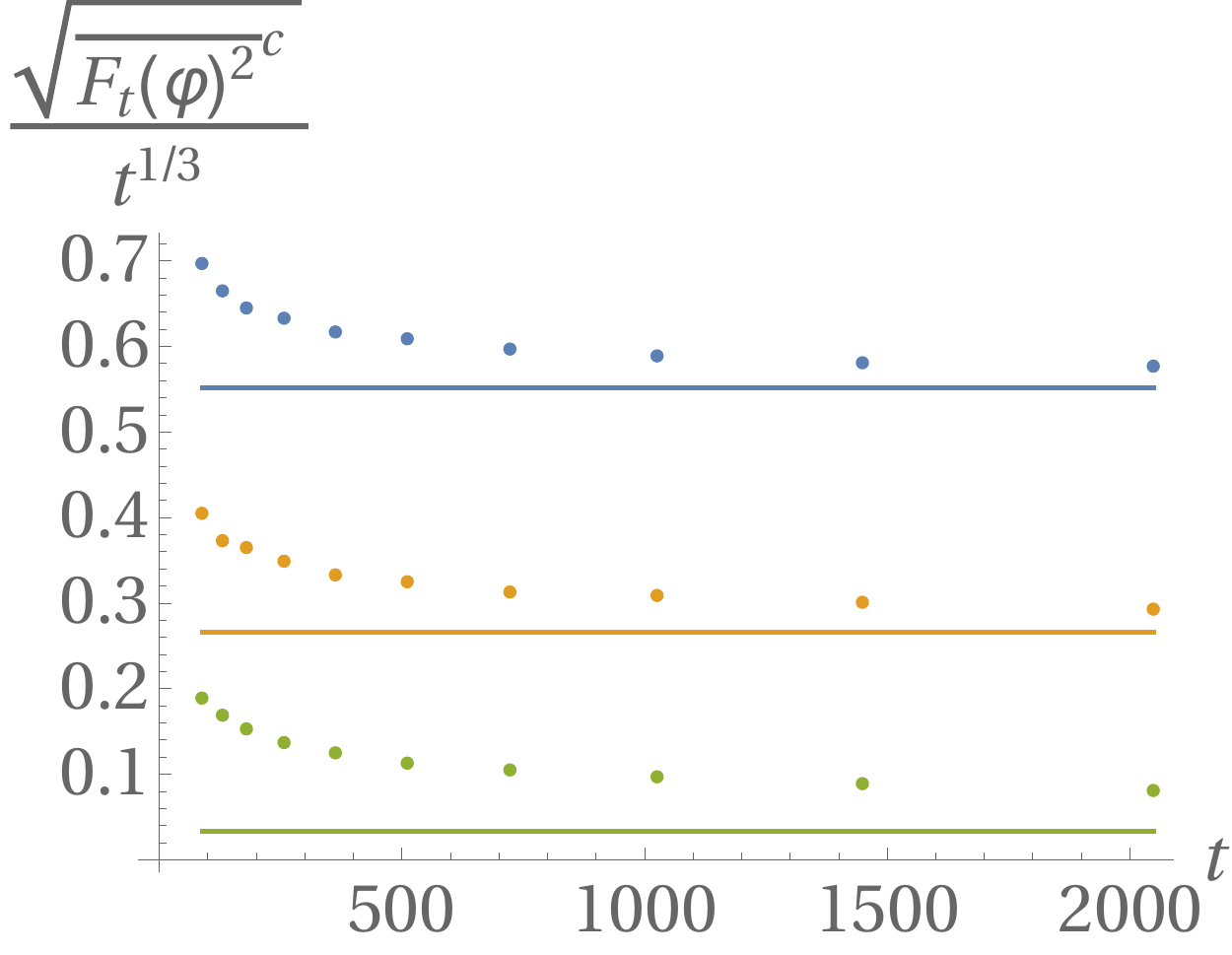} \includegraphics[width=6cm]{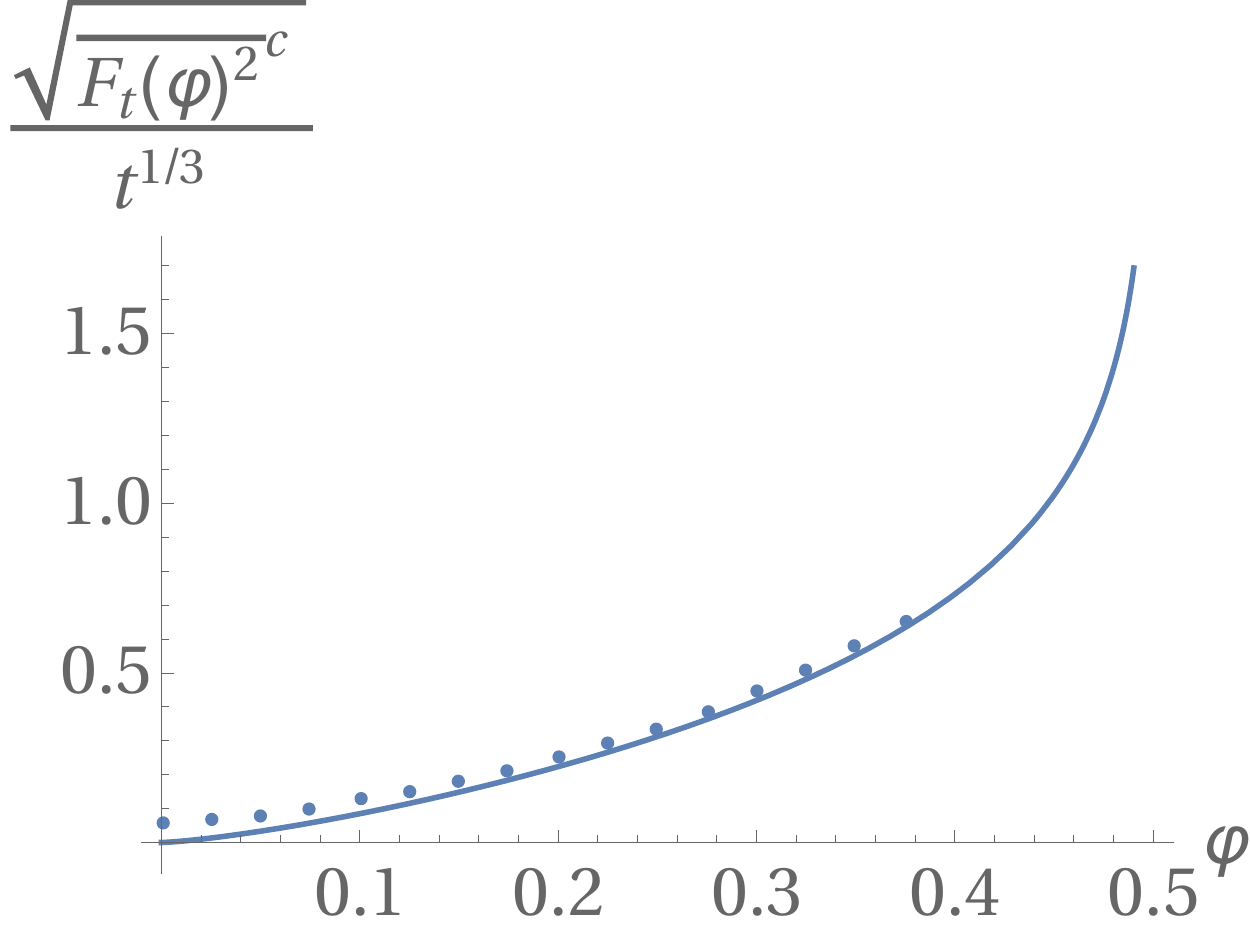}} 
\caption{Left: Dots: Empirical normalized standard deviation $\sqrt{\overline{F_t(\varphi_i t)^2}^c}/t^{\frac{1}{3} } $ as a function of $t$ for $i=2,9,14$ (green, orange and blue). Lines: asymptotic prediction (\ref{asymptNum}). Right: Dots: Empirical normalized standard deviation $\sqrt{\overline{F_t(\varphi_i t)^2}^c}/t^{\frac{1}{3} } $ as a function of $\varphi$. Line: asymptotic prediction (\ref{asymptNum}). The numerical values for the largest values of $\varphi$ are not obtained due to numerical errors, see Fig.~\ref{fig:ball1}. There are no fitting parameters.}
\label{fig:ball2}
\end{figure}

In the ballistic regime we check the result (\ref{asymptoticlim}) and (\ref{SDPeqn}). As a function of the angle $\varphi$ the predictions are notably that, noting $F_t(\varphi) = - \log Z_t(\varphi)$ the free-energy of the DP,
\bea \label{asymptNum}
\frac{\overline{F_t(\varphi)}}{t} \to_{t \to \infty} c_{\varphi}  \quad , \quad \frac{\overline{(F_t(\varphi))^2}^c}{ 8^{\frac{1}{3}} \lambda_{\varphi}^2} \to_{t \to \infty} Var (\chi_{GUE} )
\eea
where $Var( \chi_{GUE} ) \simeq 0.813$ is the variance of the Tracy-Widom GUE distribution and in the case $\alpha = \beta =1$ studied here the parameters $c_{\varphi}$ and $\lambda_{\varphi}$ given implicitly by (\ref{SDPeqn}) admits simple expressions (those were already obtain in \cite{BarraquandCorwinBeta} for the half-line to point problem):
\bea
c_{\varphi} =1 - \sqrt{1- 4 \varphi^2} \quad , \quad \lambda_{\varphi} = \left( \frac{1}{8} \frac{2 \left(1-\sqrt{1-4 \phi ^2}\right)^2}{\sqrt{1-4 \phi ^2}} \right)^{\frac{1}{3}} t^{\frac{1}{3}} \ .
\eea
These predictions are checked in Fig.~\ref{fig:ball1} and Fig.~\ref{fig:ball2} where we show both the dependence of these results on $t$ (hence the convergence to the infinite time prediction) and on the angle at fixed $t=2048$. The results are satisfying, though the convergence of the variance is slow as one approaches the optimal angle $\varphi =0$ where the Gamma-type fluctuations studied previously slowly start to dominate for finite time observations.

\medskip

\begin{figure}
\centerline{\includegraphics[width=6cm]{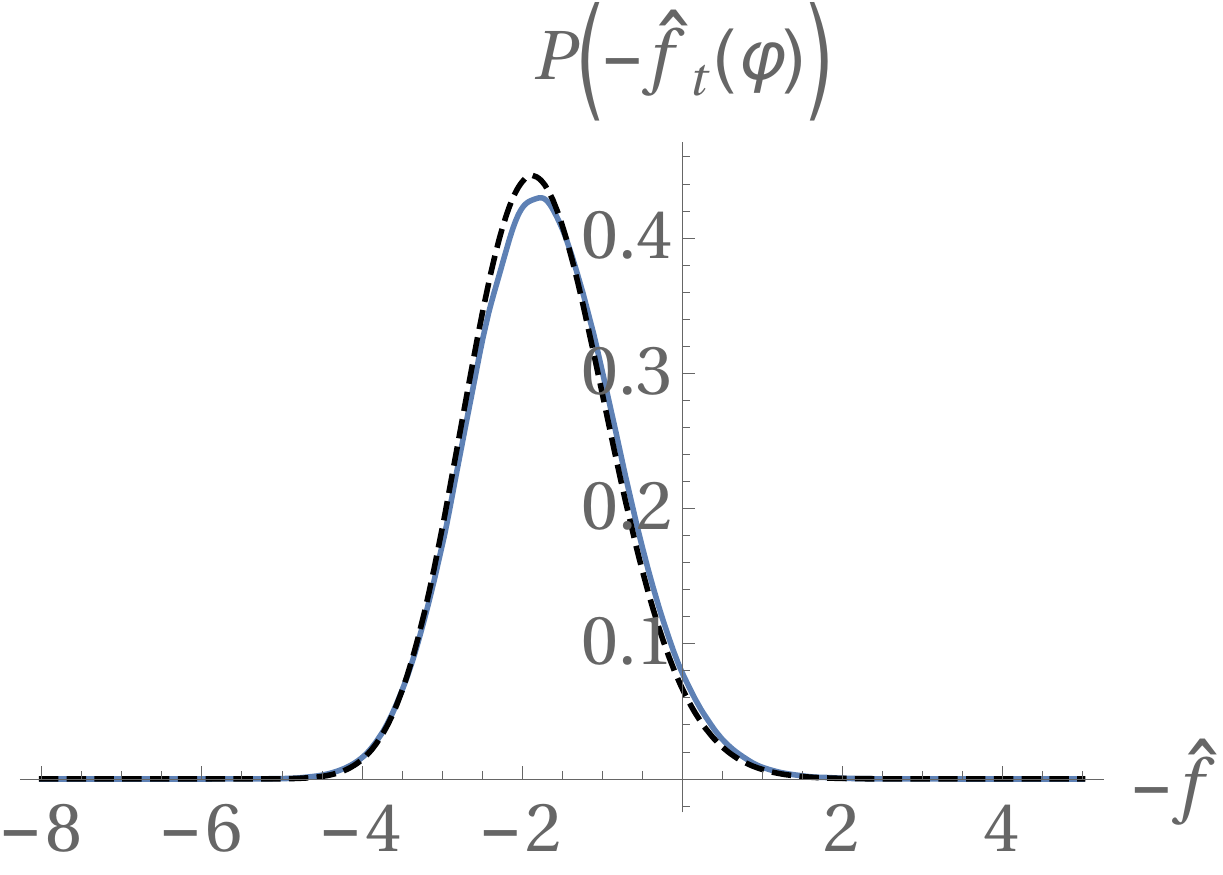} \includegraphics[width=6cm]{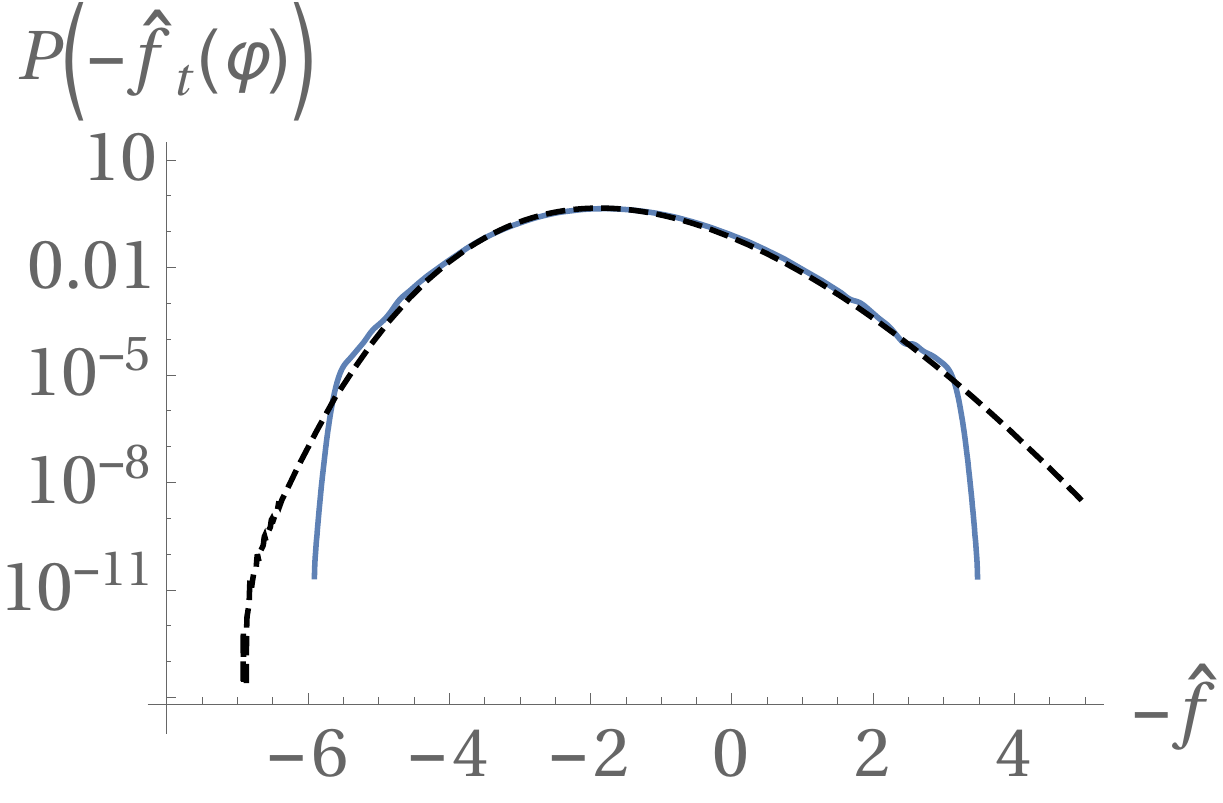} \includegraphics[width=6cm]{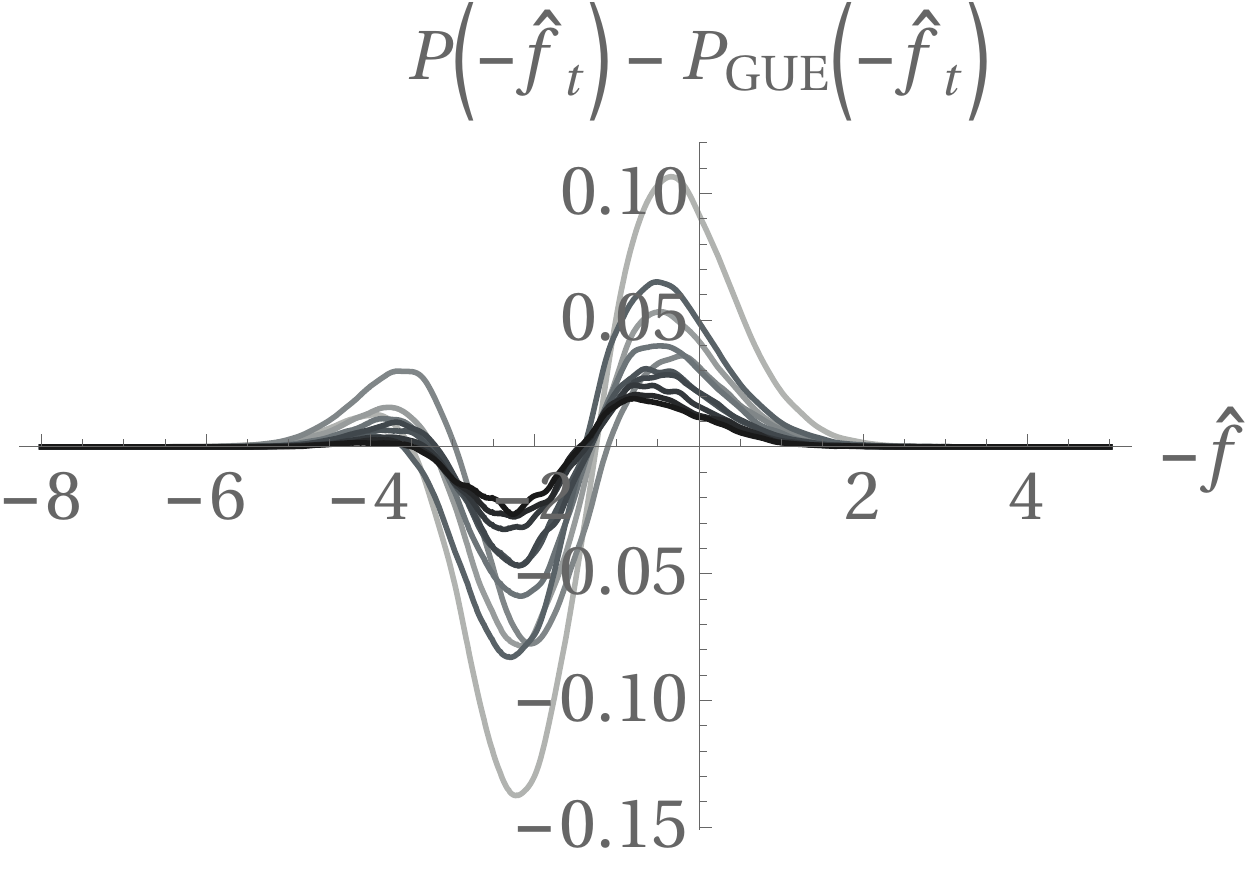}} 
\caption{Left : Blue line: Empirical PDF of $-\hat f_{t=2048}(\varphi_{13})$. Black-dashed line: PDF of a GUE Tracy-Widom distributed RV. Midlle: same as left in a logarithmic scale. Right: Difference between the empirical PDF of $-\hat f_{t_i}(\varphi_{13})$ and the PDF of a GUE Tracy-Widom distributed RV for $i=1,\cdots,10$ (from light-gray to black). There are no fitting parameters.}
\label{fig:ball3}
\end{figure}

We now check the prediction (\ref{asymptoticlim}) for the full distribution of fluctuations that we now recall: it is predicted that the rescaled free-energy
\bea
\hat f_t(\varphi) = - \frac{ - F_t(\varphi) + t c_{\varphi}}{2^{\frac{2}{3}}  \lambda_{\varphi}} \ ,
\eea
converges as $t \to \infty$ to $ - \chi_{GUE}$ where $\chi_{GUE}$ is a RV distributed with the GUE Tracy-Widom distribution. This prediction is checked in Fig.~\ref{fig:ball2} for $t=2048$ and $\varphi = \varphi_{13}$ by directly comparing the numerically obtained distribution of $- \hat f_{t=2918}(\varphi_{13})$ with the PDF of the GUE TW distribution. The agreement is satisfying. We also show in Fig.~\ref{fig:ball3} the convergence of the rescaled free-energy to the GUE Tracy-Widom distribution as a function of $t$.

\section{Conclusion}

In this paper we obtained using the Bethe ansatz, based on the results of \cite{usIBeta}, exact formulas for the statistical properties of the point to point partition sum of the Beta polymer, or equivalently for the PDF of a directed random walk in a Beta distributed random environment. These results complement the results of \cite{BarraquandCorwinBeta} where the half line to point partition sum, or equivalently the CDF of the RWRE, were considerer and a different type of Bethe ansatz approach was used.
 
We first obtained in Sec.~\ref{Sec:BA} an exact formula for the moments of the point to point partition sum (\ref{momentFormula01}). This formula was derived using the Bethe ansatz with periodic boundary conditions on a line of length $L \to \infty$. This procedure highlighted the repulsive nature of the model, an interesting property that distinguishes this model from other exactly solvable models of directed polymers, and which was related to the RWRE interpretation of the model. Based on this formula we obtained in Sec.~\ref{Sec:Diff} Cauchy-type Fredholm determinant formulas for the Laplace transform of the partition sum (\ref{kernel1}), (\ref{kernel2}), (\ref{kernel3}). Using these formulas we obtained asymptotic results for the PDF of the partition sum of the DP in the large time limit in the diffusive regime around the optimal direction of the RWRE. In this regime we showed that the distribution of a rescaled partition sum (\ref{rescaledcalZ}) converges to a Gamma distribution (\ref{ResultOptimalDirection}). This result was then extended to multi-point correlations (\ref{resrescaledcalZ}). We therefore obtained a complete picture of the fluctuations in the diffusive regime around the optimal direction, the spatial region which, in the RWRE language, actually asymptotically contains all the probability. The results in this regime of fluctuations are qualitatively new, as they are different from the KPZ scaling unveiled in BC \cite{BarraquandCorwinBeta}, and they open new perspectives in the
study of RWRE.

We then obtained in Sec.~\ref{Sec:KPZ} an alternative formula for the moments of the partitions sum of the Beta polymer (\ref{BetaWithStrings}) expressed in terms of residues calculations. This formula allowed us to formally perform the asymptotic analysis in other directions (thus in the large deviations regime of the RWRE). There we showed that the fluctuations of the free energy of the polymer scale as $t^{1/3}$ and are distributed with the Tracy-Widom GUE distribution (\ref{asymptoticlim}), a result expected from KPZ universality for point to point directed polymers. Interestingly we found that this result was formally equivalent to the result of \cite{BarraquandCorwinBeta} for the half line to point partition sum and therefore showed that the fluctuations of the PDF and of the CDF in the RWRE picture are identical up to $O(t^{1/3})$ included. Based on our results in both the large deviations regime and the optimal direction regime, we also discussed the very interesting crossover regime between them, and identified the crossover scale
$x \sim t^{3/4}$.

In Sec.~\ref{Sec:nested} we discussed the relations between our approach and the approach of BC \cite{BarraquandCorwinBeta,Barraquand-Corwin-private} and justified a posteriori some results obtained using formal computations in Sec.~\ref{Sec:KPZ}. Finally in Sec.~\ref{Sec:Numerics} we checked our main results using numerical simulations.

\medskip

For future works it would be interesting to obtain a better understanding of the crossover regime between Gamma and Tracy-Widom fluctuations in this model. Another interesting aspect would be to understand the universality of our conclusions. In this spirit this work, together with \cite{BarraquandCorwinBeta}, provides tools and results to analyze an exactly solvable model of RWRE which could serve as a testbed for future works on the subject.

\acknowledgments
We are very grateful to G. Barraquand and I. Corwin for useful remarks and
discussions and for sharing their results with us. We also thank them for useful comments on a preliminary version of this manuscript. We acknowledge hospitality from the KITP in Santa Barbara where part of this work was conducted. This research was supported in part by the National Science Foundation under Grant No. NSF PHY11-25915. We acknowledge support from
PSL grant ANR-10-IDEX-0001-02-PSL.

\appendix
\section{Fredholm determinant formula} \label{app:Fredholm}

In this appendix we show how to obtain the Fredholm determinant formula (\ref{kernel1}) starting from the moment formula (\ref{momentFormula01}). It is based on the following Cauchy determinant identity:
\bea
\prod_{1\leq i < j  \leq n} \frac{(k_i-k_j)^2}{(k_i-k_j)^2 + 1} = {\rm det} \left[ \frac{1}{ i (k_i - k_j )  + 1} \right]_{n \times n } \ .
\eea
This allows to rewrite $g_{t,x}(u)$ as, absorbing a factor $2$ in the determinant,
\bea
g_{t,x}(u) && = \sum_{n=0}^{\infty} \frac{u^n}{n!} \prod_{j=1}^{n}  \int_{\mathbb{R}} \frac{dk_j}{ \pi  } {\rm det} \left[ \frac{1}{ 2 i (k_i - k_j )  + 2} \right]_{n \times n } \prod_{j=1}^{n} 
\frac{(i k_j + \frac{\beta - \alpha}{2} )^t}{(i k_j + \frac{\alpha+\beta}{2})^{1 +x} (i k_j - \frac{\alpha+\beta}{2})^{ 1-x + t} }   \nn \\
&& =  \sum_{\sigma \in S_n} (-1)^{|\sigma|} \sum_{n=0}^{\infty} \frac{u^n}{n!} \prod_{j=1}^{n}  \int_{\mathbb{R}} \frac{dk_j}{ \pi  } \int_{v_j >0} e^{-2 v_j (  i(k_j - k_{\sigma(j)}) + 1 )}
\frac{(i k_j + \frac{\beta - \alpha}{2} )^t}{(i k_j + \frac{\alpha+\beta}{2})^{1 +x} (i k_j - \frac{\alpha+\beta}{2})^{ 1-x + t} }
\eea
where from the first to second line we have rewritten the determinant as a sum over permutations and used the identity $1/z=\int_{v >0} e^{-vz}$, valid for $Re(z)>0$. We then perform the change $\sum_j v_j k_{\sigma(j)}$ and relabel as $\sigma \to \sigma^{-1}$ to obtain
\bea
g_{t,x}(u) && = \sum_{\sigma \in S_n} (-1)^{|\sigma|} \sum_{n=0}^{\infty} \frac{u^n}{n!} \prod_{j=1}^{n}  \int_{\mathbb{R}} \frac{dk_j}{ \pi  } \int_{v_j >0} e^{-2 i k_j (v_j - v_{\sigma(j)}) - (v_j + v_{\sigma(j)} )}
\frac{(i k_j + \frac{\beta - \alpha}{2} )^t}{(i k_j + \frac{\alpha+\beta}{2})^{1 +x} (i k_j - \frac{\alpha+\beta}{2})^{ 1-x + t} } \nn \\
&& = \sum_{n=0}^{\infty} \frac{u^n}{n!} \prod_{j=1}^{n}  \int_{\mathbb{R}} \frac{dk_j}{ \pi  } \int_{v_j >0} {\rm det}\left[ e^{—2 i k_j (v_j - v_i) - (v_j + v_i) } \frac{(i k_j + \frac{\beta - \alpha}{2} )^t}{(i k_j + \frac{\alpha+\beta}{2})^{1 +x} (i k_j - \frac{\alpha+\beta}{2})^{ 1-x + t} }   \right]_{n \times n } \ ,
\eea
ans the last expression is exactly the Fredholm determinant expression associated with the kernel (\ref{kernel1}).

\section{Probability distribution of $Z_t(x)$ for finite polymers lengths.}  \label{app:Proba}

In this section we obtain, at least a formal level, a formula for the PDF of $Z_t(x)$ for arbitrary $(t,x) \in \mathbb{N}^2$. To do so, let us assume that $Z_t(x)$ can be written as the product of $3$ independent `random variables':
\bea \label{app:Proba:Eq1}
Z_t(x) = Z_1 Z_2 Z_3(t,x) \ .
\eea
Where (i) $Z_1$ is distributed as an exponential distribution, i.e. its positive integer moments are $\overline{Z_1^n} = n!$; (ii) $Z_2$ is distributed as a Gamma variable with parameters $ \alpha + \beta$, i.e. its moments are $\overline{Z_2^n} = \Gamma(\alpha+\beta+n)/\Gamma(\alpha+ \beta)$; (iii) $Z_3$ is distributed according to an unknown density (which might not be a PDF) $P(Z_3)$ that we determine self-consistently below. Using the definition of $g_{t,x}(u)$ in (\ref{defg}), we obtain
\bea
g_{t,x}(u) = \sum_{n = 0}^{\infty} \frac{(-u)^n \Gamma(\alpha + \beta)}{ n! \Gamma(\alpha +\beta + n) } \overline{Z_t(x)^n} =  \sum_{n = 0}^{\infty} (-u)^n \overline{ ( Z_3(t,x) )^n} = \int dZ_3 P(Z_3) \frac{1}{1 + u Z_3}    \ .
\eea
Assuming an analytical continuation of $g_{t,x}(u)$ we write,
\bea
g_{t,x}\left(\frac{1}{-v - i \epsilon} \right) =  \int dZ_3 P(Z_3) \frac{-v}{Z_3 - v - i \epsilon}  \ .
\eea 
And we obtain
\bea \label{app:Proba:Eqf}
P(Z_3) = \frac{1}{2 i \pi v} \lim_{\epsilon \to 0^+} \left(g_{t,x}\left(\frac{1}{-v + i \epsilon}\right)-g_{t,x}\left(\frac{1}{-v - i \epsilon}\right) \right) \ .
\eea
Hence, using the Fredholm determinant formulas for $g_{t,x}(u)$ (\ref{kernel1bis}), (\ref{kernel2bis}), (\ref{kernel3bis}) or also (\ref{KernelSmallContour}), one obtains $P(Z_3)$ by computing the limit (\ref{app:Proba:Eqf}). The distribution of $Z_t(x)$ is then obtained using (\ref{app:Proba:Eq1}).

\section{Obtention of formula (\ref{BetaWithStrings}).} \label{app:puzzle}

In this appendix we detail the heuristic reasoning that led to formula (\ref{BetaWithStrings}). It is based on the comparison of two equivalent formulas for the moments of the Strict-Weak polymer that we obtain in the first part of the Appendix. These formulas are obtained using either the fact that the Strict-Weak polymer can be obtained as a limit of the Inverse-Beta polymer or as a limit of the presently studied Beta polymer. More precisely, in the following we will use that,
\bea  \label{limit1}
&& Z_t^{SW}(x) = \lim_{\gamma \to \infty} \gamma^x Z_t^{IB}(x)    \  , \\ \label{limit2}
&&  Z_t^{SW}(x)   = \lim_{\alpha \to \infty} Z_t(x) \ .
\eea
Where here,
\begin{enumerate}
\item{$Z_t^{SW}(x)$ is the point-to-point partition sum of the so-called Strict-Weak polymer with disorder on horizontal weights only. It is a model defined similarly to the Beta polymer with $u= 1$ and $v \sim Gamma(\beta)$. We refer the reader to \cite{StrictWeak,StrictWeak2} for more details on this model.}
\item{$Z_t^{IB}(x)$ is the point-to-point partition sum of the Inverse-Beta polymer with the same conventions as those of \cite{usIBeta} and parameter $\gamma>0$. We refer the reader to \cite{usIBeta} for more details on this model, in particular the proof of the limit (\ref{limit1}).}
\item{$Z_t(x)$ is the point-to-point partition sum of the Beta polymer studied in this paper. We refer the reader to \cite{BarraquandCorwinBeta} for the proof of the limit (\ref{limit2}).}
\end{enumerate}

Using these two limits, we will obtain two equivalent formulas for $ \overline{(Z_t^{SW}(x))^n}$ that will suggest a correspondence between two different types of residues expansion. 

\subsection{Two equivalent formulas for the Strict-Weak polymer}

\subsubsection{From the Inverse-Beta to the Strict-Weak}

In \cite{usIBeta} we obtained a formal expression for the moments of the Strict-Weak polymer with horizontal weights. We recall here the full discussion for completeness. Starting from the moment formula of the Inverse-Beta polymer
\bea \label{momentIBeta}
&& \!\!\!\!\!\!\!\!\!\!\!\!\!\!\!\!\!\!\!\!\!\!\!\!\!\!\!\!\!\!\!\!\! \overline{Z^{IB}_t(x)^n} = \frac{\Gamma(\gamma)}{\Gamma(\gamma-n)} n! \sum_{n_s=1}^n  \frac{1}{n_s!} \sum_{(m_1,..m_{n_s})_n} 
\prod_{j=1}^{n_s}  \int_{-
 \infty}^{+\infty} \frac{dk_j}{2 \pi}
\prod_{1\leq i < j  \leq n_s} \frac{4(k_i-k_j)^2 + (m_i - m_j)^2}{4(k_i-k_j)^2 + (m_i + m_j)^2} \nn \\
&&
\prod_{j=1}^{n_s} \frac{1}{m_j} 
 \left( \frac{  \Gamma(-\frac{m_j}{2} + \frac{ \gamma}{2} - i k_j ) }{  \Gamma(\frac{m_j}{2} + \frac{ \gamma}{2} - i k_j )  } \right)^{1 +x} \left( \frac{   \Gamma(-\frac{m_j}{2} + \frac{ \gamma}{2} +i k_j )}{ \Gamma(\frac{m_j}{2} + \frac{ \gamma}{2} + i k_j ) } \right)^{ 1-x + t} \left( \frac{ \Gamma ( \beta +i k_j+\frac{\gamma }{2}+\frac{m_j}{2})}{\Gamma( \beta +i k_j+\frac{\gamma }{2}-\frac{m_j}{2}   )}  \right)^t \ ,
\eea

We obtain a moment formula for the Strict-Weak polymer with initial condition $Z_t^{SW}(x=0) = \delta_{x,0}$ using the limit $ \overline{(Z_t^{SW}(x))^n} = \lim_{\gamma \to \infty}  \gamma^{n x} \overline{Z_t^{IB}(x)^n} $. The point-wise limit of the integrand cannot be simply taken and we need to first perform the change of variables $k_j \to k_j + i \frac{\gamma}{2}$. We obtain

\bea \label{momentSW1}
&& \overline{(Z_t^{SW}(x))^n} =  \lim_{\gamma \to \infty} \frac{\Gamma(\gamma)}{\Gamma(\gamma-n)} \gamma^{nx} n! \sum_{n_s=1}^n  \frac{1}{n_s!} \sum_{(m_1,..m_{n_s})_n} 
\prod_{j=1}^{n_s}  \int_{L^n} \frac{dk_j}{2 \pi}
\prod_{1\leq i < j  \leq n_s} \frac{4(k_i-k_j)^2 + (m_i - m_j)^2}{4(k_i-k_j)^2 + (m_i + m_j)^2} \nn \\
&&
\prod_{j=1}^{n_s} \frac{1}{m_j}  \left( \frac{  \Gamma(-\frac{m_j}{2} +\gamma - i k_j ) }{  \Gamma(\frac{m_j}{2} + \gamma - i k_j )  } \right)^{1 +x}  \left( \frac{   \Gamma(-\frac{m_j}{2}  +i k_j )}{ \Gamma(\frac{m_j}{2} + i k_j ) } \right)^{ 1-x + t} \left( \frac{ \Gamma ( \beta +i k_j+\frac{m_j}{2})}{\Gamma( \beta +i k_j-\frac{m_j}{2}   )}  \right)^t \ .
\eea
Where $L = -i \frac{\gamma}{2} + \mathbb{R}$. Since the integral over $k_j$ quickly converges as $O(1/k_j^{2 m_j})$, we can now close the different contours of integrations on the upper half plane before taking the limit $\gamma \to \infty$. This leads to:
\bea \label{form1}
&& form1:=\overline{(Z_t^{SW}(x))^n} =  n! \sum_{n_s=1}^n  \frac{1}{n_s!} \sum_{(m_1,..m_{n_s})_n} 
\prod_{j=1}^{n_s}  \int_{\tilde L^n} \frac{dk_j}{2 \pi}
\prod_{1\leq i < j  \leq n_s} \frac{4(k_i-k_j)^2 + (m_i - m_j)^2}{4(k_i-k_j)^2 + (m_i + m_j)^2} \nn \\
&&
\prod_{j=1}^{n_s} \frac{1}{m_j} \left( \frac{   \Gamma(-\frac{m_j}{2}  +i k_j )}{ \Gamma(\frac{m_j}{2} + i k_j ) } \right)^{ 1-x + t} \left( \frac{ \Gamma ( \beta +i k_j+\frac{m_j}{2})}{\Gamma( \beta +i k_j-\frac{m_j}{2}   )}  \right)^t \ ,
\eea
where $\tilde L$ is an horizontal line that stays below all the poles of the integrand. This formula is formal because the resulting integral does not converge,  but one must remember that we have formally already closed the contours of integrations. Computing the integral on $k_i$ thus just amounts at taking the sum over the residues of all the poles of the integrand with a plus sign except those of the type $k_i = k_j - i A$ where $A>0$ (since the contours have been closed on the upper half-plane).

\medskip

\subsubsection{ From the Beta to the Strict-Weak}

We now obtain an alternative formula for $\overline{(Z_t^{SW}(x))^n}$, starting instead from the moment of the Beta polymer:

\bea 
&& \overline{Z_t(x)^n} 
=(-1)^n \frac{\Gamma(\alpha+\beta+n)}{\Gamma(\alpha + \beta  )} 
\prod_{j=1}^{n}  \int_{- \infty}^{+\infty} \frac{dk_j}{2 \pi}
\prod_{1\leq i < j  \leq n} \frac{(k_i-k_j)^2}{(k_i-k_j)^2 + 1} \prod_{j=1}^{n} 
\frac{(i k_j  + \frac{\beta - \alpha}{2} )^t}{( i k_j  +  \frac{\alpha+\beta}{2})^{1 +x} ( i k_j -  \frac{\alpha+\beta}{2})^{ 1-x + t} } \nn
\\
\eea
And we use $ \overline{(Z_t^{SW}(x))^n} = \lim_{\alpha \to \infty}  \alpha^{n x} \overline{Z_t(x)^n} $. As before, the point-wise limit of the integrand cannot be taken and we need to take first $ k_j  = - i (\alpha + \beta)/2 + k'j$ where $k'_j \in i (\alpha + \beta)/2 + \mathbb{R}$, successively close all the contours on the lower-half-plane and finally take the $\alpha \to \infty$ limit. In this way we obtain again a formal formula as
\bea  \label{form2}
&&form2:= \overline{(Z_t^{SW}(x))^n} 
= (-1)^n
\prod_{j=1}^{n}  \int_{L'} \frac{dk_j}{2 \pi}
\prod_{1\leq i < j  \leq n} \frac{(k_i-k_j)^2}{(k_i-k_j)^2 + 1} \prod_{j=1}^{n} 
\frac{( i k_j  + \beta  )^t}{( i k_j )^{ 1-x + t} } \nn
\\
\eea
now $L'$ is a straight line that stays above all the poles of the integrand. This is again true only in a formal sense given by residues calculation (since we have already closed all the contours on the lower half plane): computing the formal integral on $k_i$ in (\ref{form2}) amounts at taking the sum over the residues of all the poles of the integrand with a minus sign except those of the type $k_i = k_j + i $.

\subsection{A formal formula for the moments of the Beta polymer.}

We now give the reasoning that leads to (\ref{BetaWithStrings}). The comparison of the two formal formulas (\ref{form1}) and (\ref{form2}) for the moments of the Strict-Weak polymer shows that there is a correspondence between two types of residues calculation.

Let us now do two remarks:

(i) We first note that in going from (\ref{form2}) to (\ref{form1}) the product terms in the integrand are formally changed as
\bea \label{rule}
i k_j + X \to \frac{\Gamma(ik_j + X +  \frac{m_j}{2})}{\Gamma(ik_j + X -  \frac{m_j}{2})} \ . 
\eea

(ii) We note that the difference between the formula for the moments of the Beta polymer and (\ref{form2}) is only that, for the moments of the Beta polymer, an additional 
\bea \label{additionalTerm}
\prod_{j=1}^n \left( \frac{1}{ik_j +(\alpha + \beta)} \right)^{1+x}
\eea
term is present in the product terms of the integrand. Furthermore, note that it is possible to evaluate the formula for $\overline{(Z_t(x))^n}$ using only the residues taken into account in (\ref{form2}) and not those coming from (\ref{additionalTerm}).

\medskip

The idea beneath (\ref{BetaWithStrings}) is thus just to add in (\ref{form1}) the equivalent of (\ref{additionalTerm}) in the product term of the integrand following the rule (\ref{rule}). That is we add to (\ref{form1}) the term
\bea
\prod_{j=1}^{n_s} \left( \frac{\Gamma \left( ik_j +(\alpha + \beta)  - \frac{m_j}{2} \right)}{ \Gamma \left( ik_j +(\alpha + \beta)  + \frac{m_j}{2} \right)} \right)^{1+x} \ .
\eea

And the residues taken into account in the calculation of (\ref{BetaWithStrings}) are exactly those taken into account in (\ref{form1}). Note that (\ref{BetaWithStrings}) can almost be obtained from the formula for the moments of the Inverse-Beta polymer (\ref{momentIBeta}) by using Euler's reflection formula on the first factor of Gamma function and putting $\gamma = 1 - \alpha - \beta$. In doing so however one obtains an additional unwanted $(-1)^{m}$ term.

\section{Proof of a symmetrization identity} \label{app:ProofSymmmetrization}

In this appendix we prove the formula (\ref{symmetrization}) used in the main text. Let us show that, $\forall n \in \mathbb{N}$, $n \geq 2$ and $\forall(k_1 , \cdots , k_n) \in \mathbb{C}^n$
\bea \label{appproof1}
\frac{1}{n!} \sum_{\sigma \in S_n}  \prod_{1\leq i < j  \leq n} \frac{k_{\sigma(i)}-k_{\sigma(j)}}{k_{\sigma(i)}-k_{\sigma(j)}+ i }   = \prod_{1\leq i < j  \leq n} \frac{(k_i-k_j)^2}{(k_i-k_j)^2 + 1} \ .
\eea 
We proceed by recurrence on $n$ and prove (\ref{appproof1}) $\forall (k_1, \cdots , k_n) \in \mathbb{C}^{n}$ distincts: this is sufficient since (\ref{appproof1}) is trivially true whenever two $k_i$ are equal. The identity (\ref{appproof1}) is trivial for $n=2$. Let us now see how the identity for some $n \geq 2$ imply the identity for $n+1$. Since any permutation $\sigma \in S_{n+1}$ can be written in a unique way for some $m \in \{1 , \cdots , n+1 \}$ as $\sigma = \tau_{m,n+1} \circ \tilde{\sigma}$ with $\tilde{\sigma}$ a permutation of $S_{n+1}$ such that $\tilde{\sigma}(n+1) = n+1$ (thus equivalent to a permutation of $S_{n}$) and $\tau_{m,n+1}$ the transposition $m \leftrightarrow n+1$ we have
\bea
&& \frac{1}{(n+1)!}\sum_{\sigma \in S_{n+1}}  \prod_{1\leq i < j  \leq n+1} \frac{k_{\sigma(i)}-k_{\sigma(j)}}{k_{\sigma(i)}-k_{\sigma(j)}+ i } \nn \\
&&  = \frac{1}{(n+1)!} \sum_{m=1}^{n+1} \sum_{\tilde{\sigma} \in S_{n+1} | \tilde{\sigma}(n+1) = n+1} \prod_{1\leq i < j  \leq n} \frac{k_{\tau_{m,n+1} \circ \tilde \sigma(i)}-k_{\tau_{m,n+1} \circ \tilde \sigma(j)}}{k_{\tau_{m,n+1} \circ \tilde \sigma(i)}-k_{\tau_{m,n+1} \circ \tilde \sigma(j)}+ i }  \prod_{i=1}^{n} \frac{k_{\tau_{m,n+1} \circ \tilde \sigma(i)}-k_{\tau_{m,n+1} \circ \tilde \sigma(n+1)}}{k_{\tau_{m,n+1} \circ \tilde \sigma(i)}-k_{\tau_{m,n+1} \circ \tilde \sigma(n+1)}+ i } \nn \\
&& = \frac{1}{(n+1)!} \sum_{m=1}^{n+1} \sum_{\tilde{\sigma} \in S_{n+1} | \tilde{\sigma}(n+1) = n+1} \prod_{1\leq i < j  \leq n} \frac{k_{\tau_{m,n+1} \circ \tilde \sigma(i)}-k_{\tau_{m,n+1} \circ \tilde \sigma(j)}}{k_{\tau_{m,n+1} \circ \tilde \sigma(i)}-k_{\tau_{m,n+1} \circ \tilde \sigma(j)}+ i }  \prod_{i=1, i \neq m}^{n+1  } \frac{k_{i}-k_{m} }{k_{i}-k_{m}+ i } \nn \\
&& = \frac{n!}{(n+1)!} \sum_{m=1}^{n+1}   \prod_{1\leq i < j  \leq n} \frac{(k_{\tau_{m,n+1}(i)} -k_{\tau_{m,n+1}(j)})^2}{(k_{\tau_{m,n+1}(i)}-k_{\tau_{m,n+1}(j)})^2 + 1} \prod_{i=1, i \neq m}^{n+1  } \frac{k_{i}-k_{m} }{k_{i}-k_{m}+ i }  \\ \label{appproof2}
&& = \frac{1}{n+1} \sum_{m=1}^{n+1}   \prod_{1\leq i < j  \leq n} \frac{(k_{\tau_{m,n+1}(i)} -k_{\tau_{m,n+1}(j)})^2}{(k_{\tau_{m,n+1}(i)}-k_{\tau_{m,n+1}(j)})^2 + 1} \prod_{i=1}^{m-1 } \frac{k_{i}-k_{m} }{k_{i}-k_{m}+ i }  \prod_{i=m+1}^{n+1 } \frac{k_{m} -k_{i}}{k_{m} -k_{i}- i }  \nn \\
&& = \frac{1}{n+1} \sum_{m=1}^{n+1}   \prod_{1\leq i < j  \leq n} \frac{(k_{\tau_{m,n+1}(i)} -k_{\tau_{m,n+1}(j)})^2}{(k_{\tau_{m,n+1}(i)}-k_{\tau_{m,n+1}(j)})^2 + 1} \prod_{i=1}^{m-1 } \frac{(k_{i}-k_{m})^2 }{(k_{i}-k_{m})^2+ 1 }  \prod_{i=m+1}^{n+1 } \frac{(k_{m} -k_{i})^2}{(k_{m} -k_{i})^2+1 }   \nn \\ 
&& \times \prod_{i=1}^{m-1 } \frac{k_{i}-k_{m} -i }{k_{i}-k_{m}}  \prod_{i=m+1}^{n+1 } \frac{k_{m} -k_{i} +i}{k_{m} -k_{i} }  \nn \\
&&  =\frac{1}{n+1} \sum_{m=1}^{n+1} \prod_{1\leq i < j  \leq n+1} \frac{(k_i -k_j)^2}{(k_i-k_j)^2 + 1}  \prod_{i=1}^{m-1 } \frac{k_{i}-k_{m} -i }{k_{i}-k_{m}}  \prod_{i=m+1}^{n +1} \frac{k_{m} -k_{i} +i}{k_{m} -k_{i} }  \nn \\
&& = \left(  \prod_{1\leq i < j  \leq n+1} \frac{(k_i -k_j)^2}{(k_i-k_j)^2 + 1} \right) \psi(k_1 , \cdots , k_{n+1})
\eea
Where in (\ref{appproof2}) we have used the recursion hypothesis and we have defined
\bea \label{appproof3}
\psi(k_1 , \cdots , k_{n+1}) = \frac{1}{n+1} \sum_{m=1}^{n+1}  \prod_{i=1}^{m-1 } \frac{k_{i}-k_{m} -i }{k_{i}-k_{m}}  \prod_{i=m+1}^{n +1} \frac{k_{m} -k_{i} +i}{k_{m} -k_{i} }  \ .
\eea
Note also that in the above derivation we have taken the liberty to write products on empty set: such a product is obviously interpreted as $1$. To show (\ref{appproof1}) it remains to show that $\psi$ is constant and equal to $1$ $\forall (k_1, \cdots , k_{n+1}) \in \mathbb{C}^{n+1}$ distincts. Let us first show that $\psi(k_1, \cdots , k_{n+1})$ is analytic in $k_1$ on $\mathbb{C}$. The only thing to do is to prove that $\forall j \in \{ 2 , \cdots , n+1 \}$, the residue of $\psi(k_1, \cdots , k_n+1)$ at $k_1=k_j$ is actually $0$. It is obvious that the possible pole at $k_1=k_j$ is at most of order $1$. Let us thus write $k_1 = k_j + \delta k$ and focus on the term of order $1/ \delta k$ in (\ref{appproof3}). These terms only come from the term $m=1$ and $m=j$ in the sum over $m$. We obtain
\bea
\psi(k_j +\delta k , \cdots , k_{n+1}) && = \frac{1}{n+1} \left( \prod_{i=2}^{n+1} \frac{k_{j} +\delta k -k_{i} +i}{k_{j} + \delta k  -k_{i} }  +   \frac{k_j +\delta k  - k_j-i}{k_j +\delta k - k_j} \prod_{i=2}^{j-1} \frac{k_i - k_j-i}{k_i - k_j} \prod_{i=j+1}^{n+1} \frac{k_j - k_i+i}{k_j - k_i}\right) + O(1) \nn \\
&& = \frac{i}{n+1} \left(  \prod_{i=2 , i\neq j}^{n+1}  \frac{k_{j} -k_{i} +i}{k_{j}  -k_{i} }   - \prod_{i=2}^{j-1} \frac{k_i - k_j-i}{k_i - k_j} \prod_{i=j+1}^{n+1} \frac{k_j - k_i+i}{k_j - k_i}\right) \frac{1}{\delta k } + O(1) \nn \\
&& = O(1) \ .
\eea
Where again we have used the convention that a product on an empty set is $1$. Hence $\psi(k_j +\delta k , \cdots , k_{n+1}) $ has no poles ans is analytic on $\mathbb{C}$. Finally, it is obvious that
\bea
\lim_{|k_1| \to \infty} \psi(k_1 , \cdots , k_{n+1})  = 1 \ .
\eea
Hence, $\psi(k_1 , \cdots , k_{n+1})$ is analytic and is necessarily bounded. It is thus constant and equal to its limit as $|k_1| \to \infty$ and thus $\psi(k_1 , \cdots , k_{n+1})  = 1$ $\forall (k_1, \cdots , k_{n+1}) \in \mathbb{C}^{n+1}$. This terminates the proof of (\ref{appproof1}).

\end{document}